\documentclass[twocolumn]{emulateapj}
\usepackage{amsmath}
\usepackage{float}
\usepackage{nicefrac}
\usepackage[none]{hyphenat}
\usepackage{enumitem}
\usepackage{afterpage}
\usepackage{lipsum}
\usepackage{moresize}

\def\Rsun{R$_{\odot}$}
\def\Msun{M$_{\odot}$}

\begin{document}

\title{Early-type Eclipsing Binaries  with Intermediate Orbital Periods}

\author{Maxwell Moe\altaffilmark{1} \& Rosanne Di Stefano\altaffilmark{1}}

\altaffiltext{1}{Harvard-Smithsonian Center for Astrophysics, 60 Garden Street, MS-10, Cambridge, MA, 02138, USA; e-mail: mmoe@cfa.harvard.edu}

\begin{abstract}
  
 We analyze 221 eclipsing binaries (EBs) in the Large Magellanic Cloud with B-type main-sequence (MS) primaries ($M_1$ $\approx$ 4\,-\,14 \Msun) and orbital periods $P$ = 20\,-\,50 days that were photometrically monitored by the Optical Gravitational Lensing Experiment.   We utilize our three-stage  automated pipeline to (1) classify all 221 EBs, (2) fit physical models to the light curves of 130 detached well-defined EBs from which unique parameters can be determined, and (3) recover the intrinsic binary statistics by correcting for selection effects.  We uncover two statistically significant trends with age.  First, younger EBs tend to reside in dustier environments with larger photometric extinctions, an empirical relation that can be implemented when modeling stellar populations.  Second, younger EBs generally have large eccentricities.  This demonstrates that massive binaries at moderate orbital periods are born with a Maxwellian ``thermal'' orbital velocity distribution, which indicates they formed via dynamical  interactions.  In addition, the age-eccentricity anticorrelation provides a direct constraint for tidal evolution in highly eccentric binaries containing hot MS stars with radiative envelopes.   The intrinsic fraction of B-type MS stars with stellar companions $q$ $=$ $M_2$/$M_1$ $>$ 0.2 and orbital periods $P$\,=\,20\,-\,50 days is (7\,$\pm$\,2)\%.  We find early-type binaries at $P$~=~20\,-\,50~days are weighted significantly toward small mass ratios $q$~$\approx$~0.2\,-\,0.3, which is different than the results from previous observations of closer binaries with $P$~$<$~20~days.  This indicates that early-type binaries at slightly wider orbital separations have experienced substantially less competitive accretion and coevolution during their formation in the circumbinary disk.    
\end{abstract}


\section{Introduction}

 It has long been understood that the main-sequence (MS) binary star fraction increases with primary mass \citep[][etc.]{Abt1983, Duquennoy1991,Fischer1992,Raghavan2010,Duchene2013}.  Indeed, most massive stars with $M_1$ $>$ 10\,\Msun\ will interact with a stellar companion before they explode as core-collapse supernovae \citep{Sana2012}. Throughout the decades, there have been significant advances in the detection of close and wide companions to massive stars \citep{Wolff1978,Garmany1980,Levato1987,Abt1990,Shatsky2002,Kouwenhoven2007,Sana2012,Rizzuto2013,Kobulnicky2014}.  However, the intrinsic properties of binary companions to early-type primaries, e.g. their eccentricity and mass-ratio distributions, remain elusive at intermediate orbital periods.   The major goal of this work is to help fill this particular portion of the parameter space.  

Eclipsing binaries (EBs) offer a key to the accurate measurement of the binary properties of early-type stars.  Large photometric surveys, such as the third phase of the Optical Gravitational Lensing Experiment (OGLE-III), have discovered tens of thousands of EBs \citep{Graczyk2011,Pietrukowicz2013}.  These populations of EBs are orders of magnitude larger than previous binary samples.  Despite the geometrical selection effects, we can still achieve large sample statistics to reliably infer the intrinsic binary fraction and properties at intermediate orbital periods.  We emphasize that EBs can probe a unique portion of the binary parameter space unavailable to other observational techniques. 

In \citet[][hereafter Paper~I]{Moe2013}, we incorporated OGLE catalogs of EBs in the Large and Small Magellanic Clouds (LMC and SMC, respectively) as well as Hipparcos observations of EBs in the Milky Way.  We  compared the close binary properties ($P$~$<$~20~days) of early-B MS primaries in the three different galaxies.  The Milky Way and SMC EB samples are too small to warrant an analysis of period-dependent binary properties.  The OGLE-III LMC EB catalog \citep{Graczyk2011}, on the other hand, contains $\approx$\,5\,-\,40 times more systems, is relatively complete toward shallow eclipse depths, and includes the full I-band and V-band light curves.

 In \citet[][hereafter Paper~II]{Moe2014}, we developed a three-stage automated pipeline to analyze EBs with short orbital periods in the OGLE-III LMC database. This pipeline (1)~classifies EBs according to their light curve characteristics, (2)~measures the intrinsic physical properties of detached EBs, e.g.  ages and component masses,  based on the observed radii and temperatures, and (3)~recovers the intrinsic binary statistics by correcting for selection effects.

In the present study, we utilize EBs in the OGLE-III LMC database to measure the binary fraction, mass-ratio distribution, and eccentricity distribution of B-type MS stars with intermediate orbital periods $P$ $=$~20\,-\,50~days.  We organize the rest of this paper as follows.  In~\S2, we define our selection criteria for identifying EBs with B-type MS primaries, intermediate orbital periods, and well-defined eclipse parameters.  We next describe an automated procedure we developed to fit detailed physical models to the observed EB light curves, and we present our results for the physical properties of the individual EBs (\S3).  In \S4, we explain the observed trends in the measured EB parameters, paying special attention to the empirical age-extinction and age-eccentricity anticorrelations.  We then perform Monte Carlo simulations to quantify selection effects, and present our results for the corrected binary statistics (\S5).  We summarize our main results and conclusions in \S6.

\section{EB Selection and Classification (Stage~I)}

  In Paper II, we developed a three-stage automated pipeline to fully analyze short-period EBs in the OGLE-III LMC database.  In the present study, we adapt our routine to identify intermediate-period EBs with well-defined light curves (Stage I - this section), measure their physical properties (Stage II - \S3), and correct for selection effects (Stage III - \S5).  EBs with intermediate orbital periods exhibit two major differences that must be considered.  First, the eclipse widths $\Theta_1$ and $\Theta_2$, which are expressed as a fraction of the orbital period~$P$, become narrower with increasing orbital separation.  Given the average number $\langle {\cal N}_{I} \rangle$ $\approx$ 470 of I-band measurements in the OGLE-III LMC survey \citep{Graczyk2011}, the light curves are not sufficiently sampled if either of the eclipse widths $\Theta$~$<$~$\langle {\cal N}_{I} \rangle^{-1}$~$\approx$~0.0021 are too narrow.  EBs with small MS components and long orbital periods $P$~$\gtrsim$~50~days have narrow eclipses $\Theta$ $\lesssim$ 0.002, and are therefore not Nyquist sampled.  This subsampling leads to detection incompleteness, issues with aliasing, and the inability to fully characterize their intrinsic physical properties.   Hence, it is the finite cadence of the OGLE-III observations, {\it not} geometrical selection effects, that limits our present study of EBs to $P$ = 20\,-\,50 days (see also \citealt{Soderhjelm2005}).

Second, the majority of early-type EBs at $P$~$>$~20~days are in eccentric orbits.  We must therefore adapt our physical models to simultaneously fit the eccentricity~$e$ and argument of periastron $\omega$ (\S3).  In addition, it is possible for an eccentric binary to have a certain combination of eccentricity, periastron angle, and inclination that is sufficiently offset from edge-on (e.g.,~$i$~$\lesssim$~86$^{\rm o}$) so that there is only one eclipse per orbit.  Indeed, there are many EBs with single eclipses in the OGLE-III LMC database (see below).  Unfortunately, we cannot measure the physical properties of these systems.  We therefore remove single-eclipse EBs from our well-defined sample, and we account for their removal when we correct for selection effects (see \S5).  In the following, we review our methods from Paper II, where we pay special attention to the nuances of EBs with intermediate orbital periods.

In this study, we select the ${\cal N}_{\rm B}$ $\approx$ 96,000 systems in the OGLE-III LMC catalog \citep{Udalski2008} with mean magnitudes 16.0~$<$~$\langle I \rangle$~$<$~17.6 and  observed colors $-$0.25~$<$~$\langle V - I \rangle$~$<$~0.20.   Given the distance modulus $\mu$~=~18.5 to the LMC \citep{Pietrzynski2013} and typical dust reddenings $E(V-I)$~$\approx$~0.1\,-\,0.3~mag toward hot young stars in the LMC \citep{Zaritsky2004}, these stars have luminosities and surface temperatures that correspond to B-type MS primaries.  From this sample, we analyze the 221 systems that were identified as EBs with orbital periods $P$~=~20\,-\,50~days \citep{Graczyk2011}.  In Table~1, we list the OGLE-III LMC EB identification numbers, observed colors $\langle V - I \rangle$, and numbers of I-band measurements ${\cal N}_I$ for each of these 221 EBs.

As in Paper II, we measure the intrinsic rms scatter in the $I$-band light curve outside of eclipses for each EB.  We then calculate the correction factor $f_{\sigma,I}$ $\ge$ 1.0, i.e. the ratio  between the actual rms scatter and photometric uncertainties reported in the catalog.  For each $I$-band measurement in an EB light curve, we multiply the listed photometric uncertainties by the correction factor $f_{\sigma,I}$ to determine the corrected uncertainties.

 We classify EBs based on an analytic light curve model of two Gaussians with eight total free parameters.  The orbital phase 0 $\le$ $\phi$ $<$ 1 is determined by the time of observation and two model parameters: the orbital period $P$ (in days) and epoch of primary eclipse minimum $t_{\rm o}$ (Julian~date~$-$~2450000).  The six remaining analytic model parameters are the average I-band magnitude outside of eclipses $\langle I \rangle$, primary and secondary eclipse depths $\Delta I_1$ and $\Delta I_2$, primary and secondary eclipse widths $\Theta_1$ and $\Theta_2$, and the phase of secondary eclipse $\Phi_2$. The analytic model of Gaussians is:

\begin{align}
I_{\rm G}(\phi) = \langle I \rangle &+\Delta I_1 \Big[{\rm exp}\Big(\frac{-\phi^2}{2\Theta_1^2}\Big)+
           \rm{exp}\Big(\frac{-(\phi-1)^2}{2\Theta_1^2}\Big)\Big]\nonumber \\
  &+ \Delta I_2\, \rm{exp} \Big(\frac{-(\phi-\Phi_2)^2}{2\Theta_2^2}\Big)
\end{align}
 
We fit this analytic model to each EB I-band light curve.  Specifically, we utilize an automated Levenberg-Marquardt algorithm \citep[ MPFIT,][]{Markwardt2009} to minimize the $\chi^2_{\rm G}$ statistic.  The MPFIT routine provides robust best-fit solutions and measurement uncertainties for the eight analytic model parameters.  Some of the photometric measurements are clear outliers, so we clip up to ${\cal N}_{\rm c}$~$\le$~2 data points per light curve that exceed 4$\sigma$ from the model.  This results in $\nu$~=~${\cal N}_I$\,$-$\,${\cal N}_{\rm c}$\,$-$\,8 degrees of freedom.  For each EB, we report in Table~1 the eight fitted analytic model parameters and the fit statistics.  Excluding the few EBs that exhibit variability or are evolved Roche-lobe filling systems (see below), the goodness-of-fit statistics $\chi^2_{\rm G}$/$\nu$ = 0.87\,-\,1.16 indicate the analytic models can adequately describe the EB light curves.  

  We can measure the physical properties of EBs based solely on the observed photometric light curves (see \S3) only if: (1) the binary components are detached from their Roche lobes, (2) the light curves have two well-defined eclipses, and (3) there is no superimposed variability.  To be considered well-defined, we require that the 1$\sigma$ uncertainties in the measured eclipse depths $\Delta I_1$ and $\Delta I_2$ and eclipse widths  $\Theta_1$ and $\Theta_2$ are $<$20\% their respective values.  These criteria are not satisfied for 91 of the 221 EBs due to a variety of reasons, which we discuss below:

\begin{figure*}[t!]
\centerline{
\includegraphics[trim = 0.1cm 2.1cm 0cm 2.3cm, clip=true,width=7.1in]{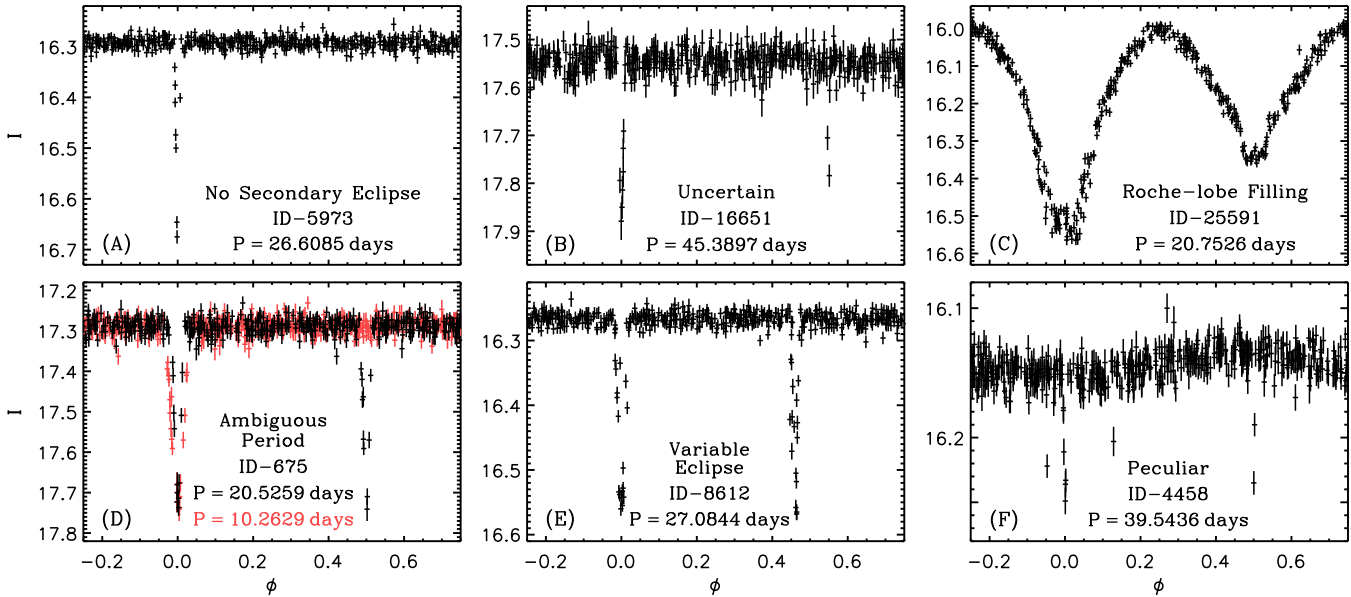}}
\caption{Six examples of the 91 EBs that have properties that are uncertain, variable, peculiar, etc., which leaves 130 EBs in our well-defined sample. Panel A: one of 16 EBs that does not have a visible secondary eclipse. Panel B: one of 32 EBs that show both eclipses but where one of them is too narrow and/or too shallow to be accurately measured.  Panel C: one of the three EBs with wide eclipses that demonstrate one or both components fill their Roche lobes.  Panel D: one of the 23 EBs with an ambiguous orbital period.  Using the catalog orbital periods (black), these systems have nearly identical eclipses separated by almost precisely 50\% in orbital phase.  The more plausible scenario is that these EBs have half the listed orbital periods (red) and therefore exhibit one eclipse per orbit such as the example shown in panel A. Panel E: one of the 15 EBs that exhibit variability.  Three of these systems are intrinsic variables.  The other 12, such as the displayed example, show changes in the eclipse properties most likely caused by orbital motion with a tertiary companion. Panel F: one of the two EBs with peculiar light curve properties.}
\end{figure*}

{\it (A) No Secondary Eclipse}.  For 16 of our EBs, there is no evidence for a secondary eclipse.   These EBs may have secondary eclipses that are too shallow and below the sensitivity of the OGLE-III LMC survey, or have eclipse widths that are too narrow and therefore not detected given the cadence of the observations.  Most likely, the EBs have a certain combination of $e$, $\omega$, and $i$ as discussed above so that there is only one eclipse per orbit.  We list these 16 systems in Category 1 of Table~1, and we show an example in panel A of Fig.~1. 

{\it (B) Uncertain}. For 32 EBs, both eclipses are observed but one or more of their measured properties are uncertain by more than 20\%.   This is because one of the eclipses is too shallow and/or too narrow.  We group these 32 systems in Category 2 of Table~1.  In panel B of Fig.~1, we display an example of a long-period $P$~$\approx$~45~day EB with a secondary eclipse at $\Phi_2$~$\approx$~0.55 that is too narrow to be accurately measured. 

{\it (C) Roche-lobe filling}. Three EBs have wide eclipses such that one or both components of the binary must be filling their Roche lobes.  We list these three systems in Category 3 of Table~1, and we show an example in panel~C of Fig.~1.

{\it (D) Ambiguous Periods}. The orbital periods of 23 of our EBs are ambiguous. 
 These 23 EBs can either have twin components $q$ $\approx$ 1.0 in nearly circular orbits $e$ $\approx$ 0.0 or have half the listed orbital periods and exhibit only one eclipse per very eccentric orbit.  Using the orbital periods listed in the OGLE-III LMC catalog, these EBs have primary and secondary eclipses that are nearly identical and separated by almost precisely 50\% in orbital phase.  Quantitatively, we identify these systems to have values of and uncertainties in eclipse depths, widths, and phases that satisfy:

\begin{subequations}
\begin{align}
  |\Delta I_1 - \Delta I_2| & \le 3\big[(\sigma_{\Delta I_1})^2 + (\sigma_{\Delta I_2})^2\big]^{\nicefrac{1}{2}}  \\
 |\Theta_1 - \Theta_2| & \le 3\big[(\sigma_{\Theta_1})^2 + (\sigma_{\Theta_2})^2\big]^{\nicefrac{1}{2}}  \\
  |\Phi_2 - \nicefrac{1}{2}| & \le 3\sigma_{\Phi_2}
\end{align}
\end{subequations}

\noindent Given the sensitivity of the data, the observed properties imply the 23 systems have large mass ratios $q$~$\gtrsim$~0.9 with extremely small eccentricities $e$~$\lesssim$~0.05 (see~\S3).  However, none of the EBs in our sample have eclipse depths that satisfy Eqn. 2a  ($q$ $\gtrsim$ 0.9) with secondary eclipse phases  3$\sigma_{\Phi_2}$ $<$ $|\Phi_2 - \nicefrac{1}{2}|$ $\le$ 10$\sigma_{\Phi_2}$ ($e$ $\approx$ 0.05\,-\,0.10).  Similarly, there is only one EB that satisfies Eqn. 2c ($e$ $\lesssim$ 0.05) with primary and secondary eclipse depths that are discrepant at the (3\,-\,10)$\sigma$ level ($q$ $\approx$ 0.8\,-\,0.9).  Hence, there are no twin systems in slightly eccentric orbits, and there is only one moderate-mass companion in a nearly circular orbit.   The prevalence of 23 twin systems in nearly circular orbits at these moderate orbital periods is therefore highly unlikely.  If there is indeed an excess of twins in circular orbits relative to twins in eccentric orbits, our study does not include them. We expect only a few of the 23 EBs that appear to be twins in circular orbits to have the listed orbital periods.  The majority of these EBs more likely have orbital periods that are half their listed values, and would therefore exhibit only one eclipse per orbit similar to the systems discussed in (A) above.  In panel~D of Fig.~1, we show one example where we fold the photometric data with the listed orbital period (in black) and the more plausible scenario that the binary has half the catalog orbital period (in red).  We list these 23 EBs in Category~4 of Table~1.  We further motivate the removal of these 23 systems in \S4 when we show the intrinsic frequency of $q$~$>$~0.6 companions with $e$ $<$ 0.2 is relatively sparse.

{\it (E) Superimposed Variability}. Fifteen of the EBs exhibit superimposed variability.  Three of these systems are intrinsic variables, two of which (ID-7651 and ID-22929) were already listed as such in the OGLE-III LMC EB catalog.  The intrinsic variability is readily apparent in the unfolded light curves.  Moreover, the measured intrinsic scatter outside of eclipses is substantially higher than the photometric errors, e.g. $f_{\sigma,I}$ $\approx$ 2.8 for ID-3414.  We note that a few additional systems with $f_{\sigma,I}$ $\approx$ 1.5\,-\,1.9 may exhibit low-amplitude variations $\delta I$ $<$ 0.01 mag, but these variations are sufficiently small so as to not to interfere with the light curve modeling.   We list the three systems that exhibit definitive intrinsic variability in Category 5 of Table~1.  The other 12 EBs exhibit variability in the eclipses themselves, only one of which (ID-17017) was identified as such in the OGLE-III LMC catalog. For these systems, it is possible that more than two bad data points occur near the eclipse.  More likely, these 12 EBs display changes in the eclipse depths and/or eclipse phases during the seven years of observations.  Apsidal motion due to tidal and relativistic effects are negligible on timescales $dt$~$\approx$~7~yrs at these wide orbital separations.  Such evolution in the eclipse parameters are most likely caused by orbital motion with a tertiary component \citep{Rappaport2013}.   We group these 12 EBs in Category~6 of Table~1, and we display an example in panel~E of Fig.~1.

{\it (F) Peculiar}. Finally, two EBs have peculiar light curves.  ID-343 exhibits a pronounced peak in the folded light curve at $\phi$ = 0.8 between eclipses.  This peak may be caused by ellipsoidal modulation in an extremely eccentric orbit.  ID-4458, which is shown in panel~F of Fig.~1, displays a sinusoidal variation between two eclipses of comparable depth.  ID-4458 may contain a hot spot and/or disk, and  is similar to the green systems in the top left corner of Fig.~3  in Paper~II. We list these two systems in Category~7 of Table~1.  

After removing these 91 systems, our well-defined sample contains 130 EBs.  We list these 130 systems in Category~8 of Table~1.  When necessary, we switch the primary and secondary eclipses to ensure $\Delta I_1$ $>$ $\Delta I_2$ in our well-defined sample.  If the epoch of primary eclipse minimum $t_{\rm o}$ substantially changed from the catalog value in order to satisfy this criterion, we place an asterisk next to our value of $t_{\rm o}$ in Table~1. 

 The 130 EBs in our well-defined sample have uncertainties in eclipse depths $\Delta I_1$ and $\Delta I_2$ and eclipse widths $\Theta_1$ and $\Theta_2$ that are $\lesssim$20\% their respective values.  The uncertainties in the Gaussian analytic fit parameters have been used only to determine which EBs have detectable and measurable eclipse properties.  These uncertainties propagate into our Monte Carlo simulations when we calculate the fraction of binaries that produce detectable eclipses (see \S5).  The uncertainties in the Gaussian analytic fit parameters are {\it not} utilized to calculate the uncertainties in the physical properties of the EBs.  Instead, we implement detailed light curve models to measure the values of and uncertainties in the physical model properties, which we now discuss.

\section{Physical Models (Stage II)}

\subsection{Algorithm}
The physical properties of EBs are routinely measured by fitting detailed models to the observational data \citep{Wilson1971,Prsa2005,Devor2006,Kallrath2009}. Normally, spectroscopic radial velocity observations are required to dynamically measure the component masses $M_1$ and $M_2$.  In the modern era of wide-field photometric surveys, the discovery of EBs is quickly outpacing our ability to obtain follow-up spectra \citep{Devor2008,Prsa2011a,Prsa2011b}. For large EB samples, the physical properties must be inferred based solely on the photometric light curves.   MS constraints \citep{Prsa2005,Kallrath2009} and isochrone fitting \citep{Devor2006} have helped ascertain EB properties from the photometric data.  In general, however, these methods lead to large systematic uncertainties and/or solutions that are highly degenerate.  

  In Paper II, we developed a technique that uniquely and accurately characterizes the intrinsic physical properties of {\it detached EBs with known distances} using only the photometric data. The distances to EBs in the LMC are known. In fact, we have already utilized the observed magnitudes $\langle I \rangle$ and colors $\langle V - I \rangle$ to select EBs with B-type MS primaries.  For detached EBs with MS primaries, both components are effectively evolving along their respective single-star evolutionary sequences.  The photospheric properties of the stellar components, e.g. effective temperatures $T_1$ and $T_2$, radii $R_1$ and $R_2$, and luminosities $L_1$ and $L_2$, therefore depend entirely on the age~$\tau$ and component masses $M_1$ and $M_2$.   The systematic uncertainties in the evolutionary tracks are relatively small, e.g. $\approx$15\% uncertainties in the masses  and $\approx$30\% uncertainties in the ages (see Paper~II and \S3.3 for further justification and a full assessment of the uncertainties).  We can therefore measure the component masses $M_1$ and $M_2$ and ages $\tau$ of detached EBs with known distances based solely on the observed light curve features (Fig.~2).

\begin{figure}[t!]
\centerline{
\includegraphics[trim = 1.3cm 5.2cm 2.0cm 7.2cm, clip=true,width=3.4in]{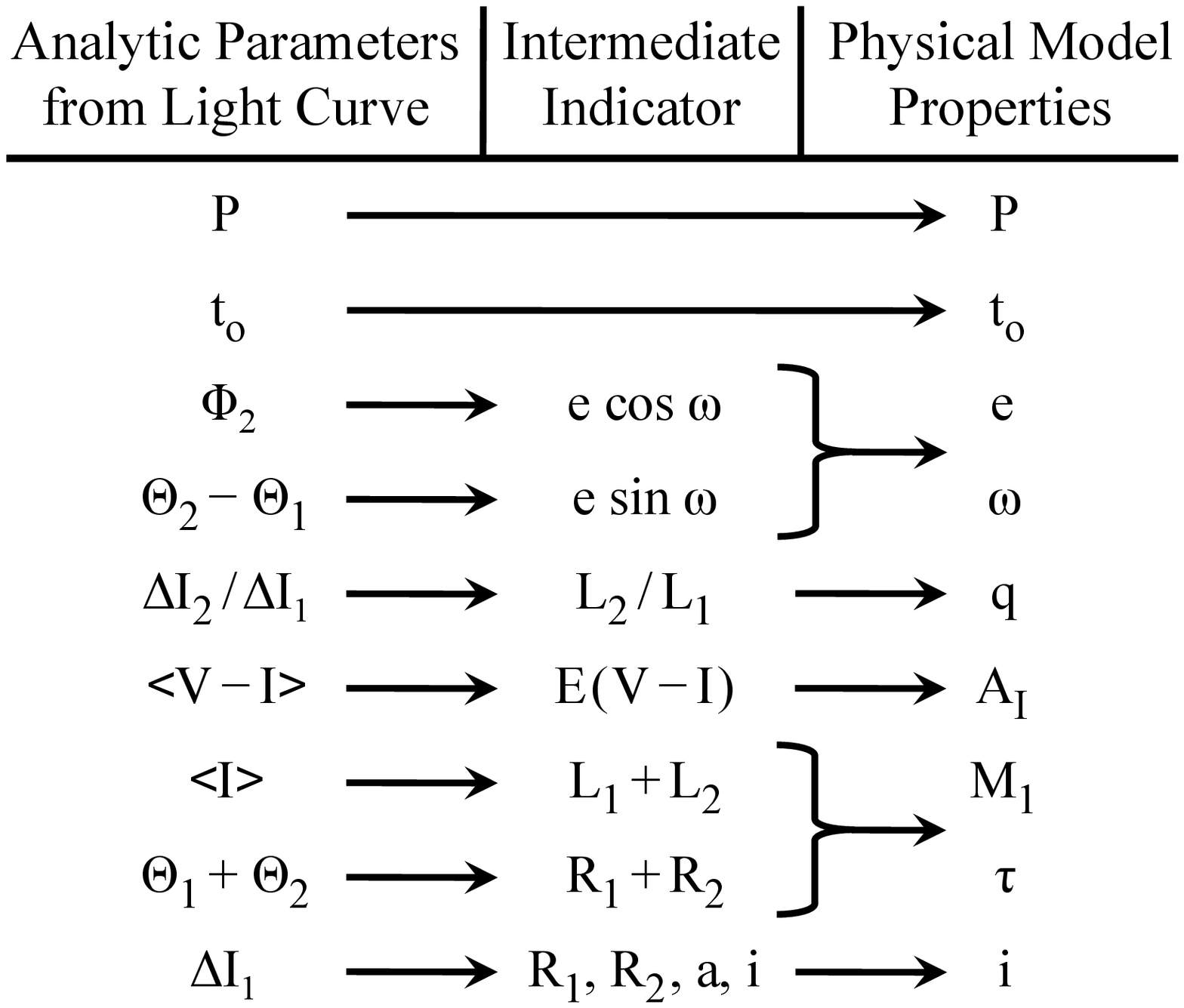}}
\caption{For detached EBs with known distances and MS primaries, the nine observed photometric light curve parameters (left) provide unique solutions for the nine independent intrinsic physical properties of the system (right). Other properties of the binary, e.g. stellar radii $R_1$ and $R_2$ and luminosities $L_1$ and $L_2$, are utilized as intermediate indicators (middle), but depend entirely on the independent properties  $M_1$, $q$ = $M_2$/$M_1$, and $\tau$ according to stellar evolutionary tracks. }
\end{figure}

In our physical models, detached EBs with B-type MS primaries can be uniquely described by nine independent properties.  These nine physical model parameters are the orbital period~$P$, epoch of primary eclipse minimum~$t_{\rm o}$, primary mass~$M_1$, secondary mass~$M_2$, age~$\tau$, inclination~$i$, eccentricity~$e$, argument of periastron~$\omega$, and I-band dust extinction~$A_I$.  Given the age $\tau$ and component masses $M_1$ and $M_2$ of the binary, we interpolate the radii $R_1$ and $R_2$, surface gravities $g_1$ and $g_2$, effective temperatures $T_1$ and $T_2$, and luminosities $L_1$ and $L_2$ from pre-MS and MS stellar evolutionary tracks with metallicity $Z$ = 0.008 \citep{Tognelli2011, Bertelli2009}.  We then use the LMC distance modulus $\mu$~=~18.5 \citep{Pietrzynski2013}, dust reddening law $E(V-I)$~=~0.7$A_I$ \citep{Cardelli1989,Fitzpatrick1999,Ngeow2005}, and temperature-dependent color indices and bolometric corrections \citep{Pecaut2013} to transform the intrinsic properties of the binary into observed magnitudes and colors.   Our physical model parameter space ($M_1$, $M_2$, $\tau$, etc.) of EBs with detached configurations, pre-MS/MS evolutionary constraints, and known distances is quite different than the typical EB parameter space ($T_2$/$T_1$, ($R_1$+$R_2$)/$a$, etc.) where the distances and evolutionary status of the components are unknown \citep[e.g.][]{Devor2006, Prsa2011a}.

Using the physical properties of a binary,  e.g., $P$, $M_1$, $M_2$, $R_1$, $R_2$, $T_1$, $T_2$, $e$, $\omega$, etc., we synthesize photometric light curves with the EB modeling package \textsc{Nightfall}\footnote{http://www.hs.uni-hamburg.de/DE/Ins/Per/Wichmann/\\Nightfall.html}. We use the same \textsc{Nightfall} model options adopted in Paper~II, e.g. a square-root limb darkening law, default gravity darkening coefficients, model atmospheres, etc., except for three notable distinctions.  First, we do not assume circular orbits for our EBs at longer orbital periods, but instead solve for both the eccentricity $e$ and periastron angle $\omega$.  Second, we set the albedo of the secondary to $A_2$ = 0.7 and implement one iteration of reflection effects.  Considering reflection effects are minuscule for our wider EBs in this study, different treatments of reflection have negligible effects on the synthesized light curves.  Finally, we simulate an EB light curve at 1,000 uniformly-spaced discrete orbital phases to ensure narrow eclipses are sufficiently sampled.

Most of our EBs with intermediate orbital periods have eccentric orbits and narrow eclipses. \textsc{Nightfall} and all other EB software packages that account for tidal effects are computationally expensive for eccentric binaries.  This is because the three-dimensional photospheric surfaces of the stars need to be recalculated at each of the 1,000 discretely sampled orbital phases.  We therefore adapt our algorithm from Paper II to guarantee fast, automated convergence.  Namely, we choose initial values for our nine physical model properties that are sufficiently close to the true values to ensure $\chi^2$ minimization converges quickly to the global solution. The major goal of our algorithm is to synthesize light curves with \textsc{Nightfall} as few times as possible.  Our routine can easily be adapted for any population of detached EBs with known distances, and can be used in combination with any EB light curve modeling software.  

We decompose our algorithm into three steps.\footnote{The three steps discussed in this section are not to be confused with the three full stages of our automated pipeline,  which classifies EBs (Stage~I~-~\S2), fits physical models to the light curves (Stage~II~-~\S3), and corrects for selection effects (Stage~III~-~\S5).  The three steps regarding physical models are all included in Stage~II.} In Step~1, we select initial values for our nine physical model properties based on the observed light curve features quantified in \S2.  In Step 2, we make small adjustments in the physical model properties until the analytic model parameters of the synthesized light curve matches those of the observed light curve. In Step 3, we utilize a Levenberg-Marquardt technique, as done in Paper II, to minimize the $\chi^2$ statistic between the observed and simulated light curves.  We elaborate on these three steps below.  To help illustrate this procedure, we display in Fig.~3 the light curve of an example EB, ID-2142, and the solutions at the end of each of these three steps.

{\bf Step 1.} We use the eight analytic model parameters ($P$, $t_{\rm o}$, $\langle I \rangle$, $\Delta I_1$, $\Theta_1$, $\Phi_2$, $\Delta I_2$, $\Theta_2$) and observed color $\langle V - I \rangle$ from Table 1 to estimate initial solutions for the nine physical model properties.  In Fig.~2, we show how the nine observed light curve features can be used to approximate the nine physical properties of the binary. We select the physical parameters $P$ and $t_{\rm o}$ to match the analytic model values.  We then estimate $e$ and $\omega$ according to the observed phase of the secondary eclipse and the difference in eclipse widths \citep[][Eqn. 3.1.24 and 3.1.26, see our Fig.~2]{Kallrath2009}:

\begin{subequations}
\begin{align}
  e\,{\rm cos}\,\omega & \approx \frac{\pi}{2}(\Phi_2 - \nicefrac{1}{2}) \\
  e\,{\rm sin}\,\omega & \approx \frac{\Theta_2 - \Theta_1}{\Theta_2+ \Theta_1} 
\end{align}
\end{subequations}

\noindent In this study, $\omega$ = 90$^{\rm o}$ if periastron coincides with the observed primary eclipse.  For our example ID-2142, $\Theta_1$~$\approx$~$\Theta_2$ and $\Phi_2$~$\approx$~0.16, indicating $\omega$ $\approx$ 180$^{\rm o}$ and $e$~$\approx$~0.5\,-\,0.6.

The intrinsic colors of B-type MS stars span a narrow interval $-$0.3 $\lesssim$ $\langle V - I \rangle_{\rm o}$ $\lesssim$ $-$0.1 \citep{Pecaut2013}.  We therefore initially assume the intrinsic color of an EB to be:

\begin{equation}
 \langle V - I \rangle_{\rm o} \approx -0.22 + 0.08(\langle I \rangle -17)
\end{equation}

\noindent where we have accounted for the fact that more luminous B-type MS stars tend to be more massive, hotter, and bluer. The dust extinction $A_I$ is simply estimated from the observed color $\langle V - I \rangle$ and our adopted dust reddening law $E(V-I)$ = $\langle V - I \rangle$ $-$ $\langle V - I \rangle_{\rm o}$ = 0.7$A_I$ (see Fig.~2). 

We then use the eclipse depths $\Delta I_1$ and $\Delta I_2$ to approximate the mass ratio $q$ = $M_2$/$M_1$.  For a MS\,+\,MS binary in a circular orbit, the ratio of eclipse depths $\Delta I_2$/$\Delta I_1$ provides an accurate indicator of the luminosity contrast $L_2$/$L_1$ (Fig.~2).  This luminosity contrast can then be used to infer the mass ratio $q$ according to a MS mass-luminosity relation (see Fig.~3 in Paper I).  For eccentric orbits, however, the eclipse depth ratio can be modified because the projected distances during primary and secondary eclipses can be different.  Nonetheless, deeper eclipses still suggest larger mass ratios.  For example, $\Delta I_2$~$>$~0.4~mag requires $q$ $>$ 0.7, regardless of the eccentricity or whether the secondary is a MS or pre-MS star.  We use a linear combination of these methods to estimate the mass ratio:

\begin{equation}
 q \approx 0.6\Delta I_1 + 0.5\Delta I_2 + 0.5 \frac{\Delta I_2}{\Delta I_1}
\end{equation} 

\noindent where the eclipse depths are in magnitudes.  
 
We next use the observed mean magnitude $\langle I \rangle$ and sum of eclipse widths $\Theta_1 + \Theta_2$ to simultaneously measure the primary mass $M_1$ and age $\tau$. Assuming non-grazing eclipses and standard limb darkening coefficients, the sum of eclipse widths $\Theta_1 + \Theta_2$ directly provides the relative sum of the radii ($R_1$+$R_2$)/$a$.  Our EBs occupy a narrow range of magnitudes 16.0 $<$ $\langle I \rangle$ $<$ 17.6 and therefore span a small interval of total masses $M$~=~$M_1$\,$+$\,$M_2$.  The orbital separation $a$ $\propto$ $P^{\nicefrac{2}{3}} M^{\nicefrac{1}{3}}$ therefore derives mainly from the known period $P$.  We can now use $\Theta_1 + \Theta_2$ and $P$ to determine $R_1$+$R_2$.   For EBs with B-type MS primaries, we find the following approximation:

\begin{equation}
  R_1 + R_2 \approx 7\,{\rm R}_{\odot} \frac{\Theta_1+\Theta_2}{0.01}\Big(\frac{P}{30\,{\rm days}}\Big)^{\nicefrac{2}{3}}
\end{equation}

\noindent Given our estimates for $q$ and $A_I$ above, we interpolate the stellar evolutionary tracks to determine the primary mass $M_1$ and age $\tau$ that reproduce the sum of the radii $R_1 + R_2$ according to Eqn.~6 and the observed combined magnitude $\langle I \rangle$.  Although  $\langle I \rangle$ and $\Theta_1 + \Theta_2$ both depend on $M_1$ and $\tau$, they are sufficiently non-degenerate so that we can calculate a unique solution.  Namely, the primary mass $M_1$ largely dictates the luminosity and therefore the observed magnitude $\langle I \rangle$, while the age $\tau$ primarily determines the radii $R_1 + R_2$ and therefore the observed eclipse widths $\Theta_1 + \Theta_2$ (see Fig.~2).

Finally, we select an inclination $i$ that approximately reproduces the observed primary eclipse depth $\Delta I_1$ (Fig.~2).  From our estimates of $M_1$, $M_2$ = $q$\,$M_1$, and $\tau$, we interpolate the radii $R_1$ and $R_2$ and effective temperatures $T_1$ and $T_2$ from stellar evolutionary tracks.  In this step only, we ignore limb darkening and colors of the two stars, and instead assume the stars are uniformly illuminated grey disks (see Paper I). We assume the surface brightnesses of the disks are proportional to the stellar temperatures, i.e. the Rayleigh-Jeans law, because we are observing at relatively long wavelengths in the near-infrared I-band.  Using these approximations, we calculate the eclipsed area $A_{\rm o}$ of the primary at the time of primary eclipse $t_{\rm o}$ based on the observed primary eclipse depth:

\begin{equation}
 \Delta I_1 \approx -2.5\,{\rm log}\,\Big(1 - \frac{A_{\rm o} T_1}{\pi (R_1^2 T_1 + R_2^2 T_2)}\Big)
\end{equation}

\noindent Given the eclipsed area $A_{\rm o}$ and stellar radii $R_1$ and $R_2$, we then determine the projected distance $d_{\rm o}$ between the two stars at $t_{\rm o}$. The actual physical separation at primary eclipse is already known via \citep[][Eqn. 3.1.36 evaluated at geometric phase $\theta$ = 0$^{\rm o}$]{Kallrath2009}:

\begin{equation}
 r_{\rm o} = a\,\frac{1-e^2}{1+e\,{\rm sin}\,\omega}
\end{equation}

\noindent where $a$ derives from our estimates of $M_1$, $M_2$, and $P$ according to Kepler's third law, and  $e$ and $\omega$ are approximated from Eq. 3.  Hence, the inclination simply derives from cos\,$i$~=~$d_{\rm o}$/$r_{\rm o}$.   We limit our initial approximation of the inclination to the interval $i$~=~86.5$^{\rm o}$\,-\,89.5$^{\rm o}$.

  We now have initial estimates for the nine physical model properties.  We emphasize that Eqns. 3\,-\,7 are simple approximations, and that the true values of $e$, $\omega$, $A_I$, $M_1$, $M_2$, $\tau$, and $i$ may substantially differ from the initial values estimated here. We simply use these estimates as initial parameters in our fitting routine in order to minimize the number of iterations and accelarate convergence toward the final solution  (see below and \S3.2).

In the top panel of Fig.~3, we compare the I-band and V-band light curves of ID-2142 to a simulated \textsc{Nightfall} model using the values of the nine physical model properties at the end of Step 1.  The model matches key features of the observed light curve, but there are three noticeable differences.  First, the simulated phase of the secondary eclipse does not match the observations; recall that Eqn. 3 is an approximation.   Second, the simulated color is bluer than the observed $\langle V - I \rangle$, indicating we underestimated the dust reddening $A_I$ in our initial step.  Finally, the simulated eclipses are slightly deeper than the observed because the more accurate \textsc{Nightfall} model accounts for limb darkening and color effects.  This suggests the actual inclination is smaller and/or the mass ratio is slightly different.  We correct for these visible discrepancies in the following step.

{\bf Step 2.} Using the nine physical properties from Step~1, we synthesize an I-band light curve with \textsc{Nightfall}.   We then fit the simple analytic model of Gaussians (Eqn. 1) as done in \S2  to the simulated \textsc{Nightfall} light curve.  In this manner, we measure the analytic parameters of the \textsc{Nightfall} model, e.g. $\Delta I_{1, {\rm mod}}$, $\Theta_{1, {\rm mod}}$,  $\Phi_{2, {\rm mod}}$, etc. 

\begin{figure}[t!]
\centerline{
\includegraphics[trim = 4.5cm 1.1cm 5.6cm 0.2cm, clip=true,width=3.4in]{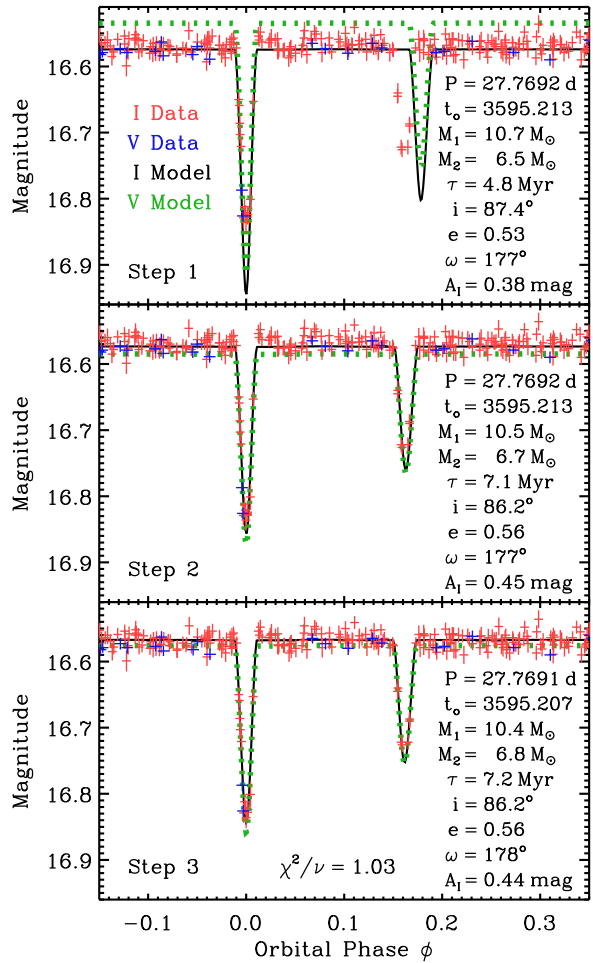}}
\caption{Observed and model light curves for ID-2142.  We compare the I-band (red) and V-band (blue) OGLE-III LMC data to the synthesized I-band (black) and V-band (dotted green) light curves at the end of the three steps in our automated procedure. We display only the interval $-$0.15 $<$ $\phi$ $<$ 0.35 that encompass the eclipses. Note how the physical model parameters vary only slightly between our initial estimate and final solution. }
\end{figure}

We adjust  the properties in our physical models according to the differences between the simulated and observed analytic model parameters. The adjustments are motivated as follows.  If the modeled eclipse widths $\Theta_{1, {\rm mod}}$\,$+$\,$\Theta_{2, {\rm mod}}$ are wider than the observed $\Theta_1$\,$+$\,$\Theta_2$, we select a slightly younger age~$\tau$ (and vice versa).  We increase the dust extinction $A_I$ if the simulated color $\langle V - I \rangle_{\rm mod}$ is too blue. If the modeled primary eclipse $\Delta I_{1,{\rm mod}}$ $>$ $\Delta I_1$ is too deep while the modeled secondary eclipse $\Delta I_{2,{\rm mod}}$ $\le$ $\Delta I_2$ matches observations or is too shallow, we increase the mass ratio $q$ and decrease the inclination $i$.  However, if both simulated eclipses are too deep (or both too shallow), we only decrease (increase) the inclination $i$. Finally, we adjust  $e$ according to the position of and differences in the secondary eclipse phases $\Phi_2$ and  $\Phi_{2, {\rm mod}}$.  In this step, we fix $P$, $t_{\rm o}$, and $\omega$ to the values determined in Step 1.  Finally, we interpolate $M_1$ from the stellar evolutionary tracks based on the observed mean magnitude $\langle I \rangle$ and the revised values for $\tau$, $q$, and $A_I$.  

  When adjusting our physical model properties, we choose step sizes that scale with the differences between the observed and simulated analytic model parameters.  After making these adjustments, we synthesize another I-band light curve with \textsc{Nightfall}.  We iterate this step until all the analytic model parameters of the simulated and observed light curves match within a small tolerance level.  In the middle panel of Fig.~3, we show our solution for ID-2142 at the end of Step 2 after five iterations. We therefore required only six \textsc{Nightfall} light curve simulations during this middle step.  

{\bf Step 3.} This final step is essentially the procedure outlined in Paper II.  We calculate the photometric correction factors $f_{\sigma, I}$ and $f_{\sigma, V}$ in both bands. Starting with initial model properties determined at the end of Step 2, we utilize a Levenberg-Marquardt technique \citep[MPFIT,][]{Markwardt2009} to minimize the $\chi^2$ statistic between the simulated and observed light curves. The Levenberg-Marquardt MPFIT algorithm operates by independently varying each of the nine physical model properties from the previous solution.  The routine then measures the resulting deviations between the data and models, and then calculates a new solution. This step therefore requires ten \textsc{Nightfall} simulations per iteration.  As in Paper~II, we simultaneously fit the I-band and V-band light curves.  We clip up to ${\cal N}_{{\rm c}, I}$ $+$ ${\cal N}_{{\rm c}, V}$ $\le$ 3 data points that exceed 4$\sigma$ from the best-fit model.  This results in $\nu$~=~${\cal N}_I$\,$+$\,${\cal N}_V$\,$-$\,${\cal N}_{{\rm c}, I}$\,$-$\,${\cal N}_{{\rm c}, V}$\,$-$\,9 degrees of freedom. 

In the bottom panel of Fig.~3, we display our final solution for ID-2142 after four iterations of the Levenberg-Marquardt MPFIT routine.  We therefore simulated light curves with \textsc{Nightfall} a total of 40 times in Step~3.  The physical model properties changed only slightly during this final step.  In fact, for ID-2142, the variations were all within the uncertainties of the physical model parameters.  We emphasize that Steps 1 and 2 were crucial in guarenteeing rapid convergence toward the final solution in Step 3.  Without them, this last step would have required many additional iterations or may have converged to a local minimum.

We utilize this automated procedure for all 130 detached EBs in our well-defined sample.  We present our fitted model parameters, physical properties, and fit statistics for these systems in Table~2. For MS binaries in circular orbits, the deeper primary eclipse $\Delta I_1$ at  time $t_{\rm o}$ always corresponds to the smaller, cooler, less massive secondary passing in front of the larger, hotter, more massive primary.  For eccentric orbits, however, the situation can be reversed depending on the combination of $e$, $\omega$, and $i$.  Indeed, for 18 EBs in our well-defined sample, we determined solutions such that the less massive component was eclipsed at time $t_{\rm o}$.   To avoid confusion in nomenclature, we list properties in Table 2 according to the primary ``p'' and secondary ``s'' eclipse features.  Namely, $M_{\rm p}$, $R_{\rm p}$, and $T_{\rm p}$ correspond to the component that was eclipsed at the epoch of primary eclipse $t_{\rm o}$, and $M_{\rm s}$, $R_{\rm s}$, and $T_{\rm s}$ correspond to the component that was eclipsed at the secondary eclipse phase $\Phi_2$.  In the text, we refer to primary mass $M_1$~=~max\{$M_{\rm p}$,\,$M_{\rm s}$\}, secondary mass $M_2$~=~min\{$M_{\rm p}$,\,$M_{\rm s}$\}, mass ratio $q$ = $M_2$/$M_1$, etc.

We measure primary masses $M_1$~=~3.6\,-\,13.9\,\Msun, which nearly encompasses the full mass range of B-type MS stars.  We determine mass ratios across the interval $q$~=~0.20\,-\,1.00, which confirms the OGLE-III observations are sensitive to EBs with low-mass companions.  Our measured dust extinctions cover $A_I$~=~0.10\,-\,0.58~mag, which is consistent with the range of extinctions found in Paper~II.  Finally, we determine ages $\tau$~=~0.5\,-\,190~Myr that span more than two orders of magnitude.  We further discuss the EB physical properties, and their interrelations, in \S4.  

 Eleven of the 130 EBs have modest fit statistics $\chi^2$/$\nu$~=~1.10\,-\,1.14, i.e. probabilities to exceed $\chi^2$ of $p$~$\approx$~0.01\,-\,0.05 given $\nu$~$\approx$~530 degrees of freedom.  Seven of these EBs are extremely young with estimated ages $\tau$~$\lesssim$~0.8~Myr  (IDs 5153, 7560, 10422, 13418, 16711, 22691, and 22764).  The components in these EBs have small radii, as demonstrated by their narrow eclipses (Eqn. 6), and are therefore consistent with the zero-age MS.  The systematic uncertainties in the stellar evolutionary tracks are larger at these young ages, especially considering some of the secondaries may still be pre-MS stars (see Paper II).  Three of the 11 EBs with modest fit statistics have primaries at the tip of the MS (IDs 91, 20746, and 21518), as indicated by their wide eclipses.  Again, the stellar evolutionary tracks are uncertain at the tip of the MS just prior to the rapid expansion toward the giant phase.  The one last EB with a poor physical model fit (ID-17569) has $\chi^2/\nu$~=~1.11 and $p$~$\approx$~0.02.  Considering our large sample of 130 EBs, we naturally expect 1\,-\,3 of these EBs with modest fit statistics.  The remaining 119 EBs in our well-defined sample have good fit statistics 0.93 $<$ $\chi^2$/$\nu$ $<$ 1.09.  This is testament that the nine independent physical model properties can adequately describe detached EBs with known distances and MS primaries.  

\subsection{Comparison between Initial Estimates\\and Final Solutions}

In the following, we compare the initial estimates for $e$, $q$, and $R_1$+$R_2$ in Step~1 according to Eqns.~3, 5, and 6, respectively, to the final solutions in Step 3 from fitting detailed \textsc{Nightfall} light curve models to the data.  We can then address the systematic uncertainties in our initial estimates and further justify the mapping between the basic EB light curve parameters to the physical model properties.

For the 130 EBs in our well-defined sample, we compare the initial values of the eccentricities~$e$ determined from the secondary eclipse phases $\Phi_2$ and eclipse widths $\Theta_1$ and $\Theta_2$ (Eqn.~3) to the final \textsc{Nightfall} solutions (top panel of Fig.~4).  The initial estimates agree quite well with the true final values.  The rms scatter between the two is only $\delta e$ = 0.03.  This validates that Eqn.~3 is more than sufficient for starting purposes in our fittng routine.  We note that the few systems that change by more than $\Delta e$~$>$~0.07 between Steps 1 and 3 have narrow, poorly sampled eclipses so that it is more difficult to precisely measure $\Theta_1$ and $\Theta_2$.

 Similarly, in the bottom panel of Fig.~4, we compare the initial estimates of the mass ratios $q$ determined from the eclipse depths $\Delta I_1$ and $\Delta I_2$ (Eqn.~5) to our final values obtained from \textsc{Nightfall} light curve fittings and $\chi^2$ minimizations.  Although the population as a whole shows rough agreement between the solutions at the ends of Steps 1 and 3, individual systems can substantially deviate from the initial estimates.  For example, an EB with an initial estimate of $q$ $\approx$ 0.6 may actually have a mass ratio anywhere in the interval $q$ = 0.3\,-\,1.0.   The rms deviation between the initial and final solutions is $\delta q$~=~0.12, or $\delta q$\,/\,$q$~$\approx$~20\% the respective values. If we had randomly chosen mass ratios $q$ in Step~1 while keeping the other initial estimates unchanged, the \textsc{Nightfall} light curve solutions in Step 3 would still converge to the same final values.  We simply find that by adopting Eqn.~5 in Step 1 to provide initial estimates for $q$, the number of iterations in Steps 2 and 3 are dramatically reduced.

\begin{figure}[t!]
\centerline{
\includegraphics[trim = 3.7cm 0.2cm 4.9cm 0.1cm, clip=true,width=3.7in]{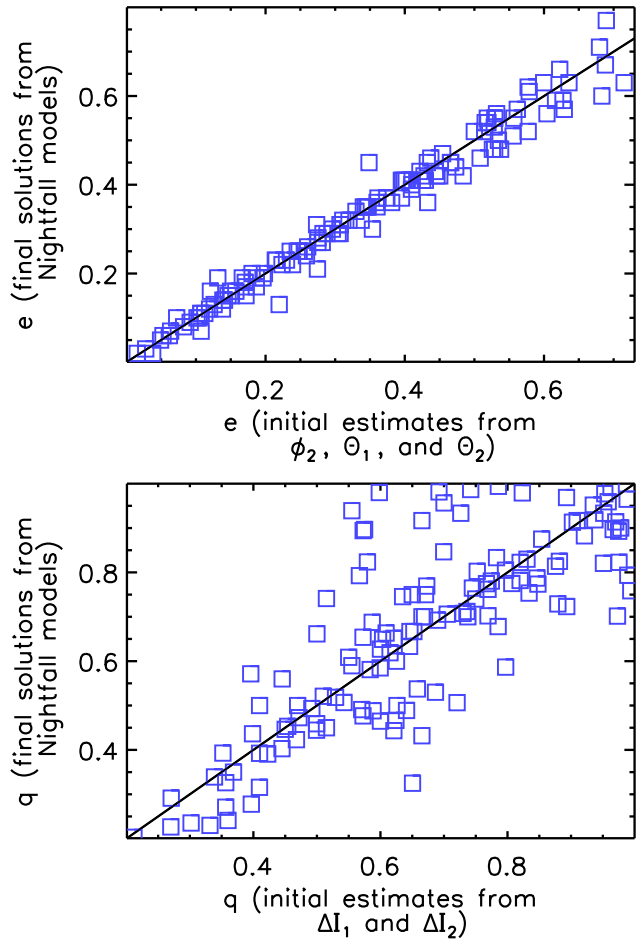}}
\caption{Comparison between the initial estimates in Step~1 of physical model properties based on the observed light curve parameters to the final solutions in Step~3 derived from fitting \textsc{Nightfall} light curve models.  Top panel: the initial eccentricites $e$ determined from the phase of the secondary eclipse $\Phi_2$ and eclipse widths $\Theta_1$ and $\Theta_1$ according to Eqn.~3 correspond quite well to the true final values.  Bottom panel: the mass ratios $q$ estimated from the eclipse depths $\Delta I_1$ and $\Delta I_2$ according to Eqn.~5 are approximate but imprecise indicators of the true mass ratios.  Nonetheless, the initial estimates are sufficient for starting purposes in our fitting routine and, on average,  dramatically reduce the number of iterations.}
\end{figure}

Finally, we evaluate the discrepancies between the values of $R_1$+$R_2$ estimated from the sum of eclipse widths $\Theta_1$+$\Theta_2$ and orbital periods $P$ according to Eqn.~6 to the final \textsc{Nightfall} solutions.  We measure an rms deviation of $\delta$($R_1$+$R_2$)~=~1.2~\Rsun, or $\delta$($R_1$+$R_2$)\,/\,($R_1$+$R_2$)~$\approx$~20\% the respective values.  The coefficient in Eqn.~6 should therefore be 7.0\,$\pm$\,1.2~\Rsun, valid only for EBs with B-type MS primaries.  As with the mass ratios $q$, the approximations for $R_1$+$R_2$ based on the observed light curve parameters are imprecise but sufficiently accurate to provide initial conditions for our fitting routine. 

As mentioned in \S3.1, $R_1$+$R_2$ is primarily an indictor of age $\tau$ of an EB in our sample rather than the component masses $M_1$ and/or $M_2$. For example, at age $\tau$ = 5 Myr, the sum of the stellar radii must be contained on the interval $R_1$+$R_2$~$\approx$~4.4\,-\,8.7~\Rsun\ given any combination of $M_1$ and $q$ $\ge$ 0.2 that satisfies our magnitude limits 16.0 $<$ $\langle I \rangle$ $<$ 17.6 and $-$0.25 $<$ $\langle V -I \rangle$ $<$ 0.20 and measured range of dust extinctions $A_I$~$\approx$~0.1\,-\,0.5~mag.  Meanwhile, at age $\tau$~=~100~Myr, the sum of the radii are systematically larger and confined to the interval $R_1$+$R_2$~$\approx$~5.3\,-\,12.3~\Rsun\ given the same photometric requirements.  Hence, EBs in our sample with $R_1$+$R_2$ $<$ 5.3 \Rsun\ must be relatively young while those with $R_1$+$R_2$ $>$ 8.7 \Rsun\ must be relatively old.  We initially estimated $R_1$+$R_2$ in Step 1 from the observed sum of eclipse widths $\Theta_1$+$\Theta_2$ according to Eqn.~6.  Although not as accurate as the final solutions, Eqn.~6 provides a model-independent measurement of $R_1$+$R_2$.  The sum of radii $R_1$+$R_2$ estimated from Eqn.~6 is therefore a robust and model-independent indicator of age $\tau$.

In Fig.~5, we display the eccentricities $e$ measured in Step 1 from Eqn. 3 as a function of the approximate sum of stellar radii $R_1$+$R_2$ estimated in Step 1 from Eqn. 6.  Both sets of parameters are model independent and based solely on the observed light curve features.  According to a Spearman rank correlation test, we find that the approximate values of $e$ and $R_1$+$R_2$ are anticorrelated ($\rho$~=~$-$0.18) at a statistically significant level ($p$~=~0.04).  This suggests that EBs with larger components, which are systematically older, favor smaller eccentricities.  We also compare the 19 EBs with approximate $R_1$+$R_2$ $<$ 5.3~\Rsun, which must be relatively young, to the 27 EBs with $R_1$+$R_2$ $>$ 8.7~\Rsun, which must be relatively old.  According to a K-S test, we find these young and old populations of EBs have distributions of eccentricities that are discrepant with each other at the $p$ = 0.02 significance level.  The anticorrelation between $R_1$+$R_2$, which is an indicator of age $\tau$, and $e$ is therefore statistically significant, robust, and model independent. In \S4, we further investigate this anticorrelation between $e$ and $\tau$ based on the more accurate final solutions obtained from the \textsc{Nightfall} light curve models.

\begin{figure}[t!]
\centerline{
\includegraphics[trim = 0.1cm 0.0cm 0.0cm 0.0cm, clip=true,width=3.4in]{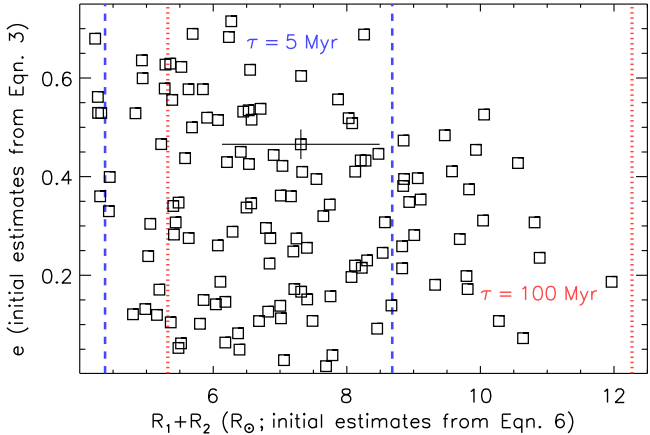}}
\caption{The approximate eccentricities $e$ vs. approximate sum of stellar radii $R_1$+$R_2$ estimated in Step 1 from the basic observed light curve parameters. For an individual system, we indicate representative error bars $\delta e$~$\approx$~0.03 and $\delta$($R_1$+$R_2$)~$\approx$~1.2~\Rsun. Young EBs with $\tau$ = 5 Myr that satisfy our photometric selection criteria must have $R_1$+$R_2$~=~4.4\,-\,8.7~\Rsun\ (dashed blue), while older EBs with $\tau$ = 100 Myr must have $R_1$+$R_2$~=~5.3\,-\,12.3~\Rsun\ (dotted red).  The values of $R_1$+$R_2$, which is an indicator of age $\tau$, and $e$ are anticorrelated at a statistically significant level.  This anticorrelation is not caused by selection effects, and is a robust and model-independent result.}
\end{figure}

\subsection{Uncertainties}

We now analyze the uncertainties in the final solutions of our \textsc{Nightfall} light curve models (see also Paper II).  For each system, we utilize MPFIT \citep{Markwardt2009} at the end of Step 3 (\S3.1) to calculate the measurement uncertainties.  For all 130 well-defined EBs, the nine physical model parameters have unique solutions and finite measurement uncertainties. Some of the model parameters, however, have solutions that are correlated with each other.  In addition, uncertainties in the dust reddening law, stellar evolutionary tracks, bolometric corrections, and \textsc{Nightfall} light curve models can lead to large systematic uncertainties in the physical model parameters.  In the following, we fully investigate the measurement uncertainties, parameter correlations, and systematic uncertainties in the context of a specific example EB, ID-2142.  We then determine the median total uncertainties of each model parameter (Eqns.~9\,-\,17) for the entire population of 130 well-defined EBs.

\begin{figure*}[t!]
\centerline{
\includegraphics[trim = 0.3cm 0.0cm 2.4cm 0.4cm, clip=true,width=7.1in]{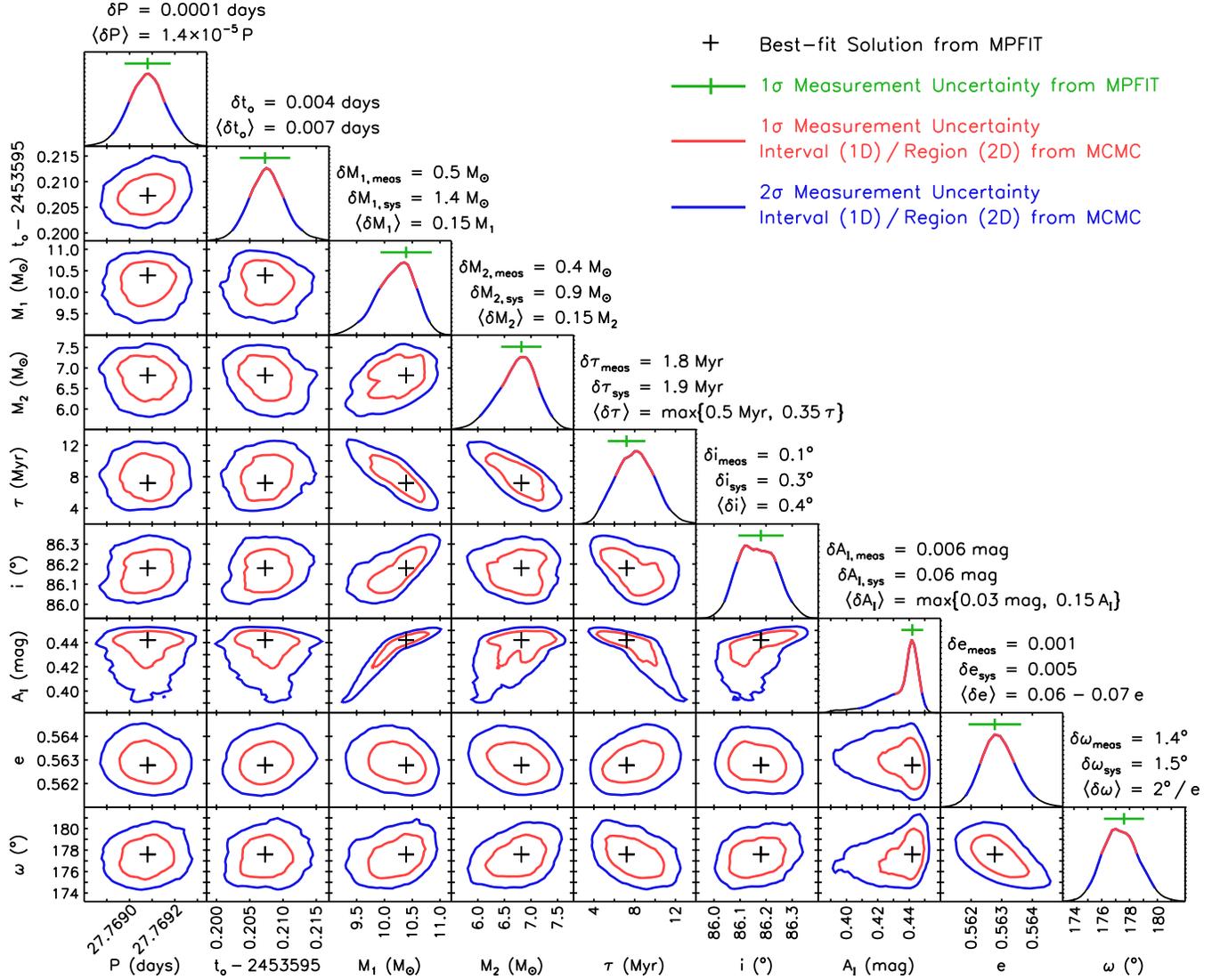}}
\caption{The probability density functions (diagonal panels) and joint probability distributions (off-diagonal panels) of the nine physical model parameters for ID-2142.  We emphasize the confidence intervals/regions in each panel only account for measurement uncertainties.  We compare the best-fit solutions and 1$\sigma$ measurement uncertainties based on the MPFIT routine (black +'s and green intervals) to the 68\% (red) and 95\% (blue) confidence intervals/regions determined from our MCMC technique.  For each physical model parameter, we list the measurement uncertainty for ID-2142, the systematic uncertainty for ID-2142 (if any), and the median total uncertainty for all 130 well-defined EBs.  These panels demonstrate: (1) the solutions are unique, (2) the uncertainties in $P$, $t_o$, $e$, and $\omega$ are small and primarily dictated by the sensitivity and cadence of the OGLE-III LMC observations, and (3) the measurement uncertainties for $M_1$, $M_2$, $\tau$, $i$ and $A_I$ are correlated, but the systematic uncertainties in the bolometric corrections, stellar evolutionary tracks, and dust reddening law dominate the total uncertainties.}
\end{figure*}

For our example EB, ID-2142, we explore the physical parameter space via a Markov chain Monte Carlo (MCMC) technique.  Starting with our final solution at the end of Step 3, we implement a Metropolis-Hastings ``random walk'' MCMC algorithm to generate and select steps in our phase space of nine physical model parameters.  At each proposed step,  we synthesize a \textsc{Nightfall} light curve model given the proposed nine physical model parameters.  The probability $p$ $\propto$ $e^{-\Delta \chi^2/2}$ of accepting the proposed step is determined by evaluating the difference in the $\chi^2$ statistic between the proposed step and the current solution. Obviously, if $\Delta \chi^2$ $<$ 0, the proposed step is always taken.  If the proposed step is rejected, the step length is effectively zero, i.e., the previous solution is counted again.  We generate proposed steps according to a Gaussian distribution with a fixed standard deviation for each of the nine physical model parameters.  We choose the standard deviation in the step sizes so that approximately one-third of the proposed steps are accepted.  We simulate 32,000 proposed steps and light curves with \textsc{Nightfall}, which exceeds the total number of models generated in \S3.1 used to fit solutions for all 130 well-defined EBs! It is therefore quite computationally expensive to calculate robust measurement uncertainties and correlations between model parameters for an individual EB with this MCMC algorithm.  The distribution of the $\approx$12,000 accepted steps and $\approx$20,000 repeated solutions provide the nine-dimensional joint probability distribution for the physical models.  For each of the nine physical model parameters, we marginalize across the other eight parameters to calculate the one-dimensional probability density function.  We also compute the two-dimensional joint probability distributions for each of the $_9C_2$~=~36 parameter combinations. 

In Fig.~6, we display the one-dimensional probability distributions for the nine physical model parameters (diagonal panels) and the two-dimensional joint probability distributions for the 36 parameter combinations (off-diagonal panels). Although some of the parameters are mildly to significantly correlated with each other, the measurement uncertainties are finite for all nine physical model parameters.  The MCMC technique confirms the uniqueness and non-degeneracy of the physical model solutions.  Moreover, the measurement uncertainties we determined from the robust MCMC algorithm are consistent with the measurement uncertainties we evaluated with the MPFIT routine.  We can therefore rely on the MPFIT measurement uncertainties we calculated for all 130 well-defined EBs.

The uncertainties in the orbital parameters $P$ and $t_{\rm o}$ are solely due to the measurement uncertainties and dictated by the sensitivity and cadence of the OGLE-III LMC observations. The solutions for $P$ and $t_{\rm o}$ are therefore independent of the other seven model parameters (note the fairly circular contours in the first and second columns of panels in Fig.~6).  For ID-2142, we measure 1$\sigma$ uncertainties of $\delta P$ $\approx$ 0.0001 days and $\delta t_{\rm o}$~$\approx$~0.004~days.  We find the median 1$\sigma$ uncertainties for the entire population of 130 well-defined EBs to be:

\begin{equation}
 \langle \delta P \rangle \approx 1.4\times10^{-5} P \approx 0.0004~{\rm days}
\end{equation}

\begin{equation}
 \langle \delta t_{\rm o} \rangle \approx 0.007~{\rm days}
\end{equation}

\noindent  Note that our example ID-2142 has slightly smaller uncertainties than average because it is relatively bright and its eclipses are well sampled.

As with $P$ and $t_{\rm o}$, the uncertainties in $e$ and $\omega$ are primarily determined by the sensitivity and cadence of the OGLE-III LMC observations.  The solutions for $e$ and $\omega$ are therefore independent of the other parameters, but are slightly correlated with each other (see last two rows in Fig.~6).  The eclipses are sufficiently sampled to easily break this degeneracy.  For ID-2142, we calculate a 95\% confidence interval of $\omega$~=~175$^{\rm o}$\,-\,180$^{\rm o}$. Note that we measured eclipse widths $\Theta_1$~=~0.0047~$\lesssim$~$\Theta_2$~=~0.0050, also indicating $\omega$ $\lesssim$ 180$^{\rm o}$ according to the approximations in Step~1 (Eqn.~3).  Based on the \textsc{Nightfall} light curve models, we calculate formal 1$\sigma$ measurement uncertainties of $\delta e_{\rm meas}$ $\approx$ 0.001 and $\delta \omega_{\rm meas}$ $\approx$ 1.4$^{\rm o}$ for ID-2142.

For ID-2142 and some other EBs in our sample, the measurement uncertainties $\delta e_{\rm meas}$~$\lesssim$~0.005 and $\delta \omega_{\rm meas}$~$\lesssim$~1.5$^{\rm o}$ are extremely small. \textsc{Nightfall} treats each stellar component as a three-dimensional polyhedral mesh with a finite number of flat surfaces.  We suspect this finite resolution limits the true sensitivity to systematic uncertainties of $\delta e_{\rm sys}$ $\approx$ 0.005 and $\delta \omega_{\rm meas}$ $\approx$ 1.5$^{\rm o}$. In any case, the measurement uncertainties $\delta e_{\rm meas}$ and $\delta \omega_{\rm meas}$ increase and dominate the total uncertainties as the eccentricities $e$ decrease.  We measure median total uncertainties of $\delta e$~$\approx$~0.02 and $\delta \omega$~$\approx$~4$^{\rm o}$ for $e$~$\gtrsim$~0.5, $\delta e$~$\approx$~0.03 and $\delta \omega$~$\approx$~10$^{\rm o}$ for $e$~$\approx$~0.3, and $\delta e$~$\approx$~0.05 and $\delta \omega$~$\approx$~20$^{\rm o}$ for $e$~$\approx$~0.1. Obviously, the periastron angle $\omega$ is not defined, and therefore not constrained, if the orbits are circular.  For the entire population of 130 well-defined EBs, we find the following relations adequately describe the median total uncertainties:

\begin{equation}
 \langle \delta e \rangle \approx 0.06 - 0.07 e
\end{equation}

\begin{equation}
 \langle \delta \omega \rangle \approx \frac{2^{\rm o}}{e}
\end{equation}

Solutions for the remaining five parameters $M_1$, $M_2$, $\tau$, $i$, and $A_I$ are all correlated with each other (see Fig. 6).  Moreover, unlike $P$, $t_{\rm o}$, $e$, and $\omega$, which have relatively symmetric Gaussian errors, the probability density functions of $M_1$, $M_2$, $\tau$, $i$, and $A_I$ are mildly to significantly asymmetric.  The three parameters $M_1$, $\tau$, and $A_I$ are especially correlated along the observed magnitude $\langle I \rangle$.  In other words, solutions with more massive primaries $M_1$ require younger ages $\tau$ and higher extinctions $A_I$ to produce the same observed I-band flux.  The secondary mass $M_2$ is also anticorrelated with $\tau$.  Finally, the inclination $i$ mildly depends on the three parameters $M_1$, $\tau$, and $A_I$ that are significantly correlated with each other.

Although $M_1$, $M_2$, $\tau$, $i$ and $A_I$ are correlated with each other, there is sufficient information in the observed light curves and our constraints (e.g., distance, evolutionary tracks, dust reddening law) to break the degeneracies and provide unique solutions (see also \S3.1).  For example, if we were to fix the primary mass at $M_1$ = 11.0~\Msun\ (i.e., the 2.5$\sigma$ upper limit according to the probability density function in Fig.~6), the other parameters would converge to $M_2$ = 7.1 \Msun, $\tau$ = 3.6 Myr, $i$ = 86.35$^{\rm o}$, and $A_I$ = 0.45 mag with a fit statistic that is $\Delta \chi^2$~=~6.7 larger than the best-fit solution.  For this larger primary mass, there is no combination of $\tau$ and $A_I$ that can satisfactorally reproduce the observed magnitude $\langle I \rangle$ and color $\langle V - I \rangle$.  Similarly, if we were to fix the primary mass at $M_1$ = 9.2~\Msun\ (i.e., the 2.5$\sigma$ lower limit), the other parameters would converge to $M_2$ = 6.1~\Msun, $\tau$~=~13.3~Myr, $i$~=~86.08$^{\rm o}$, and $A_I$ = 0.39 mag with a fit statistic that is $\Delta \chi^2$ = 5.8 larger than the best-fit solution.  In this case, the component masses both decrease by $\approx$12\% (to maintain the same ratio of eclipse depths $\Delta I_1/\Delta I_2$), and so the orbital separation $a$ decreases by 4\% according to Kepler's third law.  The relative sum of the radii ($R_1 + R_2$)/$a$, which derives directly from the sum of eclipse widths $\Theta_1 + \Theta_2$, is measured to 1\% precision in our \textsc{Nightfall} light curve models.  If $a$ decreases by 4\%, then $R_1 + R_2$ must also decrease by $\approx$4\%.  According to the MS stellar evolutionary tracks, if $M_1$ and $M_2$ decrease by 12\%, then the radii $R_1$ and $R_2$ decrease by 9\% given the same age $\tau$ = 7.2 Myr.  Hence, the age must increase to $\tau$ = 13.3~Myr so that the sum of radii $R_1 +R_2$ only decreases by 4\%.  If the masses decrease, the radii decrease, and the age increases, then the temperatures $T_1$ and $T_2$ both decrease according to the stellar evolutionary constraints.  However, if $R_1$ and $T_1$ both decrease, it is difficult to maintain the same values of $\langle I \rangle$ and $\langle V - I \rangle$ with only one free extra paramter $A_I$.  Hence, there is no combination of $M_2$, $\tau$, and $A_I$ that can satisfactorially reproduce the observed values of  $\Delta I_1/\Delta I_2$, $\Theta_1 + \Theta_2$, $\langle I \rangle$, and $\langle V - I \rangle$ if $M_1$~=~9.2~\Msun.  This line of reasoning holds for all EBs in our sample, and so the physical model parameters will always have unique solutions with finite measurement uncertainties.  For ID-2142, we measure formal 1$\sigma$ measurement uncertainties of $\delta M_{1,{\rm meas}}$~$\approx$~0.04$M_1$~$\approx$~0.5~\Msun, $\delta M_{2,{\rm meas}}$~$\approx$~0.07$M_2$~$\approx$~0.4~\Msun, $\delta \tau_{\rm meas}$~$\approx$~0.25$\tau$~$\approx$~1.8~Myr, $\delta i_{\rm meas}$~$\approx$~0.1$^{\rm o}$, and $\delta A_{I,{\rm meas}}$~$\approx$~0.01$A_I$~$\approx$~0.006~mag.  We find similar percentage measurement uncertainties in these parameters for the 130 well-defined EBs in our sample.

The systematic uncertainties in $M_1$, $M_2$, $\tau$, $i$ and $A_I$ can be considerably larger and derive from a variety of sources.  We first investigate the systematic uncertainties in the adopted bolometric corrections.  Our B-type MS primaries and secondaries span a large range of temperatures $T$ $\approx$ 10,000\,-\,30,000 K and therefore a broad interval of bolometric corrections $BC$~=~M$_{\rm bol}$~$-$~M$_V$~$\approx$~$-$3.0\,-\,$-$0.3~mag \citep{Pecaut2013}.  For the hottest stars in our sample with $T$~$\approx$~30,000~K and $BC$~$\approx$~$-$3.0, the bolometric corrections are uncertain by $\delta BC$~$\approx$~0.2~mag, i.e. $\delta BC$/$BC$~$\approx$~7\% \citep{Bertelli2009,Pecaut2013}.  To propagate this systematic uncertainty into our solution for ID-2142, we decrease the absolute magnitudes of the bolometric corrections by 7\% and repeat our fitting routine from \S3.1.  The \textsc{Nightfall} light curve models now converge to a final solution of $M_1$~=~9.3~\Msun, $M_2$~=~6.3~\Msun, $\tau$~=~7.5~Myr, $i$~=~86.35$^{\rm o}$, and $A_I$~=~0.40~mag.  The main effect of decreasing $|BC|$ is to decrease the masses $M_1$ and $M_2$.  This is because more of the flux is radiated in the optical and so the component luminosities need to be reduced to maintain the same observed magnitude $\langle I \rangle$.   Fortunately, the mass ratio $q$~=~$M_2$/$M_1$ is not significantly affected by the uncertainties in the bolometric corrections.  The decrease in masses lead to slightly longer ages (to maintain the observed eclipse widths), higher inclinations (to maintain the observed eclipse depths), and lower extinctions (to maintain the observed color).   The systematic uncertainties in the physical model parameters due to the uncertainties in the bolometric corrections are therefore $\delta M_{1,{\rm BC}}$~=~0.11$M_1$~=~1.1~\Msun, $\delta M_{2,{\rm BC}}$~=~0.08$M_2$~=~0.5~\Msun, $\delta \tau_{\rm BC}$~=~0.04$\tau$~=~0.3~Myr, $\delta i_{\rm BC}$~=~0.18$^{\rm o}$, and $\delta A_{I,{\rm BC}}$~=~0.1$A_I$~=~0.04~mag.  Because the primary mass $M_1$ is mainly dictated by the observed $\langle I \rangle$ and the bolometric corrections, we expect similar percentage systematic uncertainties in  $M_1$, $M_2$, $\tau$, $i$, and $A_I$ for the other EBs in our sample.

We next propagate the uncertainties in the intrinsic colors, observed colors, and dust reddening law.  The uncertainty in the intrinsic colors of B-type MS stars are $\approx$0.01\,-\,0.02 mag \citep{Pecaut2013}, the color calibrations of stars in the OGLE-III LMC database are also uncertain by $\approx$0.01\,-\,0.02 mag \citep{Udalski2008}, and the coefficient in our adopted dust reddening law $E(V-I)$ = 0.70$A_I$ has a $\approx$10\% uncertainty \citep{Cardelli1989,Schlegel1998,Fitzpatrick1999,Ngeow2005}.  The systematic uncertainty in the dust extinction $A_I$ due to dust/color uncertainties is therefore $\delta A_{I,{\rm dust/color}}$ = max\{0.02 mag, 0.1$A_I$\}. To confirm this estimate, we replace the dust reddening law with $E(V-I)$ = 0.63$A_I$ in our models for ID-2142 and repeat our fitting routine from \S3.1.  As expected, we measure $A_I$~=~0.48~mag, i.e. the dust extinction increased by $\delta A_I$~=~0.1$A_I$~=~0.04~mag, while the other parameters do not vary beyond the measurement uncertainties.

We finally investigate the uncertainties in the stellar evolutionary tracks, including the effects of metallicity and rotation.  We replace the Z\,=\,0.008 tracks from \citet{Bertelli2009} with the Z\,=\,0.006 non-rotating models from \citet{Georgy2013}.  We refit ID-2142 and measure $M_1$~=~10.5~\Msun, $M_2$~=~6.8~\Msun, $\tau$~=~9.0~Myr, $i$~=~86.38$^{\rm o}$, and $A_I$~=~0.45~mag.  Hence, the systematic uncertainties in the stellar evolutionary models, including our ability to interpolate between the tracks, dominates the uncertainty in the age $\delta \tau_{\rm track}$~=~0.26$\tau$~=~1.9~Myr and inclination $\delta i_{\rm track}$~=~0.21$^{\rm o}$. We then replace the evolutionary tracks with the Z\,=\,0.006 tracks from \citet{Georgy2013} that are rotating on the zero-age MS at $v/v_{\rm crit}$ = 50\% the critical break-up velocity.  We note that $\approx$80\% of B-type MS stars are rotating at $v$ $\lesssim$ 0.5$v_{\rm crit}$ $\approx$ 250 km s$^{-1}$ \citep{Abt2002,Levato2013}, and our EBs with intermediate orbital periods $P$ = 20\,-\,50 days may have tidally evolved toward slower rotational velocities.  B-type MS stars initially rotating at $v$/$v_{\rm crit}$ = 0.5 have equatorial radii that are only $\approx$(3\,-\,4)\% larger than their polar radii, but MS lifetimes $\tau_{\rm MS}$ that are 20\% longer \citep{Georgy2013}.  It is therefore the differences in the evolutionary tracks of stars with rotation, not the distortions in their shapes, that can significantly affect our model solutions.  We refit ID-2142 with the rotating non-synchronized stellar models, and measure $M_1$~=~10.5~\Msun, $M_2$~=~6.9~\Msun, $\tau$~=~8.6~Myr, $i$~=~86.30$^{\rm o}$, and $A_I$~=~0.45~mag.  For ID-2142, the differences between the non-rotating and rotating tracks from \citet{Georgy2013} are within the measurement uncertainties.  This is because the tracks with $v$/$v_{\rm crit}$ = 0.5 do not significantly deviate from their non-rotating counterparts until the ages reach $\tau$~$>$~0.8$\tau_{\rm MS}$ the non-rotating MS lifetimes. For ID-2142 and the majority of EBs in our sample with primary ages $\tau$ $<$ 0.8$\tau_{\rm MS}$, the uncertainties due to the effects of rotation are negligible.  For the few systems that are extremely young ($\tau$~$<$~1~Myr) or old ($\tau$~$>$~0.8$\tau_{\rm MS}$), we expect slightly larger systematic uncertainties in the ages and masses.

By adding the measurement uncertainties and various systematic uncertainties above in quadrature, we estimate the total median 1$\sigma$ uncertainties for the 130 well-defined EBs to be:

\begin{equation}
 \langle \delta M_1 \rangle \approx 0.15 M_1
\end{equation}

\begin{equation}
 \langle \delta M_2 \rangle \approx 0.15 M_2
\end{equation}

\begin{equation}
 \langle \delta \tau \rangle \approx {\rm max}\{0.5~{\rm Myr},~0.35 \tau \}
\end{equation}

\begin{equation}
 \langle \delta i \rangle \approx 0.4^{\rm o}
\end{equation}

\begin{equation}
 \langle \delta A_I \rangle \approx {\rm max}\{0.03~{\rm mag},~0.15 A_I \}
\end{equation}

\noindent For these five parameters, the uncertainties are dominated by the systematic uncertainties in the bolometric corrections, dust reddening law, and evolutionary tracks.  For older EBs with primary ages $\gtrsim$80\% their MS lifetimes,  the total uncertainties in the masses $\langle \delta M_1 \rangle$ $\approx$ 0.2$M_1$ and $\langle \delta M_2 \rangle$ $\approx$ 0.2$M_2$ and ages  $\langle \delta \tau \rangle$ $\approx$ 0.45$\tau$ are slightly larger due to the effects of rotation.

Some of our EB light curve model solutions can be biased due to contamination with a third light source, e.g., a tertiary companion or a background/foreground object along similar lines of sight.  In Papers~I~and~II, we estimated that only $\approx$10\% of our B-type MS EBs in the LMC can be contaminated by a third light source that is bright enough to significantly contribute to the systematic uncertainties.  Unlike the previously discussed sources of systematic uncertainties that contribute to all 130 well-defined EBs, contamination by a third light source affects only a small subset of our sample.  

In addition to calculating the uncertainties for the nine independent physical model parameters, we also estimate the uncertainties in the dependent physical properties.  The total uncertainties in $M_1$ and $M_2$ are $\approx$15\% but slightly correlated with each other (see above).  The total median uncertainty in the mass ratio is therefore $\langle \delta q \rangle$~=~max\{0.03,~0.12$q$\}.   Because the quantity ($R_1$+$R_2$)/$a$ is precisely constrained from the observed eclipse widths, the uncertainties in $R_1$, $R_2$, and $a$ mainly derive from the uncertainties in $M_1$ and $M_2$ according to Kepler's third law.  We measure $\langle \delta R_1 \rangle$~$\approx$~0.07$R_1$~$\approx$~0.3\Rsun, $\langle \delta R_2 \rangle$~$\approx$~0.07$R_2$~$\approx$~0.2~\Rsun, and $\langle \delta a \rangle$~$\approx$~0.06$a$~$\approx$~6~\Rsun.  Finally, given the $\approx$(20\,-\,30)\% uncertainties in the luminosities (primarily due to uncertainties in the bolometric corrections) and the $\approx$7\% uncertainties in the radii, the uncertainties in the temperatures are $\approx$8\% according to the Stefan-Boltzmann law.  Hence, the total median uncertainties are $\langle \delta T_1 \rangle$~$\approx$~0.08$T_1$~$\approx$~1,500~K and $\langle \delta T_2 \rangle$~$\approx$~0.08$T_2$~$\approx$~1,100~K.

\section{EB Trends}

In our sample of 130 EBs, several trends and correlations exist among the nine physical model properties.  Most of these trends are caused by geometrical and evolutionary selection effects in our magnitude-limited sample of EBs.  We correct for these selection effects in the third stage of our pipeline (\S5).  Two correlations, however, are intrinsic to the population of binaries with B-type MS primaries. In this section, we first discuss these two empirical relations we uncovered from the data, and then we explain the trends that are caused by selection effects.

In Fig.~7, we display the measured I-band dust extinctions $A_I$ as a function of age $\tau$ for the 130 well-defined EBs.  These two parameters are anticorrelated (Spearman rank correlation coefficient $\rho$~=~$-$0.34) at a statistically significant level (probability of independence $p$~=~8$\times$10$^{-5}$). We fit a log-linear trend to the total population of 130 EBs (green line in Fig.~7):

\begin{equation}
  A_{I,\rm{total}}\,({\rm mag}) = 0.39 - 0.07\,{\rm log}\Big(\frac{\tau}{1\,{\rm Myr}}\Big)
\end{equation} 

The slope in the above relation may be biased toward negative values due to a  photometric selection effect in our magnitude-limited sample.  Specifically, EBs that are intrinsically bluer and more luminous systematically contain younger, short-lived, more massive primaries.  These blue, luminous, younger EBs may therefore require larger dust extinctions and reddenings to satisfy our photometric selection criteria (and vice versa).  In Fig.~8, we show the measured absolute magnitudes $M_I$ and intrinsic colors $\langle V-I \rangle_{\rm o}$ as a function of dust extinction $A_I$ for our 130 well-defined EBs.  We also display our photometric selection criteria based on the observed magnitudes 16.0 $<$ $\langle I \rangle$ $<$ 17.6 and observed colors $-$0.25~$<$~$\langle V-I \rangle$~$<$~0.20 (green lines).    Indeed, there are several intrinsically red, low-luminosity, older EBs with $M_I$~$\approx$~$-$1.2 that are in our sample only because they have small dust extinctions $A_I$ $\approx$ 0.2 mag.  If they were to have slightly higher dust extinctions, they would fall below our selection limit of $\langle I \rangle$ $=$ 17.6 (see Fig.~8).

\begin{figure}[t!]
\centerline{
\includegraphics[trim = 0.5cm 0.3cm 0.3cm 0.4cm, clip=true,width=3.3in]{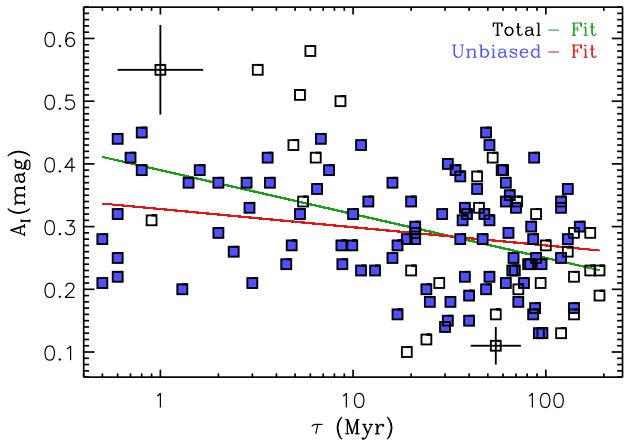}}
\caption{Measured I-band dust extinctions $A_I$ and ages~$\tau$ for the 130 EBs in our well-defined sample (black squares).  We display representative uncertainties for two systems in opposite corners of this parameter space.  We also display our unbiased subsample of 98 EBs that is relatively free from photometric selection effects (filled blue).  The dust extinctions clearly diminish with age, even within our unbiased subsample,  demonstrating the dust content in stellar environments systematically decreases with time.  We fit a log-linear relation to the total population (green) and unbiased subsample (red).  The latter is an empirical age-extinction relation that can be implemented when modeling other stellar populations.}
\end{figure}

In \S5, we account for our photometric selection criteria when analyzing all 130 EBs in our well-defined sample.  Here, we correct for photometric selection effects by further culling our sample according to the intrinsic properties of $M_I$ and $\langle V-I \rangle_{\rm o}$.  To obtain an unbiased subsample, we can choose EBs across any interval of $M_I$ and $\langle V-I \rangle_{\rm o}$ that also satisfies our selection criteria on observed magnitudes and colors.  To retain most of the sample, we select the regions enclosed by $-$2.63~$<$~$M_I$~$<$~$-$1.35, $-$0.341~$<$~$\langle V-I \rangle_{\rm o}$~$<$~$-$0.115, and 0.13~$<$~$A_I$\,(mag)~$<$~0.45 (red lines in Fig.~8).  The 98 EBs that satisfy these extra selection criteria (filled blue systems in Fig.~8) represent an unbiased sample relatively free from photometric selection effects.  

\begin{figure}[t!]
\centerline{
\includegraphics[trim = 3.2cm 0.3cm 4.5cm 0.3cm, clip=true,width=3.15in]{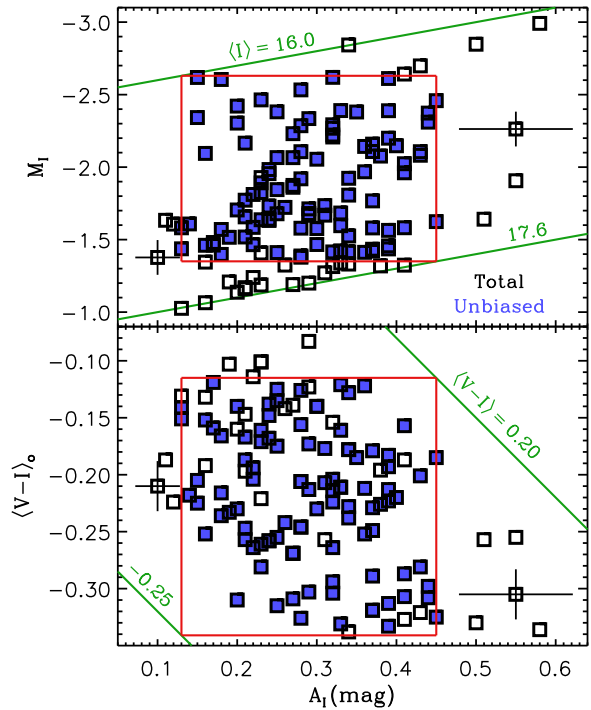}}
\caption{Measured absolute magnitudes $M_I$ (top) and intrinsic colors $\langle V-I \rangle_{\rm o}$ (bottom) as a function of dust extinction $A_I$ for our 130 well-defined EBs (black squares; representative errors shown for two systems).  We also display the limits on observed magnitudes $\langle I \rangle$ and observed colors $\langle V-I \rangle$ imposed by our photometric selection criteria (green lines).  The 98 EBs (filled blue) that are enclosed by both red regions are relatively free from photometric selection effects.  Even within our unbiased sample, intrinsically bluer EBs that contain hotter, more massive, younger primaries favor larger dust extinctions.}
\end{figure}

Even within this unbiased sample of 98 EBs, the intrinsic colors $\langle V-I \rangle_{\rm o}$ and dust extinctions $A_I$ are still anticorrelated ($\rho$ = $-$0.25) at a statistically significant level ($p$ = 0.02).  As can be observed in the bottom panel of Fig.~8, there are relatively few intrinsically blue systems $\langle V-I \rangle_{\rm o}$ $\approx$ $-$0.30 with small dust extinctions $A_I$~$\approx$~0.2~mag.  Similarly, there are few intrinsically redder EBs $\langle V-I \rangle_{\rm o}$ $\approx$ $-$0.15 with large dust extinctions $A_I$ $\approx$ 0.4 mag.  Intrinsically bluer EBs contain hot primaries that are systematically more massive, short-lived, and younger.  Hence, the anticorrelation between age $\tau$ and dust extinction $A_I$ is real.

In Fig.~8, we also display $\tau$ and $A_I$ for the 98 EBs (filled blue) in our unbiased subsample.  Although not as prominent, the ages $\tau$ and dust extinctions $A_I$ for the 98 EBs in our unbiased sample are still anticorrelated ($\rho$~=~$-$0.23) at a statistically significant level ($p$~=~0.02).  For example, there is a complete absence of EBs with $A_I$~$<$~0.2~mag at $\tau$ $<$ 15 Myr.  In contrast, there are many EBs in our unbiased sample with $A_I$~=~0.1\,-\,0.2~mag at $\tau$~$>$~15 Myr.  The intrinsic anticorrelation between $A_I$ and $\tau$ in our unbiased sample demonstrates a relationship between dust content and ages of stellar environments. Young EBs, and young B-type MS stars in general, with $\tau$~$\approx$~1~Myr are embedded in dusty envelopes and/or molecular clouds with photometric extinctions $A_I$~$\approx$~0.33~mag.  Meanwhile, older EBs with $\tau$~$\approx$~100~Myr reside in less attenuating environments with $A_I$~$\approx$~0.26~mag. We fit a log-linear trend to the unbiased sample of 98 EBs:

\begin{equation}
  A_{I,\rm{unbiased}}\,({\rm mag}) = 0.33 - 0.03\,{\rm log}\Big(\frac{\tau}{1\,{\rm Myr}}\Big)
\end{equation} 

\noindent  valid for 0.5~Myr~$<$~$\tau$~$<$~200~Myr (red line in Fig.~8). Even after accounting for selection effects, the value of and measurement uncertainty in the slope $-$0.029\,$\pm$\,0.011 is still inconsistent with zero at the 2.6$\sigma$ confidence level.  This is a similar probability of significance based on the Spearman rank test above (probability of no correlation $p$ = 0.02 between $A_I$ and $\langle V - I \rangle_{\rm o}$). The $\approx$30\% systematic uncertainty in the ages $\tau$ and $\approx$10\% systematic uncertainty in the extinctions $A_I$ propagate into Eqn.~19.  The values of and total uncertainties are therefore $-$0.029\,$\pm$\,0.014 for the slope and 0.33\,$\pm$\,0.04 mag for the mean dust extinction at $\tau$ $\approx$ 1 Myr. The rms in the measured dust extinctions $A_I$ around the above relation is $\sigma$~=~0.08~mag.

\begin{figure}[t!]
\centerline{
\includegraphics[trim = 0.6cm 0.3cm 0.2cm 0.4cm, clip=true,width=3.3in]{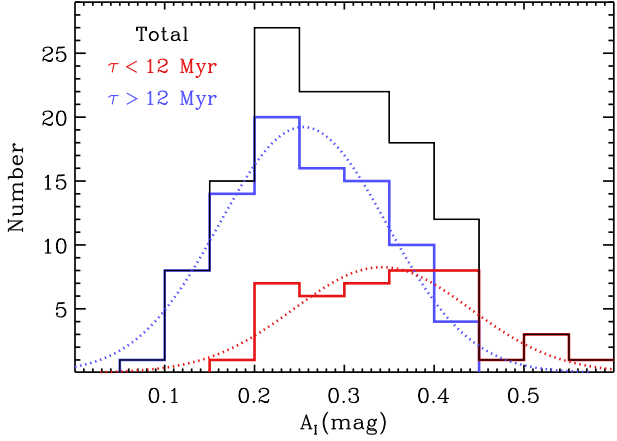}}
\caption{Distribution of I-band dust extinctions $A_I$ for the total population of 130 EBs (black), 42 EBs with ages $\tau$~$\le$~12~Myr (red), and 88 EBs with $\tau$~$>$~12~Myr (blue). Although the total population peaks at $A_I$ $\approx$ 0.25 mag with a long tail toward high dust extinctions, the young and old subsamples can each be accurately described with Gaussian distributions (dotted) centered at  $A_I$~$\approx$~0.34~mag and  $A_I$~$\approx$~0.25~mag, respectively.}
\end{figure}

It had been previously known that younger early-type stars in the LMC experience slightly higher dust extinctions than late-type stars \citep{Zaritsky1999, Zaritsky2004}.  In the present study, we have measured the relationship between age $\tau$ and dust extinction $A_I$. Quantifying age-dependent dust extinctions is crucial when analyzing the spectral energy distributions of unresolved stellar populations in distant galaxies \citep{Panuzzo2007,Silva2011}.  Young O- and B-type stars, which dominate the ultraviolet component in star-forming galaxies, will experience systematically higher dust extinctions than the older, redder stars.  To accurately constrain the star-formation histories of these galaxies, it is imperative to account for age-dependent dust extinctions.   We note that different galaxies and stellar populations will have slightly different dust extinctions as a function of age. Nonetheless, our empirical age-extinction relation (Eqn. 19) can provide insight when calibrating models of unresolved stellar populations.

 \citet{Zaritsky2004} found that the dust extinction distribution toward young, hot stars in the LMC peaks at $A_I$ $\approx$ 0.25 mag with a long tail toward higher values.  This is consistent with our total population of 130 EBs with B-type MS primaries (see Fig.~9).  By dividing our EB population into young ($\tau$~$\le$~12~Myr) and old ($\tau$~$>$~12~Myr) subsamples,  we find that both subsamples can be fitted with simple Gaussians centered at $A_I$~$\approx$~0.34~mag and  $A_I$~$\approx$~0.25~mag, respectively.  Hence, the non-Gaussian distribution of dust extinction may be simply due to a selection effect with age.  The very young EBs, which represent a small fraction of the total population, occupy the long tail toward large dust extinctions.  Meanwhile, the long-lived EBs, which comprise the majority of the sample, form the peak in the distribution at $A_I$~$\approx$~0.25~mag.

 We now examine the second physically-genuine trend in our EB population.  In Fig.~10, we show the measured eccentricities $e$ as a function of age $\tau$ for the 130 EBs in our well-defined sample.  The eccentricities and ages are anticorrelated (Spearman rank correlation coefficient $\rho$~=~$-$0.39) at a statistically significant level (probability of no correlation $p$~=~5$\times$10$^{-6}$).  

This observed anticorrelation is primarily because eccentricities decrease with time due to tidal evolution.  The observed trend may be accentuated by a secondary effect, whereby EBs with more massive, short-lived primaries favor larger eccentricities.  However, this relation between primary mass $M_1$ and eccentricity $e$ cannot fully explain the observed anticorrelation between $\tau$ and $e$.  For example, the eccentricities and ages of the 32 EBs with massive primaries $M_1$ $\approx$ 8.5\,-\,13.9\,\Msun\ are still anticorrelated ($\rho$~=~$-$0.44) at a statistically significant level ($p$~=~0.01).  Similarly, the 98 less massive EBs with $M_1$~$\approx$~3.6\,-\,8.5\,\Msun\ have eccentricities and ages that are anticorrelated ($\rho$~=~$-$0.28) at a statistically significant level ($p$~=~0.005).  Although EBs with early-B primaries may be born with systematically larger eccentricities, the anticorrelation between age $\tau$ and eccentricity $e$ is dominated by tidal evolution and is observed in both early-B and late-B MS subsamples.

\begin{figure}[t!]
\centerline{
\includegraphics[trim = 0.8cm 0.3cm 0.2cm 0.4cm, clip=true,width=3.4in]{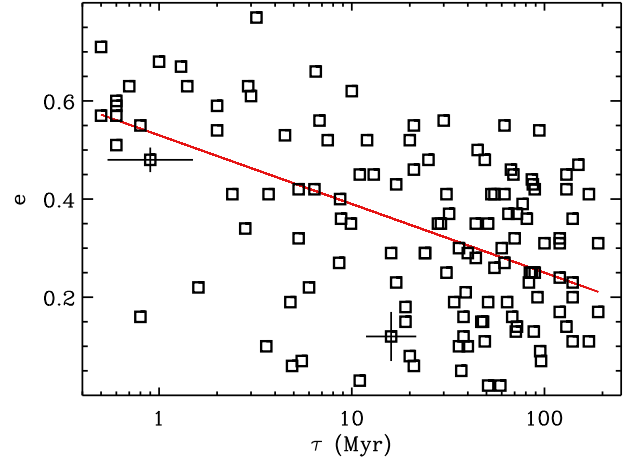}}
\caption{Measured eccentricities $e$ and ages~$\tau$ for the 130 EBs in our well-defined sample (black squares; representative uncertainties shown for two systems).   Binaries with B-type MS primaries and intermediate orbital periods are preferentially born with large eccentricities, which suggest they formed via dynamical interactions and/or tidal capture.  Moreover, the observed slope (red line) in the age-eccentricity anticorrelation provides a constraint for dynamical tides in hot MS stars with radiative envelopes.}
\end{figure}

For late-type stars with $M$ $\lesssim$ 1.3 \Msun, orbital energy is most efficiently dissipated into the interior of the stars via convective eddies in the stellar atmospheres \citep{Zahn1977,Hut1981,Zahn1989a,Hurley2002}.  This equilibrium tide model for convective damping has been tested against observations of late-type binaries in various environments with different ages \citep{Meibom2005}.  For more massive stars $M$ $>$ 1.3 \Msun\ with radiative envelopes, such as our B-type MS stars, tides operate dynamically via oscillations in the stellar interiors \citep{Zahn1975, Hurley2002}.  By estimating the ages of 130 early-type EBs, we have measured the evolution of binary eccentricities due to dynamical tides with radiative damping.

The slope of the observed age-eccentricity anticorrelation provides insight into the tidal evolution of highly eccentric binaries.  We fit a log-linear trend to the observations (red line in Fig.~10):

\begin{equation}
  e_{\rm fit} = 0.53 - 0.14\,{\rm log}\Big(\frac{\tau}{1\,{\rm Myr}}\Big)
\end{equation}

\noindent The value of and measurement uncertainty in the slope is $-$0.14\,$\pm$\,0.03.  Hence, the slope is negative at the 5$\sigma$ confidence level, similar to the statistical significance determined from the Spearman correlation test above.   Again, systematic uncertainties in the ages $\tau$ and eccentricities $e$ contribute to the uncertainties in the coefficients in Eqn.~20.  After calculating the total uncertainties, we find the mean eccentricity at $\tau$~$\approx$~1~Myr is 0.53\,$\pm$\,0.05 while the slope is $-$0.14\,$\pm$\,0.05. The rms scatter in the measured eccentricities around the above relation is $\sigma_e$~$\approx$~0.16.

The intercept in Eqn.~20 implies a circularization timescale of $\tau_{\rm circ}$~$\approx$~5~Gyr for our EBs with B-type MS primaries and moderate orbital periods $P$ $\approx$ 20\,-\,50~days.  However, tidal damping is not as efficient when the orbits become less eccentric \citep{Hut1981}. The true circularization timescale may therefore be longer if the age-eccentricity relation flattens beyond $\tau$~$>$~200~Myr.  Conversely, older EBs have systematically larger components (see Fig. 5), and so tidal damping may become more efficient as the primary fills a larger fraction of its Roche lobe. In any case, these short-lived B-type MS primaries will expand beyond $R_1$ $\gtrsim$ 10\,\Rsun\ and evolve toward the giant branch long before the orbits are completely circularized. 

 Our young EBs with generally large eccentricities experience extreme tidal forces.  In fact, a few of the EBs with $e$ $>$ 0.6 in our sample have modest Roche-lobe fill-factors $RLFF$ $\approx$ 0.3 at periastron.  Tidal evolution of highly eccentric binaries is quite complicated, especially considering second-order effects and non-linear terms can become quite important \citep{Hut1981}.  A full analysis of tidal evolution in our EB sample is therefore not within the scope of the present study. Nonetheless, the observed age-eccentricity anticorrelation provides a constraint for models of tidal evolution in highly eccentric early-type binaries.  

\begin{figure}[t!]
\centerline{
\includegraphics[trim = 0.8cm 0.3cm 0.2cm 0.4cm, clip=true,width=3.4in]{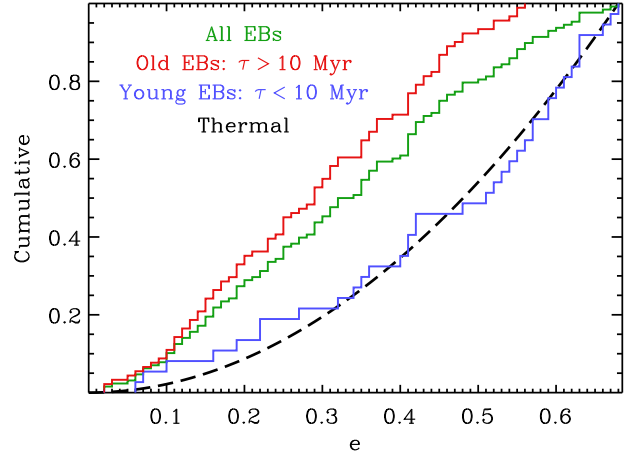}}
\caption{Cumulative distributions of eccentricities $e$ for all 128 EBs with $e$ $<$ 0.68 (green) and subsamples of 91 old EBs with $\tau$~$>$~10~Myr~(red) and 37 young EBs with $\tau$~$\le$~10~Myr~(blue).  The young population is fully consistent with a thermal eccentricity distribution (dashed black), indicating early-type binaries at intermediate orbital periods were dynamically captured. }
\end{figure}

In Fig.~11, we display the cumulative distribution function of the eccentricities for the 128 EBs with $e$~$\le$~0.68 (green).  We do not consider the two EBs with $e$~$=$~0.71 and 0.77 because highly eccentric binaries are not complete in our EB sample (see below and \S5).  Moreover, as discussed above, binaries with $P$~=~20\,-\,50~days and $e$~=~0.7\,-\,0.8 nearly fill their Roche lobes at periastron, and are expected to evolve toward smaller eccentricities on rapid timescales.  In Fig.~11, we also divide our sample into the 91 old EBs with $\tau$~$>$~10~Myr (red) and 37 young EBs with $\tau$~$\le$~10~Myr (blue).  Using a maximum likelihood method, we fit a power-law eccentricity probability distribution $p_e$~$\propto$~$e^{\eta}$ to the observed EBs.  We measure $\eta$~=~0.1\,$\pm$\,0.2, $-$0.1\,$\pm$\,0.2, and 0.8\,$\pm$\,0.3 for the total, old, and young EB samples, respectively.  Our total population of EBs ($\eta$ = 0.1\,$\pm$\,0.2) is consistent with the flat distribution ($\eta$ = 0) observed by \citet{Abt2005} for his sample of binaries with B-type MS primaries and intermediate orbital periods.

If the orbital velocities and energies of a binary population follow a Maxwellian ``thermal'' probability distribution, then the eccentricity probability distribution $p_e$~=~$2e$\,$de$ will be weighted toward large eccentricities  \citep{Ambartsumian1937}. Such a population of eccentric and thermalized binaries would suggest the binaries formed through dynamical interactions, either through tidal / disk capture, dynamical perturbations in a dense cluster, three-body exchanges, and/or Kozai cycles with a tertiary companion \citep{Heggie1975,Pringle1989,Turner1995,Kroupa1995,Kiseleva1998,Naoz2014}.  Surprisingly, the observed population of 37 young EBs ($\eta$ = 0.8\,$\pm$\,0.3) is fully consistent with a thermal eccentricity probability distribution ($\eta$ = 1; dashed black line in Fig.~11).  This indicates that massive binaries with intermediate orbital periods formed via dynamical interactions on rapid timescales $\tau$ $<$ 5 Myr.

Previous observations of spectroscopic \citep{Duquennoy1991} and visual \citep{Harrington1977}  solar-type binaries have indicated a thermal eccentricity distribution.  However, these studies recovered the thermal eccentricity distribution only after applying large and uncertain correction factors for incompleteness.  In both the spectroscopic and visual binary surveys, the raw samples were weighted significantly toward smaller eccentricities relative to the thermal distribution. In addition, more recent and complete observations of solar-type \citep{Abt2006,Raghavan2010} and early-type \citep{Abt2005} binaries at intermediate orbital periods have revealed a uniform eccentricity distribution that is clearly discrepant with a thermal distribution.

Our raw sample of young early-type EBs is only slightly biased toward small eccentricities.  In fact, the small excess of young EBs with $e$ $\approx$ 0.1\,-\,0.3 relative to the thermal distribution in Fig.~11 would be reduced  after correcting for selection effects.  In other words, we expect even better agreement between our sample of young early-type EBs and the thermal eccentricity distribution after considering observational biases (\S5). By choosing only the EBs with young ages, we have probed the initial binary properties of massive stars shortly after their formation.  For the first time, we have directly observed the theoretical thermal eccentricity distribution before tides have dramatically reduced the eccentricities.

\begin{figure}[t!]
\centerline{
\includegraphics[trim = 0.3cm 0.3cm 0.2cm 0.4cm, clip=true,width=3.45in]{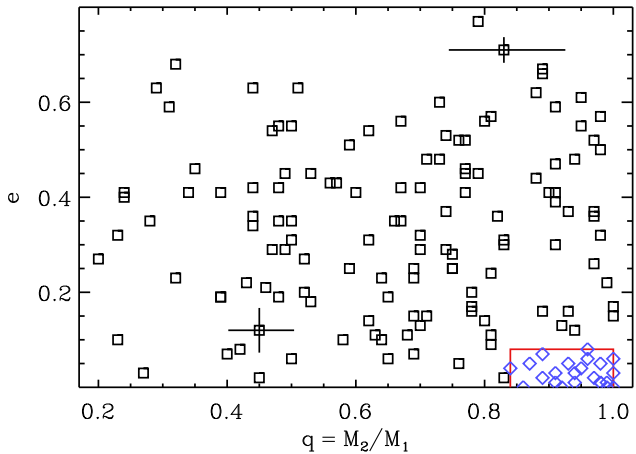}}
\caption{Measured eccentricities $e$ versus mass ratios $q$ = $M_2$/$M_1$ for the 130 well-defined EBs (squares; representative errors shown for two systems).  There is no evidence for a statistically significant correlation between $q$ and $e$ in our sample. We removed 23 EBs that have ambiguous orbital periods (see item D in \S2).  If we were to fit these 23 systems using the listed periods, they would all have $q$~$>$~0.84 and $e$~$<$~0.08 (blue diamonds enclosed within red lines). Such a dense population in this corner of the parameter space is highly unlikely, so it was justifiable to exclude these 23 systems from our well-defined sample. }
\end{figure}

In the following, we compare other physical model parameters and examine additional trends that could be caused by observational biases.  We use these observed distributions to further justify our selection criteria in \S2. We also motivate the necessity for incompleteness corrections and Monte Carlo simulations, which we perform in \S5.

We display the measured eccentricities $e$ as a function of mass ratio $q$ in Fig.~12.  A Spearman rank test reveals no statistically significant correlation ($p$ = 0.25).  The mass ratios $q$ of early-type binaries are independent of their eccentricities $e$ at intermediate orbital periods $P$~=~20\,-\,50~days.  

  In \S2 (see item D), we removed 23 EBs with nearly identical primary and secondary eclipses separated by $\approx$50\% in orbital phase. We concluded the majority of these systems have half their listed orbital periods, and therefore exhibit only one eclipse per orbit.  If we were to fit physical models to these systems assuming the listed orbital periods, they would all have $q$~$>$~0.84 and~$e$ $<$~0.08 (blue diamonds within red region of Fig.~12).  A concentration of 23 EBs in this small corner of the parameter space is highly unlikely considering the density of systems in the surrounding phase space is substantially smaller.  We expect only 3\,-\,5 of the 23 EBs to be twins in nearly circular orbits with the listed orbital periods.  Given only the photometric data, however, we cannot easily determine which of the systems are truly twins with small eccentricities and which have half the listed orbital periods.  In \S2, we simply excluded all 23 EBs with ambiguous periods, and we account for the incidental removal of the 3\,-\,5 genuine systems in \S5.  We emphasize that most of the 23 EBs with ambiguous orbital periods have half the listed values, and therefore it was appropriate to remove these systems.

In Fig.~13, we compare the measured eccentricities~$e$ to the arguments of periastron $\omega$ for the 130 well-defined EBs.  Assuming random orientations, the periastron angle should be uniformly distributed across 0$^{\rm o}$~$\le$~$\omega$~$<$~360$^{\rm o}$. However, the observed systems are not evenly concentrated across all $\omega$ and $e$.  We notice two observational biases in the data, both of which are due to geometrical selection effects.  

\begin{figure}[t!]
\centerline{
\includegraphics[trim = 0.3cm 0.2cm 0.2cm 0.3cm, clip=true,width=3.3in]{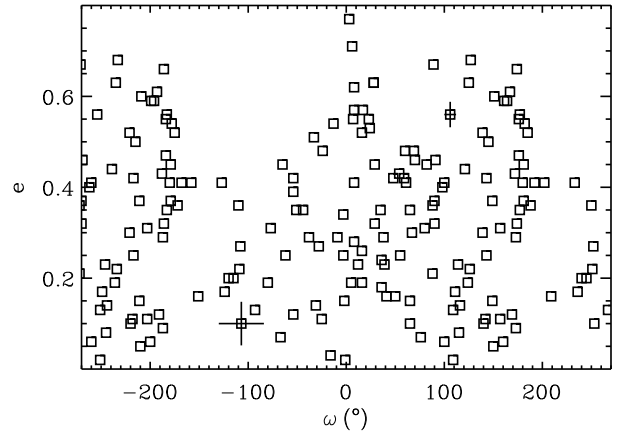}}
\caption{Measured eccentricities $e$  as a function of periastron angle $\omega$ for the 130 well-defined EBs (squares; representative errors shown for two systems). Note that we display the interval $-$270$^{\rm o}$~$<$~$\omega$~$<$~270$^{\rm o}$, so that some of the systems are repeated.  For $e$~$\gtrsim$~0.4, the concentration of EBs at $\omega$~=~0$^{\rm o}$ and $\omega$~=~$-$180$^{\rm o}$~=~180$^{\rm o}$ as well as deficit at $\omega$~=~$-$90$^{\rm o}$~=~270$^{\rm o}$ are due to geometrical selection effects.}
\end{figure}

First, for modest to large eccentricities $e$~$>$~0.4, the EBs cluster near $\omega$~=~0$^{\rm o}$ and $\omega$~=~180$^{\rm o}$.  In fact, the two systems with $e$~$\approx$~0.7\,-\,0.8 have $\omega$~$\approx$~0$^{\rm o}$.  For EBs with $\omega$~$=$~90$^{o}$ or $\omega$~$=$~270$^{o}$, one of the eclipses would occur at periastron while the other at apastron.  The eclipse at periaston would be quite narrow according to Kepler's second law, and may be too narrow to be accurately measured given the cadence of the OGLE-III data (see item B in \S2).  If the inclination is not sufficiently close to edge-on, e.g. $i$ $\approx$ 87$^{\rm o}$, then the eclipse at apastron may be too shallow to be accurately measured (again, see item B in \S2).  If the inclination was even smaller, e.g. $i$~$\approx$~85$^{\rm o}$, the projected separation at apastron could be large enough so there would be no secondary eclipse.  These systems would exhibit only one eclipse per orbit such as those presented in item A of \S2. Considering the above, it is extremely difficult to observe and measure highly eccentric EBs with eclipses that occur near periastron and apastron. As the eccentricity increases, well-defined EBs are only detected as the argument of periastron approaches $\omega$ $\approx$ 0$^{\rm o}$ or $\omega$ $\approx$ 180$^{\rm o}$.

Second, there is an overabundance of EBs with $\omega$~$\approx$~90$^{\rm o}$ relative to those with $\omega$~$\approx$~270$^{\rm o}$.  Quantitatively, there are 90 EBs with 0$^{\rm o}$~$<$~$\omega$~$<$~180$^{\rm o}$ and only 40 EBs with 180$^{\rm o}$~$<$~$\omega$~$<$~360$^{\rm o}$.  These two values are discrepant at the 4.4$\sigma$ level according to Poisson statistics.  This observational bias is due to our definition of the primary eclipse minimum $t_{\rm o}$, which determines the reference frame for $\omega$.  Recall the primary eclipse $\Delta I_1$ $>$ $\Delta I_2$ at $t_{\rm o}$ must be deeper than the secondary eclipse. If $e$ $\gtrsim$ 0.2, $i$ $\lesssim$ 89$^{\rm o}$, and the primary $M_1$ $>$ $M_2$ is eclipsed closer to apastron, then the eclipse of the most massive luminous component $M_1$ may actually coincide with the secondary eclipse $\Delta I_2$~$<$~$\Delta I_1$.  Indeed, we found 18 EBs in such a configuration whereby $M_1$ is eclipsed at $\Phi_2$ and $M_2$ is eclipsed at $t_{\rm o}$ (see \S3 and Table 2).  Sixteen of these 18 EBs have 0$^{\rm o}$~$<$~$\omega$~$<$~180$^{\rm o}$.  If we were to define $\omega$ according to $M_1$ instead of in terms of $\Delta I_1$, then 74 EBs would have 0$^{\rm o}$~$<$~$\omega$~$<$~180$^{\rm o}$ and 58 EBs would have 180$^{\rm o}$~$<$~$\omega$~$<$~360$^{\rm o}$. These two values are now consistent with each other, i.e. they only differ at the 1.4$\sigma$ significance level.

\begin{figure}[t!]
\centerline{
\includegraphics[trim = 0.3cm 0.3cm 0.2cm 0.4cm, clip=true,width=3.55in]{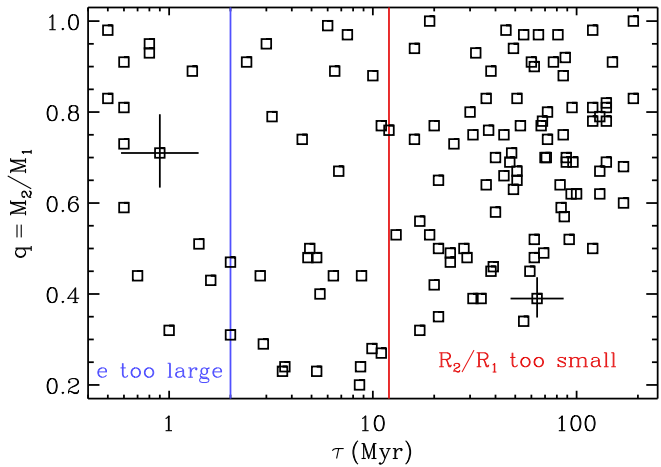}}
\caption{Measured mass ratios $q$ = $M_2$/$M_1$ as a function of age $\tau$ for the 130 well-defined EBs (squares; representative errors shown for two systems).  At young ages $\tau$~$<$~2~Myr (left of blue line), most EBs are in highly eccentric orbits with $e$~$\approx$~0.6.  Because of geometrical selection effects, it is difficult to detect these EBs, especially if they have small, low-mass companions $q$~$<$~0.3.  At older ages $\tau$~$>$~12~Myr (right of red line), the primaries are systematically larger.  The primary eclipse depths $\Delta I_1$, which are largely determined by $R_2$/$R_1$, are therefore shallower.  Given the sensitivity of the OGLE-III data, EBs with low-mass companions $q$~$<$~0.3 become undetectable as the primary evolves toward the upper MS.}
\end{figure}

As indicated above and discussed in \S2, we suspect the majority of the 48 EBs we removed in items A and B of \S2 have $e$~$>$~0.4 and either 20$^{\rm o}$~$<$~$\omega$~$<$~160$^{\rm o}$ or 200$^{\rm o}$~$<$~$\omega$~$<$~340$^{\rm o}$.  We test this hypothesis using the statistics of the measured systems in our well-defined sample.  Of the 53 EBs with $e$~$>$~0.4, 22 have $\omega$~$<$~20$^{\rm o}$, 160$^{\rm o}$~$<$~$\omega$~$<$~200$^{\rm o}$, or $\omega$~$>$~340$^{\rm o}$.  If these 22 systems are complete across the specified intervals of $\omega$, which total 80$^{\rm o}$, and if the intrinsic distribution of periastron angles is uniform, then we expect 22\,$\times$\,360$^{\rm o}$/80$^{\rm o}$~=~99~EBs with $e$~$>$~0.4.  We detected only 53 EBs with $e$~$>$~0.4, implying 99\,$-$\,53~=~46~EBs did not satisfy our selection criteria.  These 46~EBs most likley have secondary eclipses that are too narrow, too shallow, or completely absent.  This prediction of 46 missing EBs nearly matches the 48 EBs we removed in items A and B of \S2.  This consistency further demonstrates that geometrical selection effects are understood in our sample and the removal of EBs in \S2 were well-motivated.  

\begin{figure}[t!]
\centerline{
\includegraphics[trim = 0.3cm 0.3cm 0.2cm 0.4cm, clip=true,width=3.55in]{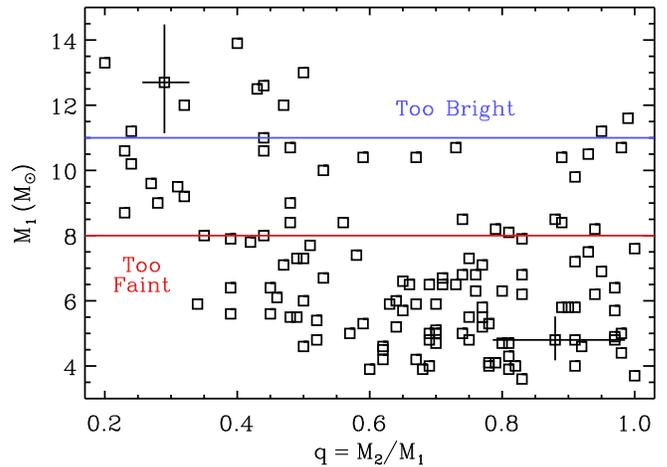}}
\caption{Measured primary masses $M_1$ versus mass ratios $q$~=~$M_2$/$M_1$ for the 130 well-defined EBS (squares; representative errors shown for two systems). Massive primaries $M_1$ $\gtrsim$ 11\,\Msun\ (above blue line) with luminous companions $q$ $>$ 0.6 are too bright to be contained in our magnitude-limited sample. Similarly, low-mass primaries $M_1$ $<$ 8 \Msun\ (below red line) with $q$ = 0.2\,-\,0.3 companions are either too faint to satisfy our photometric selection criteria and/or too old and large to produce detectable eclipses.}
\end{figure}

We compare the mass ratios $q$ to the ages $\tau$ of our 130 EBs in Fig.~14.  There is a lack of extreme mass-ratio binaries $q$ $<$ 0.3 at young ($\tau$ $<$ 2 Myr) and old ($\tau$ $>$ 12 Myr) ages.  The former is due to geometrical selection effects.  At extremely young ages, we have shown early-type binaries with intermediate orbital periods favor large eccentricities.  In fact, the median eccentricity of the 16 EBs with $\tau$~$\le$~2~Myr is $\langle e \rangle$~=~0.59.  At these large eccentricities, the eclipse that occurs closest to apastron will have a larger projected distance, and may therefore have a shallower eclipse (see above).  Shallow eclipses are easily missed given the sensitvity and cadence of the OGLE-III observations, especially if the EB contains a small, low-mass companion $q$ $<$ 0.3.

The bias against low-mass companions $q$ $<$ 0.3 at older ages is primarily due to an evolutionary selection effect.  As the primary evolves and expands, the ratio of radii $R_2$/$R_1$ decreases and the primary eclipse depth $\Delta I_1$ becomes shallower (see Fig.~5 in Paper I).  At $\tau$~$\approx$~15~Myr, only companions with $q$~$>$~0.3 produce eclipses $\Delta I_1$~$\gtrsim$~0.15~mag that are deep enough to be detected given the sensitivity of the OGLE-III data.  If the primary is near the tip of the MS, then $q$~$>$~0.45 is required to produce a visible and well-defined eclipse. Considering the above, only EBs with ages $\tau$~$\approx$~2\,-\,12 are sensitive toward low-mass companions with $q$~$\approx$~0.2\,-\,0.3.

We compare the primary masses $M_1$ to the mass ratios $q$ = $M_2$/$M_1$ in Fig.~15.  There is a clear observational bias such that massive primaries $M_1$ = 12\,-\,14\,\Msun\ contain only small mass ratios $q$ = 0.2\,-\,0.5 while late-B MS primaries with $M_1$ = 3.6\,-\,4.5\,\Msun\ include only large mass ratios $q$ = 0.6\,-\,1.0.  This trend is simply due to the magnitude limits imposed by our photometric selection criteria.  Massive MS primaries $M_1$ $\gtrsim$ 12\,\Msun\ with luminous $q$ $\gtrsim$ 0.6 MS companions will be brighter than our selection limit of $\langle I \rangle$ = 16.0.  Similarly, low-mass primaries with $M_1$ $\lesssim$ 5\,\Msun will be fainter than our detection limit of $\langle I \rangle$ = 17.6 unless there is a bright companion $q$ $\gtrsim$ 0.6 that increases the total luminosity of the system.

\begin{figure}[t!]
\centerline{
\includegraphics[trim = 0.6cm 0.3cm 0.2cm 0.4cm, clip=true,width=3.55in]{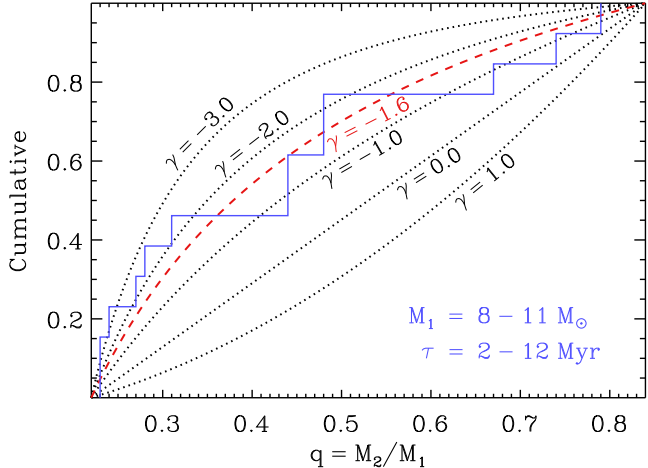}}
\caption{Cumulative distribution of mass ratios $q$~=~$M_2$/$M_1$ for the 13 EBs with $M_1$~=~8\,-\,11\,\Msun, $\tau$ = 2\,-\,12 Myr, and $q$~=~0.22\,-\,0.84 that are relatively free from selection effects (blue solid line).  Assuming the mass-ratio probability distribution $p_q$ $\propto$ $q^{\gamma}$ can be described by a power-law, we display curves for $\gamma$ = $-$3, $-$2, $-$1, 0, and 1 (dotted black).  For the 13 unbiased EBs, we measure $\gamma$ = $-$1.6\,$\pm$\,0.4 (dashed red), demonstrating binaries with massive primaries $M_1$ $\approx$ 10\,\Msun\ and orbital periods $P$ = 20\,-\,50 days are weighted toward small mass ratios $q$ $\approx$ 0.2\,-\,0.3.}
\end{figure}

The precise mass versus mass-ratio cutoffs in our sample also depend on the age of the binary.  For example, older primaries with $M_1$ $\approx$ 5\,-\,7\,\Msun\ on the upper MS will be bright enough $\langle I \rangle$ $<$ 17.6 to satisfy our photometric selection criteria.  As stated above, EBs with small mass ratios $q$ = 0.2\,-\,0.3 produce visible well-defined eclipses with $\Delta I_1$ $\gtrsim$ 0.15 mag only when the primary is relatively small and young.  However, young modererate-mass primaries $M_1$ $\approx$ 5\,-\,7\,\Msun\ with low-luminosity companions are fainter than our detection limit of $\langle I \rangle$ = 17.6.  Hence, our EB sample is sensitive to extreme mass ratios $q$ $\approx$ = 0.2\,-\,0.3 only if $M_1$ $\gtrsim$ 7\,\Msun.  To be conservative, we consider only the primary mass interval $M_1$ = 8\,-\,11\,\Msun\ to be sensitive to companions across the entire interval $q$ = 0.2\,-\,1.0 (distinguished by red and blue lines in Fig.~15).

Considering the above, our EB sample is relatively unbiased across the mass-ratio interval $q$ = 0.22\,-\,0.84 (Fig.~12), age interval $\tau$ = 2\,-\,12 Myr (Fig.~14), and primary mass interval $M_1$ = 8\,-\,11\,\Msun\ (Fig.~15). The 13 EBs that are contained in this cube of the three-dimensional phase space therefore represent a small subsample relatively free from geometrical, evolutionary, and photometric selection effects.  In Fig.~16, we display the cumulative distribution of mass ratios $q$ for these 13 EBs in our unbiased subsample.  Using a maximum liklihood technique, we fit a power-law mass-ratio probability distribution $p_q$ $\propto$ $q^{\gamma}$ to these 13 EBs. We measure $\gamma$ = $-$1.6\,$\pm$\,0.4, demonstrating binaries with massive primaries favor extreme mass ratios $q$ $\approx$ 0.2\,-\,0.3 at intermediate orbital periods $P$~=~20\,-\,50~days.

We emphasize this statistic is based on the small unbiased subsample of the 13 EBs, and therefore valid only for early-B MS primaries with $M_1$ $\approx$ 10\,\Msun.  The median primary mass in our total sample of 130 EBs is $\langle M_1 \rangle$ $\approx$ 6\,\Msun. We therefore utilize our total sample to derive more accurate statistics as well as probe the companion properties of late-B MS stars.  In the following section, we correct for selection effects so that we can make full use of all EBs in our well-defined sample. 

\section{Corrected Binary Statistics (Stage III)}

  For the final stage of our pipeline, we recover the intrinsic binary statistics and distributions by correcting for selection effects.  As done in Paper II, we first determine the probability density functions that describe the nine physical model parameters of our EBs (\S5.1).  We then calculate simple estimates for the detection efficiencies (\S5.2), and then synthesize a large population of EBs via a Monte Carlo technique (\S5.3). In \S5.4, we present our results for the intrinsic binary fraction and mass-ratio distribution.

\subsection{Probability Density Functions}

We utilize probability density functions similar to those in Paper~II.  For example, we assume random epochs of primary eclipse minima $t_{\rm o}$, and that the logarithmic orbital periods log~$P$ are uniformly distributed across $P$~=~20\,-\,50~days \citep[i.e., \"{O}pik's law;][]{Abt1983}.  We select primary masses $M_1$~=~3\,-\,30\,\Msun\ and ages $\tau$~=~0\,-\,320\,Myr according to the initial mass function (IMF) and star-formation history, respectively, measured for the OGLE-III LMC footprint in Paper~II.  In short, we fitted an IMF slope $\alpha$~=~$-2.4$ consistent with the Salpeter value and a star-formation history such that the present-day star-formation rate is approximately double the rate at earlier epochs $\tau$ = 40\,-\,320 Myr.  We assume random orientations, i.e. cos\,$i$ = 0\,-\,1 and $\omega$ = 0$^{\rm o}$\,-\,360$^{\rm o}$ are both uniformly distributed across their respective intervals.

In the present study, we account for the empirical age-extinction and age-eccentricity anticorrelations.  Given an age $\tau$, we select dust extinctions $A_I$  according to a Gaussian distribution:

\begin{equation}
 p_{A_I} \propto {\rm exp} \Big(-\frac{[A_I -  A_{I,\rm{unbiased}}(\tau)]^2}{2 \sigma_{A_I}^2}\Big)
\end{equation}

\noindent for 0 $<$ $A_I$\,(mag) $<$ 1 and where $A_{I,\rm{unbiased}}$\,($\tau$) and $\sigma_{A_I}$~=~0.08~mag derive from the fit to the unbiased subsample in Eqn.~19.  We also choose eccentricities $e$ from an age-dependent Gaussian distribution:

\begin{equation}
 p_e \propto {\rm exp} \Big(-\frac{[e -  e_{\rm fit}(\tau)]^2}{2 \sigma_{e}^2}\Big)
\end{equation}

\noindent for 0.0 $<$ $e$ $<$ 0.8 and where $e_{\rm fit}$\,($\tau$) and $\sigma_{e}$~=~0.16 derive from Eqn.~20. Finally, we consider the detection efficiencies as a continuous function of mass ratio $q$~=~0.2\,-\,1.0.  

\subsection{Simple Estimates}

Before we conduct detailed Monte Carlo simulations, we perform simple calculations to estimate the probabilities of detecting EBs with $P$ = 20\,-\,50 days.  For $q$ = 0.8\,-\,1.0 companions, the detection efficiencies are primarily dictated by two geometrical selection effects.  First, the orientations must be sufficiently close to edge-on.  About 90\% of our well-defined EBs have $i$ $>$ 86.6$^{\rm o}$, implying the probability of having the necessary inclinations to produce observable eclipses is ${\cal P}_i$ = cos\,(86.6$^{\rm o}$) = 0.06.  

Second, EBs with longer orbital periods are more likely to be missed.  Not only do binaries with longer periods require larger inclinations $i$ $\gtrsim$ 87$^{\rm o}$ to produce eclipses, but the eclipse widths can also become too narrow to be detected given the cadence of the OGLE-III LMC observations.   We found 73 well-defined EBs with $P$~=~20\,-\,30~days.  Assuming the intrinsic distribution of log~$P$ is uniform, then we would expect $\approx$92 EBs with $P$~=~30\,-\,50~days.  Our well-defined sample includes only 57 EBs with $P$~=~30\,-\,50~days, suggesting 35 systems were missed due to narrow and/or shallow eclipses.  Note that this is consistent with the 32 EBs we removed in item B of \S2 with uncertain eclipse parameters.  The probability that EBs have orbital periods that are sufficiently short is therefore ${\cal P}_P$ = 130 / (130 + 35) $\approx$ 0.8.  Considering these two factors, the probability of detecting well-defined EBs with $q$ $\approx$ 0.8\,-\,1.0 is ${\cal P}_i$\,${\cal P}_P$ = 0.06\,$\times$\,0.8 $\approx$ 5\% (red line in Fig.~17).  

For $q$ = 0.2\,-\,0.3 companions, we must also consider evolutionary and photometric selection effects.  As discussed in \S4, well-defined EBs satisfy our photometric selection criteria and are sensitive to low-mass companions only if the primaries are relatively young with ages $\tau$ = 2\,-\,12 Myr (Fig.~14) and massive with $M_1$~$\approx$~8\,-\,11\,\Msun\ (Fig.~15).  Given a typical MS lifetime of $\tau_{\rm MS}$~$\approx$~30~Myr for  $M_1$ $\approx$ 8\,-\,11\,\Msun\ primaries, then the probability of having the necessary ages $\tau$ = 2\,-\,12~Myr is ${\cal P}_{\tau}$ $\approx$ (12\,$-$\,2)/30 $\approx$ 0.3.  The smallest primary mass in our well-defined EB sample is 3.6\,\Msun.  Assuming our adopted IMF, the probability that an EB contains a massive primary $M_1$ $>$ 8\,\Msun compared to the probability of having any B-type MS primary with $M_1$ $>$ 3.6\,\Msun\ is ${\cal P}_{M_1}$ = 0.2.  Combining these additional factors, then the probability of detecting well-defined EBs with $q$~$\approx$~0.2\,-\,0.3 is ${\cal P}_i$\,${\cal P}_P$\,${\cal P}_{\tau}$\,${\cal P}_{M_1}$ = 0.06\,$\times$\,0.8\,$\times$\,0.3\,$\times$\,0.2 $\approx$ 0.3\% (blue line in Fig.~17).

\subsection{Monte Carlo Simulations}

We utilize the same technique from Paper~II to correct for incompleteness across a continuous function of mass ratios $q$.  For a given $q$, we select $M_1$, $\tau$, and $A_I$ from their respective probability density functions.  If the simulated binary does not satisfy our photometric selection criteria, we generate a new binary.  Otherwise, we keep the binary and consider its contribution toward the total number ${\cal N}_{\rm sim}$ of simulated binaries.  We then select the other physical parameters, i.e. $t_{\rm o}$, $P$, $i$, $e$, and $\omega$, from their respective probability density functions.

With the nine physical model parameters for our simulated binary, we synthesize an I-band light curve with \textsc{Nightfall}.  We match the cadence and sensitivity of the OGLE-III LMC survey.  Specifically, we sample the simulated light curve at $\langle {\cal N}_I \rangle$ = 470 random epochs and add Gaussian noise according to:

\begin{equation}
  \sigma_{I} = \big[1+10^{(I-17.0)/2}\big]\,\times\,0.0075~{\rm mag}
\end{equation}

\noindent  This equation derives from fitting the relation between the I-band magnitudes and corrected photometric uncertainties for all 221 OGLE-III LMC EBs with intermediate orbital periods $P$~=~20\,-\,50~days.

We then fit our analytic model of Gaussians (Eqn.~1) to the simulated \textsc{Nightfall} light curve.  As in \S2,  we utilize the MPFIT \citep{Markwardt2009} Levenberg-Marquardt routine to measure the values of and uncertainties in the eight analytic model parameters, e.g. $\Delta I_1$, $\Phi_2$, etc.  To be considered well-defined, we impose the same selection criteria adopted in \S2.  Namely, we require the uncertainties in the eclipse depths $\Delta I_1$ and $\Delta I_2$ and eclipse widths $\Theta_1$ and $\Theta_2$ to be $<$20\% their respective values.  We also require the fitted orbital periods to be unambiguous according to Eqn. 2. If the synthesized binary satisfies these selection criteria, we consider its contribution toward the total number ${\cal N}_{\rm well}$ of well-defined EBs.  Using a Monte Carlo technique, we repeat the above procedure until we simulate ${\cal N}_{\rm well}$ = 50 well-defined EBs for each value of $q$.  The probability of detecting well-defined EBs is simply ${\cal P}$~=~${\cal N}_{\rm well}$\,/\,${\cal N}_{\rm sim}$. 

By creating this mock data set of EBs, we find the measurement uncertainties in the eclipse depths and widths calculated by MPFIT are themselves uncertain by $\approx$20\%.  Hence, a simulated light curve with $\approx$6$\sigma$ confidence in the analytic light curve parameters may be accidentally rejected, while a system with $\approx$4$\sigma$ detections in the eclipse depths and widths may be included as a well-defined EB.  To account for this systematic uncertainty in our selection criteria, we simulate two additional sets of EB populations.  We first relax our criteria and consider EBs as well-defined if the MPFIT uncertainties in the eclipse depths $\Delta I_1$ and $\Delta I_2$ and eclipse widths $\Theta_1$ and $\Theta_2$ are $<$25\% their respective values.  For our final set of simulations, we impose a more stringent requirement that the MPFIT relative uncertainties are $<$15\%.  For each value of $q$, we therefore simulate a total of 3\,$\times$\,${\cal N}_{\rm well}$ = 150 well-defined EBs.  In this manner, we have determined the values of and uncertainties in ${\cal P}(q)$.

In Fig.~17, we display the probabilities ${\cal P}$ (and their uncertainties) of detecting well-defined EBs as a function of mass ratio $q$.  As expected, the ability to detect EBs with extreme mass ratios $q$ = 0.2\,-\,0.3 is substantially smaller than the ability to observe EBs with mass ratios near unity.  At large $q$ $>$ 0.6, the relative uncertainties in the probabilities are $\delta{\cal P}$/${\cal P}$ $\approx$ 11\%, which is only slightly larger than that expected from Poisson statistics 150$^{-\nicefrac{1}{2}}$ = 8\%.  Essentially, the majority of EBs with $q$ $>$ 0.6 have deep and accurately measured eclipse properties,  and so the precise definition of our selection criteria does not significantly affect which EBs are considered well-defined. At the smallest mass ratios $q$~=~0.2, however, the relative uncertainties more than double to $\delta{\cal P}$/${\cal P}$ $\approx$ 25\%.  All extreme mass-ratio EBs have shallow eclipses (see \S3) and are close to the detection limit, and so the probabilities ${\cal P}$ of detecting well-defined EBs are more uncertain. The probabilities ${\cal P}$ we calculate from our Monte Carlo simulations are consistent with the simple estimates derived in \S5.2.  This demonstrates the selection effects are well-understood and the probabilities  ${\cal P}$ are reliably measured.  

\begin{figure}[t!]
\centerline{
\includegraphics[trim = 0.6cm 0.3cm 0.2cm 0.4cm, clip=true,width=3.4in]{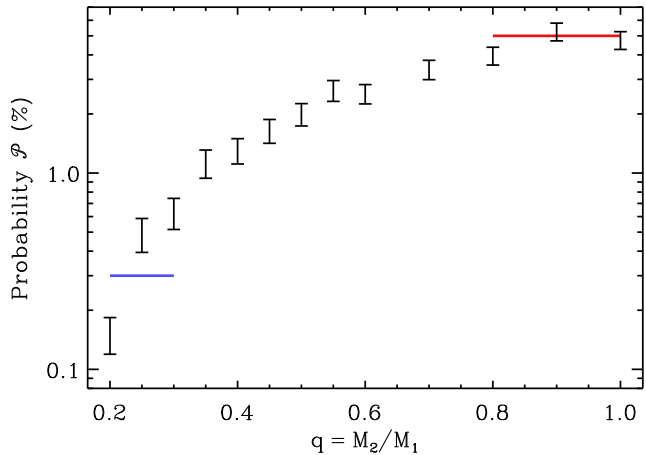}}
\caption{Probability ${\cal P}$ of detecting well-defined EBs with $P$~=~20\,-\,50~days as a function of mass ratio $q$~=~$M_2$/$M_1$.  The results of our detailed Monte Carlo simulations (black) are consistent with our simple estimates (blue and red). In addition to orientation effects, evolutionary and photometric selection effects in our magnitude-limited sample substantially reduce the detection efficiencies for well-defined EBs with extreme mass ratios $q$~$\approx$~0.2\,-\,0.3.}
\end{figure}

\subsection{Corrected Binary Fraction}

The intrinsic binary statistics are determined by weighting each well-defined EB by the inverse of their respective probability ${\cal P}$($q$) of detection as displayed in Fig.~17.  The total number of B-type MS stars with companions $q$ = 0.2\,-\,1.0 at $P$~=~20\,-\,50~days is simply ${\cal N}_{\rm comp}$~=~$\sum_{j=1}^{130} [{\cal P}(q_j)]^{-1}$~$\approx$\,6,500.  Given ${\cal N}_{\rm B}$~=~96,000 B-type MS primaries in our photometric sample, then ${\cal F}$~=~${\cal N}_{\rm comp}$\,/\,${\cal N}_{\rm B}$~=~6,500\,/\,96,000~$\approx$~6.7\% of B-type MS stars have companions with $P$~=~20\,-\,50~days and $q$~=~0.2\,-\,1.0.  The uncertainty in this fraction derives from a variety of sources.  First, the predicted number ${\cal N}_{0.2<q<0.3}$~$\approx$~2,000 of low-mass companions with $q$ = 0.2\,-\,0.3 is relatively large but also uncertain. In our sample of 130 well-defined EBs, only 8 systems have mass ratios $q$ = 0.2\,-\,0.3, and so the measurement uncertainty from Poisson statistics is 8$^{-\nicefrac{1}{2}}$~$\approx$ 35\%. The systematic uncertainty at $q$ = 0.2\,-\,0.3 is $\approx$25\% due to the uncertainty in the probabilities ${\cal P}$ of detection (see above).  The total relative uncertainty in the number of companions with $q$ = 0.2\,-\,0.3 is therefore $\approx$43\%, and so ${\cal N}_{0.2<q<0.3}$~=~2,000\,$\pm$\,900.  We repeat this calculation for the other mass ratio intervals, and find the total relative uncertainty in the number of companions is $\approx$31\%, i.e. ${\cal N}_{\rm comp}$~=~6,500\,$\pm$\,2,000.  Finally, in our Monte Carlo simulations, we account for the removal of the 74 systems represented in panels A\,-\,D of Fig.~1 from our total initial sample of 221 EBs.  We did not, however, account for the 17 EBs represented in panels E\,-\,F that exhibited variable or peculiar eclipse properties.  These systems contribute a small relative uncertainty of 17/221 = 8\%.  Hence, the total relative uncertainty in the number of companions is $\approx$33\%.  The fraction of B-type MS stars that have companions with $P$~=~20\,-\,50~days and $q$~=~0.2\,-\,1.0 is therefore ${\cal F}$~=~(6.7\,$\pm$\,2.2)\%.

Surveys for double-lined spectroscopic binaries (SB2s) with early-type primaries are generally complete for modest mass ratios $q$~$>$~0.25 and short orbital periods $P$~$<$~20~days \citep{Abt1990,Sana2012}.  In a sample of 109 B-type MS stars, \citet{Abt1990} found seven SB2s with $q$ $>$ 0.25 and $P$ =  2\,-\,20 days.  Similarly, in a sample of 71 O-type stars, \citet{Sana2012} identified 18 SB2s across the same mass-ratio and period intervals.  These statistics imply $f_{\rm logP}$ = 7\,/\,109 = 0.06\,$\pm$\,0.02 and $f_{\rm logP}$ = 18\,/\,71 = 0.25\,$\pm$\,0.06 companions with~$q$~$>$~0.25 per decade of orbital period at log\,$P$\,(days) = 0.8 for B-type and O-type stars, respectively. As discussed in Paper~II and in \citet{Chini2012}, the close binary fraction dramatically increases with primary mass.

Based on our B-type MS EBs, we measure a (5.6\,$\pm$\,1.4)\% corrected binary fraction across $P$~=~20\,-\,50~days and $q$~$>$~0.25.  This results in $f_{\rm logP}$ = (0.056\,$\pm$\,0.014)\,/\,(log\,50\,$-$\,log\,20) = 0.14\,$\pm$\,0.04 companions with $q$~$>$~0.25 per decade of orbital period centered at log\,$P$\,(days) = 1.5.  This value is consistent with the early-type spectroscopic binary fraction measured at short orbital periods $P$ $<$ 20 days, implying the intrinsic period distribution of early-type binaries closely resembles \"{O}pik's law \citep{Abt1983,Abt1990}.  

\subsection{Corrected Mass-ratio Distribution}

In Fig.~18, we display the cumulative distribution of mass ratios $q$ after weighting each well-defined EB by the inverse of their respective probability ${\cal P}$ of producing observable eclipses.  By fitting a power-law probability distribution $p_q$ $\propto$ $q^{\gamma}$ to the corrected mass-ratio distribution, we measure $\gamma$ = $-$1.1\,$\pm$\,0.3.  This is consistent with our estimate in \S4 (Fig.~16) of $\gamma$~=~$-$1.4\,$\pm$\,0.3 based on a relatively unbiased subsample of 13 young EBs with early-B MS primaries $M_1$~$\approx$~10\,\Msun.  In both cases, binaries with B-type MS primaries and orbital periods $P$~=~20\,-\,50~days favor small mass ratios $q$~=~0.2\,-\,0.3.  

\begin{figure}[t!]
\centerline{
\includegraphics[trim = 0.6cm 0.3cm 0.2cm 0.4cm, clip=true,width=3.45in]{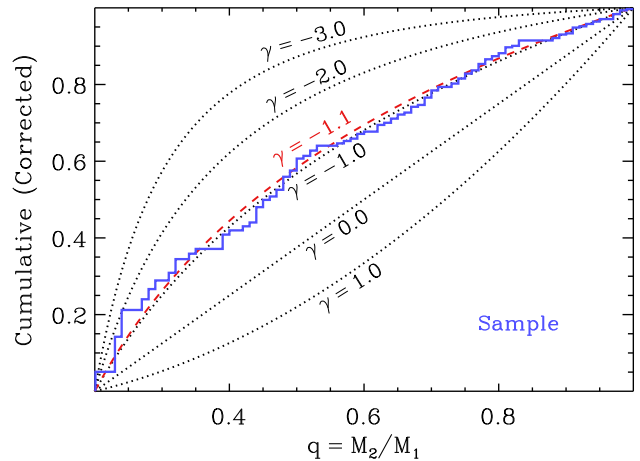}}
\caption{Cumulative distribution of mass ratios $q$~=~$M_2$/$M_1$~=~0.2\,-\,1.0 for the 130 EBs in our well-defined sample (blue) after weighting each system by the inverse of their respective detection probability ${\cal P}(q)$.  Assuming the mass-ratio probability distribution can be described by a power-law $p_q$~$\propto$~$q^{\gamma}$, we display curves for exponents $\gamma$ = $-$3, $-$2, $-$1, 0, and 1 (dotted from top to bottom). After correcting for selection effects, we measure $\gamma$~=~$-$1.1\,$\pm$\,0.3 (dashed red), demonstrating early-type binaries with intermediate orbital periods are weighted toward extreme mass ratios.}
\end{figure}

Observations of early-type spectroscopic and eclipsing binaries with short orbital periods $P$ = 2\,-\,20 days reveal a mass-ratio probability distribution that is only slightly weighted toward small values, e.g. $\gamma$ $\approx$ $-$0.9\,-\,$-$0.2  \citep[][Papers I and II]{Abt1990,Sana2012}.  In addition to the power-law component, close massive binaries with $P$ $<$ 20 days exhibit a small excess of twins with $q$ $\gtrsim$ 0.9 \citep[][Paper I]{Tokovinin2000,Pinsonneault2006}. The preponderance of close binaries with moderate mass ratios and excess of twins suggest early-type binaries with $P$ $<$ 20 days coevolved via fragmentation and competitive accretion in the circumbinary disk \citep{Abt1990,Tokovinin2000,Bonnell2005,Kouwenhoven2009}.

In contrast to close massive binaries with $P$~$<$~20~days, we find early-type binaries at moderate orbital periods $P$ = 20\,-\,50 days are even further weighted toward extreme mass ratios, i.e. $\gamma$ = $-$1.1\,$\pm$\,0.3 for our total sample and $\gamma$ = $-$1.6\,$\pm$\,0.4 for early-B MS primaries.  In addition, there is no evidence for an excess population of twins at intermediate orbital periods.  Previous spectroscopic surveys have indicated that the mass-ratio probability distribution becomes weighted toward smaller values with increasing orbital period \citep{Abt1990,Kobulnicky2014}.  However, this result is primarily based on the smaller frequency of SB2s at intermediate orbital periods, especially when compared to the frequency of single-lined spectroscopic binaries.  In the present study, we have measured the mass-ratio probability distribution at intermediate orbital periods. Our results indicate that early-type binaries at slightly longer orbital periods $P$~=~20\,-\,50~days have experienced less coevolution. In a future paper (Moe et al., in prep.), we will analyze early-type binaries discovered through other observational techniques, e.g. long-baseline interferometry, adaptive optics, common proper motion, etc., and investigate this anticorrelation between $P$ and $q$ in a more thorough and self-consistent manner.

\section{Summary}

{\it Eclipsing Binary Sample (\S2)}.  We analyzed the 221 EBs in the OGLE-III LMC database with B-type MS primaries and orbital periods $P$ = 20\,-\,50 days.  After fitting analytic models of Gaussians to the observed light curves, we identified 130 detached EBs that exhibit two well-defined eclipses per orbit.  The remaining 90 EBs have uncertain, peculiar, and/or variable eclipse properties, including 12 systems that displayed changes in the secondary eclipse parameters most likely due to orbital motion with a tertiary companion. 

{\it Physical Models (\S3)}. We developed an automated procedure to robustly and quickly fit detailed physical models to the EB light curves.  Our algorithm can be adapted for any population of detached EBs with known distances and MS primaries.  We implemented our procedure on our 130 detached well-defined EBs to measure their intrinsic physical properties, including their ages $\tau$, component masses $M_1$ and $M_2$, dust extinctions $A_I$, and eccentricities $e$.  We incorporated various techniques to demonstrate the uniqueness and robustness of the model solutions as well as the accuracy of the model parameters.    

{\it Age-Extinction Anticorrelation (\S4)}.  Even after considering selection effects, we find the ages $\tau$ and dust extinctions $A_I$ are anticorrelated ($\rho$ = $-$0.23) at a statistically significant level ($p$ = 0.02). This suggests young stars with $\tau$~$<$~10~Myr are embedded in dusty envelopes and/or molecular clouds with $A_I$~$\approx$~0.35~mag, while older stars with $\tau$~$>$~100~Myr reside in less attenuating environments with $A_I$~$\approx$~0.25~mag.  This empirical relation between $\tau$ and $A_I$ should prove beneficial when modeling stellar populations.  

{\it Age-Eccentricity Anticorrelation (\S4)}.  We also discover the ages $\tau$ and eccentricities $e$ are anticorrelated ($\rho$ = $-$0.39) at a statistically significant level ($p$~=~5\,$\times$\,10$^{-6}$) due to tidal evolution.  The slope in the observed trend provides a diagnostic for the radiative damping constant via dynamical tides in highly eccentric binaries with hot MS components.  We note the tidal circularization timescales $e$/$\dot{e}$ in highly eccentric binaries with $e$ $\approx$ 0.5\,-\,0.8 may be orders of magnitude shorter than the circularization timescales when the eccentricities $e$ $\lesssim$ 0.4 are already small.  

{\it Initial Eccentricity Distribution (\S4)}.  We find that massive binaries at $P$ = 20\,-\,50 days are initially born with larger eccentricities $\langle e \rangle$ $\approx$ 0.6.  Assuming a power-law eccentricity probability distribution $p_e$~$\propto$~$e^{\eta}$, we measure $\eta$~=~0.8\,$\pm$\,0.3 for our young early-type EBs with $\tau$~$\le$~10~Myr.  This is consistent with a Maxwellian ``thermal'' eccentricity distribution ($\eta$~=~1), which indicates massive binaries with intermediate orbital periods formed via dynamical interactions, either through tidal / disk capture,  dynamical perturbations in a dense cluster, three-body exchanges, and/or Kozai cycles with a tertiary companion.  

{\it Binary Fraction (\S5)}. After utilizing a Monte Carlo technique to correct for selection effects, we measure that (6.7\,$\pm$\,2.2)\% of B-type MS stars have companions with $P$ = 20\,-\,50 days and $q$ = 0.2\,-\,1.0.  The frequency of companions per decade of orbital period at log~$P$\,(days)~=~1.5 is consistent with spectroscopic observations of close massive binaries at log~$P$\,(days)~=~0.8.  This suggests the intrinsic period distribution of binary companions to B-type MS stars closely resembles \"{O}pik's law for $P$~$<$~50~days.    

{\it Mass-ratio Distribution (\S5)}.  In our corrected binary sample with B-type MS primaries $\langle M_1 \rangle$ = 6\,\Msun, we measure a mass-ratio probability distribution $p_q$ $\propto$ $q^{\gamma}$ weighted toward small values ($\gamma$ = $-$1.1\,$\pm$\,0.3).  There is a slight indication that binaries with early-B MS primaries $\langle M_1 \rangle$ = 10\,\Msun\ are even further skewed toward extreme mass ratios ($\gamma$ = $-$1.6\,$\pm$\,0.4).  Close massive binaries with $P$ $<$ 20 days favor moderate mass ratios and exhibit a small excess of twin components $q$ $\gtrsim$ 0.9. This indicates our early-type MS binaries with intermediate orbital periods $P$~=~20\,-\,50~days have experienced substantially less coevolution via fragmentation and competitive accretion in the circumbinary disk.

This work was supported in part by support from NSF AST-1211843, AST-0708924 and AST-0908878 and NASA NNX12AE39G and  AR-13243.01-A.  We thank the anonymous referee for the useful comments that improved the quality of the manuscript. We also acknowledge use of publicly available data from the Optical Gravitational Lensing Experiment \citep{Udalski2008,Graczyk2011}.

\bibliographystyle{apj}                       
\bibliography{bibliography}

\afterpage{

\begin{figure*}[t!]\footnotesize
{\small {\bf Table 1:} Analytic model parameters that describe the basic light curve features for the 221 EBs with OGLE-III~LMC catalog properties 16.0~$<$~$\langle I \rangle$~$<$~17.6, $-$0.25 $<$ $\langle V - I \rangle$ $<$ 0.20, and $P$ = 20\,-\,50 days.  Based on the measured analytic model parameters, we divide the total sample into eight categories: (1) EBs without secondary eclipses, (2) EBs with uncertain eclipse parameters, (3)~Roche-lobe filling EBs, (4) EBs with ambiguous orbital periods, (5) intrinsic variables, (6) EBs with variable eclipses, (7)~peculiar EBs, and (8) detached well-defined EBs.  For each category, we list the OGLE-III LMC catalog properties \citep{Graczyk2011} including the identification number, mean color $\langle V-I \rangle$, and number ${\cal N}_I$ of I-band measurements.  We then list the eight best-fit analytic model parameters: orbital period $P$ (days), epoch of primary eclipse minimum $t_{\rm o}$ (Julian date\,$-$\,2450000), mean magnitude $\langle I \rangle$, primary and secondary eclipse depths $\Delta I_1$ and $\Delta I_2$ (mag), orbital phase of secondary eclipse $\Phi_2$, and eclipse widths  $\Theta_1$ and $\Theta_2$ (fraction of the orbital period).  Finally, we list the fit statistics, including the correction factor $f_{\sigma,I}$ in the photometric errors, number ${\cal N}_{\rm c}$ of clipped data points, and goodness-of-fit statistic $\chi^2_{\rm G}/\nu$. } 

~

{\small Category 1: list of 16 EBs without visible secondary eclipses.  These EBs most likely have a certain combination of $e$, $\omega$, and $i$ so there is only one eclipse per orbit.}

~
\vspace*{-0.47cm}
\begin{center}
\begin{tabular}{|r|r|c|c|l|c|c|c|c|c|c|c|c|c|}
\hline
\multicolumn{3}{|c|}{Catalog Properties} & \multicolumn{8}{c|}{Analytic Model Parameters} & \multicolumn{3}{c|}{Fit Statistics}\\
\hline
ID~  & $\langle V-I \rangle$ & ${\cal N}_I$ & $P$ & ~~~~~$t_{\rm o}$ &
     $\langle I \rangle$ & $\Delta I_1$ & $\Delta I_2$ & $\Phi_2$ & $\Theta_1$ & $\Theta_2$ &
     $f_{\sigma,I}$ & ${\cal N}_{\rm c}$ & $\chi^2_{\rm G}/\nu$ \\
\hline
  316 &    0.01~ &  211 & 24.9947 & 3538.528  & 17.34 & 0.22 &   -  &   -   & 0.0074 &    -   & 1.00 & 0 & 1.01 \\ 
\hline
 2164 & $-$0.04~ &  440 & 43.0029 & 3569.803  & 17.34 & 0.40 &   -  &   -   & 0.0029 &    -   & 1.09 & 1 & 1.01 \\ 
\hline
 3091 & $-$0.08~ &  424 & 27.4499 & 3602.602  & 17.14 & 0.21 &   -  &   -   & 0.0054 &    -   & 1.14 & 0 & 1.09 \\ 
\hline
 4652 &    0.02~ &  445 & 23.7441 & 3572.959  & 16.61 & 0.10 &   -  &   -   & 0.0040 &    -   & 1.06 & 2 & 1.03 \\ 
\hline
 5548 & $-$0.13~ &  466 & 23.5235 & 3585.619  & 17.23 & 0.33 &   -  &   -   & 0.0027 &    -   & 1.09 & 0 & 1.02 \\ 
\hline
 5704 & $-$0.12~ &  421 & 25.2059 & 3566.122  & 16.25 & 0.18 &   -  &   -   & 0.0044 &    -   & 1.20 & 0 & 1.04 \\ 
\hline
 5973 & $-$0.05~ &  468 & 26.6085 & 3635.421  & 16.29 & 0.42 &   -  &   -   & 0.0039 &    -   & 1.20 & 0 & 1.08 \\ 
\hline
 9850 &    0.03~ &  422 & 37.3082 & 3634.111  & 17.39 & 0.49 &   -  &   -   & 0.0025 &    -   & 1.11 & 1 & 1.03 \\ 
\hline
12084 & $-$0.03~ &  477 & 29.8000 & 3588.434  & 16.90 & 0.18 &   -  &   -   & 0.0037 &    -   & 1.15 & 0 & 1.02 \\ 
\hline
12913 & $-$0.08~ &  493 & 25.2768 & 3589.836  & 16.08 & 0.09 &   -  &   -   & 0.0050 &    -   & 1.23 & 0 & 1.04 \\ 
\hline
13991 & $-$0.02~ &  449 & 30.6988 & 3592.784  & 17.16 & 0.41 &   -  &   -   & 0.0029 &    -   & 1.08 & 2 & 1.02 \\ 
\hline
17232 & $-$0.03~ &  454 & 26.9222 & 3639.974  & 16.24 & 0.12 &   -  &   -   & 0.0023 &    -   & 1.37 & 2 & 1.07 \\ 
\hline
21007 & $-$0.04~ &  367 & 33.7279 & 3600.454  & 16.40 & 0.09 &   -  &   -   & 0.0097 &    -   & 1.14 & 1 & 1.04 \\ 
\hline
22467 & $-$0.14~ &  437 & 20.2896 & 3593.730  & 16.46 & 0.14 &   -  &   -   & 0.0028 &    -   & 1.06 & 0 & 1.03 \\ 
\hline
23086 &    0.08~ &  434 & 25.0090 & 3608.833  & 17.30 & 0.21 &   -  &   -   & 0.0017 &    -   & 1.09 & 0 & 1.02 \\ 
\hline
25112 & $-$0.04~ &  391 & 39.9125 & 3581.880  & 16.07 & 0.09 &   -  &   -   & 0.0083 &    -   & 1.49 & 0 & 1.04 \\ 
\hline
\end{tabular}
\end{center}

~

{\small Category 2: list of 32 EBs with uncertain eclipse parameters.  These EBs generally have eclipses that are too narrow ($\Theta$~$\lesssim$~0.003) and/or too shallow ($\Delta I$ $\lesssim$ 0.15 mag) to be accurately measured.}

~

\vspace*{-0.47cm}
\begin{center}
\begin{tabular}{|r|r|c|c|l|c|c|c|c|c|c|c|c|c|}
\hline
\multicolumn{3}{|c|}{Catalog Properties} & \multicolumn{8}{c|}{Analytic Model Parameters} & \multicolumn{3}{c|}{Fit Statistics}\\
\hline
ID~  & $\langle V-I \rangle$ & ${\cal N}_I$ & $P$ & ~~~~~$t_{\rm o}$ &
     $\langle I \rangle$ & $\Delta I_1$ & $\Delta I_2$ & $\Phi_2$ & $\Theta_1$ & $\Theta_2$ &
     $f_{\sigma,I}$ & ${\cal N}_{\rm c}$ & $\chi^2_{\rm G}/\nu$ \\
\hline
  219 &    0.01~ &  405 & 49.9100 & 3626.670  & 17.53 & 0.42 & 0.56 & 0.390 & 0.0090 & 0.0011 & 1.00 & 1 & 0.99 \\ 
\hline
 1450 &    0.11~ &  465 & 28.5096 & 3573.275  & 17.52 & 0.38 & 0.17 & 0.426 & 0.0045 & 0.0048 & 1.17 & 0 & 1.01 \\ 
\hline
 1924 &    0.06~ &  464 & 31.0128 & 3607.649  & 16.89 & 0.25 & 0.09 & 0.651 & 0.0053 & 0.0070 & 1.15 & 0 & 1.00 \\ 
\hline
 2539 & $-$0.06~ &  457 & 38.1274 & 3585.424  & 17.07 & 0.14 & 0.13 & 0.756 & 0.0033 & 0.0021 & 1.03 & 1 & 1.02 \\ 
\hline
 2843 & $-$0.05~ &  440 & 26.0647 & 3599.963  & 17.57 & 0.17 & 0.46 & 0.746 & 0.0096 & 0.0013 & 1.07 & 0 & 0.98 \\ 
\hline
 3492 &    0.06~ &  457 & 20.8220 & 3598.229  & 17.58 & 0.48 & 0.09 & 0.769 & 0.0033 & 0.0074 & 1.15 & 0 & 1.02 \\ 
\hline
 3745 &    0.03~ &  225 & 26.5908 & 3575.121  & 17.58 & 0.41 & 0.53 & 0.527 & 0.0094 & 0.0015 & 1.11 & 0 & 1.08 \\ 
\hline
 4095 & $-$0.05~ &  432 & 33.9572 & 3615.866  & 17.39 & 0.19 & 0.15 & 0.936 & 0.0061 & 0.0022 & 1.14 & 2 & 1.00 \\ 
\hline
 4396 & $-$0.03~ &  358 & 25.3069 & 3598.022  & 17.14 & 0.14 & 0.09 & 0.542 & 0.0060 & 0.0059 & 1.20 & 1 & 1.10 \\ 
\hline
 5257 & $-$0.02~ &  463 & 35.1506 & 3575.663  & 17.29 & 0.16 & 0.07 & 0.472 & 0.0044 & 0.0036 & 1.05 & 2 & 0.99 \\ 
\hline
 6494 & $-$0.12~ &  421 & 45.3346 & 3623.790  & 16.96 & 0.18 & 0.14 & 0.606 & 0.0030 & 0.0017 & 1.25 & 0 & 1.01 \\ 
\hline
 7832 & $-$0.07~ &  476 & 26.6117 & 3613.760  & 17.35 & 0.12 & 0.06 & 0.468 & 0.0059 & 0.0016 & 1.05 & 0 & 1.02 \\ 
\hline
 7954 & $-$0.05~ &  435 & 27.1161 & 3606.847  & 17.25 & 0.20 & 0.32 & 0.709 & 0.0044 & 0.0030 & 1.11 & 1 & 1.00 \\ 
\hline
 8824 &    0.01~ &  437 & 37.6744 & 3617.323  & 17.56 & 0.18 & 0.18 & 0.277 & 0.0052 & 0.0037 & 1.14 & 1 & 1.06 \\ 
\hline
10248 & $-$0.14~ &  557 & 43.9229 & 3633.148  & 16.36 & 0.35 & 0.72 & 0.258 & 0.0047 & 0.0005 & 1.49 & 2 & 1.00 \\ 
\hline
11655 & $-$0.02~ &  477 & 30.0005 & 3595.741  & 16.29 & 0.10 & 0.07 & 0.431 & 0.0056 & 0.0150 & 1.23 & 0 & 1.00 \\ 
\hline
12065 & $-$0.08~ &  459 & 29.0408 & 3563.836  & 16.81 & 0.20 & 0.21 & 0.806 & 0.0048 & 0.0025 & 1.09 & 1 & 1.06 \\ 
\hline
12202 & $-$0.17~ &  477 & 41.4788 & 3572.545  & 16.76 & 0.35 & 0.45 & 0.217 & 0.0048 & 0.0013 & 1.11 & 2 & 1.03 \\ 
\hline
12696 & $-$0.09~ &  457 & 40.3637 & 3587.966  & 16.76 & 0.14 & 0.05 & 0.373 & 0.0050 & 0.0033 & 1.46 & 1 & 1.02 \\ 
\hline
13076 & $-$0.08~ &  493 & 33.6509 & 3637.571  & 16.01 & 0.06 & 0.03 & 0.436 & 0.0036 & 0.0023 & 1.07 & 1 & 1.01 \\ 
\hline
14307 &    0.01~ &  540 & 22.3483 & 3601.494  & 17.58 & 0.12 & 0.23 & 0.290 & 0.0055 & 0.0041 & 1.00 & 1 & 1.01 \\ 
\hline
16651 & $-$0.10~ &  449 & 45.3897 & 3684.769  & 17.54 & 0.34 & 0.69 & 0.549 & 0.0045 & 0.0015 & 1.06 & 0 & 1.02 \\ 
\hline
16922 &    0.02~ &  580 & 41.6338 & 3670.541  & 17.58 & 0.13 & 0.30 & 0.186 & 0.0037 & 0.0018 & 1.22 & 2 & 1.00 \\ 
\hline
17204 & $-$0.01~ &  473 & 31.9060 & 3650.939  & 17.60 & 0.23 & 0.07 & 0.566 & 0.0032 & 0.0028 & 1.06 & 0 & 1.00 \\ 
\hline
17262 & $-$0.14~ &  626 & 33.8546 & 3650.508  & 16.50 & 0.34 & 0.19 & 0.499 & 0.0014 & 0.0016 & 1.30 & 2 & 1.02 \\ 
\hline
17957 & $-$0.11~ &  626 & 47.9194 & 3623.161  & 17.23 & 0.19 & 0.05 & 0.492 & 0.0030 & 0.0050 & 1.15 & 2 & 1.07 \\ 
\hline
18800 &    0.03~ &  470 & 22.0031 & 3582.822  & 17.47 & 0.26 & 0.17 & 0.325 & 0.0044 & 0.0050 & 1.10 & 0 & 1.02 \\ 
\hline
20667 &    0.10~ &  417 & 35.3260 & 3627.664  & 17.54 & 0.36 & 0.60 & 0.498 & 0.0022 & 0.0014 & 1.05 & 1 & 0.97 \\ 
\hline
22464 & $-$0.08~ &  437 & 20.7824 & 3575.419  & 17.38 & 0.30 & 0.10 & 0.499 & 0.0034 & 0.0036 & 1.00 & 2 & 0.99 \\ 
\hline
22512 &    0.08~ &  437 & 25.1642 & 3588.758  & 17.41 & 0.22 & 0.25 & 0.484 & 0.0091 & 0.0440 & 1.31 & 0 & 0.99 \\ 
\hline
22853 & $-$0.10~ &  430 & 23.4011 & 3637.436  & 16.94 & 0.19 & 0.06 & 0.398 & 0.0020 & 0.0049 & 1.12 & 1 & 1.02 \\ 
\hline
23330 & $-$0.02~ &  414 & 43.4464 & 3616.771  & 16.94 & 0.26 & 0.35 & 0.529 & 0.0035 & 0.0029 & 1.28 & 0 & 1.01 \\ 
\hline
\end{tabular}
\end{center}
\end{figure*}

\begin{figure*}[p]\footnotesize
{\small {\bf Table 1 (cont.): }}

~

{\small Category 3: list of 3 Roche-lobe filling EBs, as demonstrated by their wide eclipses $\Theta$ $>$ 0.06.}

~
\vspace*{-0.47cm}
\begin{center}
\begin{tabular}{|r|r|c|c|l|c|c|c|c|c|c|c|c|c|}
\hline
\multicolumn{3}{|c|}{Catalog Properties} & \multicolumn{8}{c|}{Analytic Model Parameters} & \multicolumn{3}{c|}{Fit Statistics}\\
\hline
ID~  & $\langle V-I \rangle$ & ${\cal N}_I$ & $P$ & ~~~~~$t_{\rm o}$ &
     $\langle I \rangle$ & $\Delta I_1$ & $\Delta I_2$ & $\Phi_2$ & $\Theta_1$ & $\Theta_2$ &
     $f_{\sigma,I}$ & ${\cal N}_{\rm c}$ & $\chi^2_{\rm G}/\nu$ \\
\hline
 3864 & $-$0.09~ &  473 & 26.6937 & 3602.081  & 16.41 & 0.02 & 0.02 & 0.497 & 0.0640 & 0.0670 & 1.13 & 2 & 0.89 \\ 
\hline
16199 & $-$0.04~ &  460 & 46.3408 & 3632.722  & 17.18 & 0.08 & 0.08 & 0.496 & 0.0609 & 0.0668 & 1.00 & 0 & 1.22 \\ 
\hline
25591 &    0.18~ &  423 & 20.7526 & 3588.389  & 16.01 & 0.52 & 0.31 & 0.503 & 0.0777 & 0.0856 & 1.41 & 2 & 3.61 \\ 
\hline
\end{tabular}
\end{center}

~

{\small Category 4: list of 23 EBs with ambiguous orbital periods. These systems have $\Delta I_1$ $\approx$ $\Delta I_2$, $\Phi_2$ $\approx$ 0.5, and $\Theta_1$ $\approx$ $\Theta_2$ given the listed orbital periods.  The majority of these EBs most likely have half the listed orbital periods, and therefore exhibit only one eclipse per orbit such as the systems listed in Category 1.  }

~

\vspace*{-0.47cm}
\begin{center}
\begin{tabular}{|r|r|c|c|l|c|c|c|c|c|c|c|c|c|}
\hline
\multicolumn{3}{|c|}{Catalog Properties} & \multicolumn{8}{c|}{Analytic Model Parameters} & \multicolumn{3}{c|}{Fit Statistics}\\
\hline
ID~  & $\langle V-I \rangle$ & ${\cal N}_I$ & $P$ & ~~~~~$t_{\rm o}$ &
     $\langle I \rangle$ & $\Delta I_1$ & $\Delta I_2$ & $\Phi_2$ & $\Theta_1$ & $\Theta_2$ &
     $f_{\sigma,I}$ & ${\cal N}_{\rm c}$ & $\chi^2_{\rm G}/\nu$ \\
\hline
  675 &    0.08~ &  442 & 20.5259 & 3584.169  & 17.29 & 0.44 & 0.47 & 0.500 & 0.0079 & 0.0071 & 1.08 & 0 & 1.05 \\ 
\hline
  885 & $-$0.04~ &  456 & 23.5967 & 3564.996  & 17.31 & 0.33 & 0.33 & 0.500 & 0.0028 & 0.0027 & 1.15 & 0 & 1.01 \\ 
\hline
 1478 &    0.14~ &  460 & 32.0021 & 3628.807  & 17.09 & 0.14 & 0.14 & 0.501 & 0.0020 & 0.0026 & 1.24 & 1 & 1.03 \\ 
\hline
 8321 &    0.00~ &  467 & 34.8297 & 3643.253  & 17.12 & 0.11 & 0.11 & 0.502 & 0.0031 & 0.0025 & 1.05 & 2 & 1.02 \\ 
\hline
 8376 &    0.01~ &  468 & 36.9128 & 3638.843  & 17.33 & 0.31 & 0.23 & 0.500 & 0.0025 & 0.0029 & 1.15 & 0 & 1.00 \\ 
\hline
 9146 & $-$0.04~ &  559 & 21.1796 & 3606.499  & 17.40 & 0.25 & 0.25 & 0.500 & 0.0023 & 0.0021 & 1.11 & 0 & 1.00 \\ 
\hline
11930 & $-$0.24~ &  477 & 42.0006 & 3542.808  & 16.06 & 0.21 & 0.20 & 0.500 & 0.0012 & 0.0013 & 1.26 & 1 & 1.03 \\ 
\hline
11931 &    0.06~ &  493 & 24.6953 & 3569.815  & 16.67 & 0.27 & 0.29 & 0.500 & 0.0050 & 0.0048 & 1.11 & 0 & 1.05 \\ 
\hline
14753 & $-$0.01~ &  600 & 22.4059 & 3580.305  & 17.41 & 0.29 & 0.30 & 0.500 & 0.0023 & 0.0024 & 1.16 & 1 & 1.02 \\ 
\hline
15309 &    0.06~ &  566 & 44.9548 & 3621.008  & 16.54 & 0.18 & 0.18 & 0.501 & 0.0029 & 0.0019 & 1.29 & 0 & 1.00 \\ 
\hline
17257 & $-$0.03~ &  626 & 30.1341 & 3586.993  & 16.48 & 0.30 & 0.28 & 0.500 & 0.0016 & 0.0019 & 1.19 & 2 & 1.00 \\ 
\hline
17407 & $-$0.06~ &  626 & 25.8992 & 3612.024  & 16.92 & 0.11 & 0.12 & 0.499 & 0.0053 & 0.0048 & 1.24 & 2 & 1.08 \\ 
\hline
17715 & $-$0.07~ &  626 & 35.5355 & 3616.239  & 16.73 & 0.13 & 0.13 & 0.499 & 0.0022 & 0.0024 & 1.32 & 1 & 1.03 \\ 
\hline
18138 &    0.12~ &  601 & 42.9000 & 3599.789  & 16.92 & 0.39 & 0.38 & 0.500 & 0.0024 & 0.0024 & 1.23 & 0 & 1.03 \\ 
\hline
19309 & $-$0.09~ &  612 & 32.1801 & 3593.017  & 16.13 & 0.08 & 0.09 & 0.500 & 0.0015 & 0.0018 & 1.34 & 0 & 1.01 \\ 
\hline
19582 & $-$0.11~ &  473 & 20.0314 & 3608.284  & 16.59 & 0.08 & 0.10 & 0.498 & 0.0024 & 0.0026 & 1.15 & 1 & 1.03 \\ 
\hline
19612 & $-$0.01~ &  605 & 20.8496 & 3616.468  & 17.14 & 0.46 & 0.41 & 0.500 & 0.0026 & 0.0029 & 1.18 & 0 & 1.03 \\ 
\hline
19651 & $-$0.06~ &  473 & 26.4068 & 3598.679  & 16.83 & 0.25 & 0.19 & 0.499 & 0.0026 & 0.0034 & 1.20 & 1 & 0.99 \\ 
\hline
20441 &    0.08~ &  437 & 23.5402 & 3608.235  & 17.21 & 0.12 & 0.13 & 0.498 & 0.0042 & 0.0051 & 1.00 & 0 & 0.98 \\ 
\hline
20661 &    0.07~ &  436 & 20.5829 & 3618.710  & 17.41 & 0.15 & 0.13 & 0.499 & 0.0036 & 0.0045 & 1.16 & 0 & 0.99 \\ 
\hline
21273 &    0.04~ &  213 & 20.7646 & 3598.007  & 16.77 & 0.21 & 0.25 & 0.499 & 0.0031 & 0.0037 & 1.07 & 0 & 1.02 \\ 
\hline
21477 & $-$0.02~ &  436 & 23.3476 & 3602.139  & 17.52 & 0.27 & 0.23 & 0.499 & 0.0044 & 0.0035 & 1.00 & 2 & 0.97 \\ 
\hline
24604 & $-$0.02~ &  840 & 38.0192 & 3591.216  & 17.04 & 0.21 & 0.17 & 0.500 & 0.0028 & 0.0033 & 1.34 & 0 & 0.99 \\ 
\hline
\end{tabular}
\end{center}

~

{\small Category 5: list of 3 EBs that are intrinsic variables, as indicated by their large rms scatter $f_{\sigma, I}$ $\gtrsim$ 1.6.}

~

\vspace*{-0.47cm}
\begin{center}
\begin{tabular}{|r|r|c|c|l|c|c|c|c|c|c|c|c|c|}
\hline
\multicolumn{3}{|c|}{Catalog Properties} & \multicolumn{8}{c|}{Analytic Model Parameters} & \multicolumn{3}{c|}{Fit Statistics}\\
\hline
ID~  & $\langle V-I \rangle$ & ${\cal N}_I$ & $P$ & ~~~~~$t_{\rm o}$ &
     $\langle I \rangle$ & $\Delta I_1$ & $\Delta I_2$ & $\Phi_2$ & $\Theta_1$ & $\Theta_2$ &
     $f_{\sigma,I}$ & ${\cal N}_{\rm c}$ & $\chi^2_{\rm G}/\nu$ \\
\hline
 3414 & $-$0.12~ &  479 & 38.9097 & 3607.475  & 17.02 & 0.40 & 0.50 & 0.897 & 0.0041 & 0.0021 & 2.78 & 0 & 1.02 \\ 
\hline
 7651 & $-$0.01~ &  435 & 34.8819 & 3595.802  & 16.64 & 0.14 & 0.12 & 0.267 & 0.0075 & 0.0038 & 2.25 & 0 & 0.99 \\ 
\hline
22929 & $-$0.06~ &  764 & 26.4805 & 3597.217  & 16.63 & 0.26 & 0.22 & 0.772 & 0.0068 & 0.0064 & 1.56 & 2 & 1.06 \\ 
\hline
\end{tabular}
\end{center}

~

{\small Category 6: list of 12 EBs with variable eclipses. These EBs either have more than ${\cal N}_c$ $>$ 2 bad data points near the eclipses or, more likely, exhibit variations in the eclipse parameters due to orbital motion with a tertiary companion. }

~

\vspace*{-0.47cm}
\begin{center}
\begin{tabular}{|r|r|c|c|l|c|c|c|c|c|c|c|c|c|}
\hline
\multicolumn{3}{|c|}{Catalog Properties} & \multicolumn{8}{c|}{Analytic Model Parameters} & \multicolumn{3}{c|}{Fit Statistics}\\
\hline
ID~  & $\langle V-I \rangle$ & ${\cal N}_I$ & $P$ & ~~~~~$t_{\rm o}$ &
     $\langle I \rangle$ & $\Delta I_1$ & $\Delta I_2$ & $\Phi_2$ & $\Theta_1$ & $\Theta_2$ &
     $f_{\sigma,I}$ & ${\cal N}_{\rm c}$ & $\chi^2_{\rm G}/\nu$ \\
\hline
3112 & $-$0.12~ &  456 & 20.8143 & 3569.074  & 16.96 & 0.23 & 0.33 & 0.498 & 0.0023 & 0.0012 & 1.07 & 2 & 1.16 \\ 
\hline
 3233 & $-$0.10~ &  476 & 32.8948 & 3573.927  & 16.66 & 0.15 & 0.11 & 0.272 & 0.0039 & 0.0038 & 1.34 & 2 & 1.02 \\ 
\hline
 7992 &    0.16~ &  858 & 49.8658 & 3624.125  & 16.75 & 0.21 & 0.09 & 0.477 & 0.0052 & 0.0005 & 1.35 & 2 & 1.06 \\ 
\hline
 8612 &    0.02~ &  476 & 27.0844 & 3607.949  & 16.27 & 0.30 & 0.26 & 0.462 & 0.0079 & 0.0054 & 1.08 & 2 & 2.82 \\ 
\hline
12973 & $-$0.07~ &  605 & 30.8419 & 3590.014  & 16.69 & 0.20 & 0.15 & 0.511 & 0.0045 & 0.0060 & 1.30 & 2 & 1.63 \\ 
\hline
12987 &    0.02~ &  493 & 25.9121 & 3584.179  & 17.41 & 0.18 & 0.13 & 0.363 & 0.0034 & 0.0065 & 1.03 & 2 & 1.08 \\ 
\hline
14083 &    0.01~ &  577 & 29.9824 & 3583.922  & 17.51 & 0.19 & 0.20 & 0.657 & 0.0054 & 0.0045 & 1.31 & 2 & 1.03 \\ 
\hline
17017 & $-$0.10~ &  606 & 25.7548 & 3637.727  & 16.29 & 0.12 & 0.11 & 0.519 & 0.0041 & 0.0007 & 1.20 & 2 & 1.10 \\ 
\hline
18037 &    0.00~ &  433 & 31.5140 & 3563.204  & 17.00 & 0.45 & 0.36 & 0.573 & 0.0047 & 0.0063 & 1.06 & 2 & 1.46 \\ 
\hline
19353 &    0.02~ &  437 & 22.2690 & 3587.215  & 17.14 & 0.27 & 0.34 & 0.448 & 0.0077 & 0.0058 & 1.01 & 2 & 1.24 \\ 
\hline
20313 &    0.04~ &  369 & 27.7087 & 3602.991  & 16.40 & 0.37 & 0.59 & 0.496 & 0.0012 & 0.0031 & 1.30 & 2 & 1.38 \\ 
\hline
23350 &    0.07~ &  402 & 28.3841 & 3623.853  & 17.40 & 0.46 & 0.63 & 0.369 & 0.0029 & 0.0011 & 1.25 & 2 & 1.10 \\ 
\hline
\end{tabular}
\end{center}

~

{\small Category 7: list of 2 peculiar EBs that exhibit variations between eclipses. }

~

\vspace*{-0.47cm}
\begin{center}
\begin{tabular}{|r|r|c|c|l|c|c|c|c|c|c|c|c|c|}
\hline
\multicolumn{3}{|c|}{Catalog Properties} & \multicolumn{8}{c|}{Analytic Model Parameters} & \multicolumn{3}{c|}{Fit Statistics}\\
\hline
ID~  & $\langle V-I \rangle$ & ${\cal N}_I$ & $P$ & ~~~~~$t_{\rm o}$ &
     $\langle I \rangle$ & $\Delta I_1$ & $\Delta I_2$ & $\Phi_2$ & $\Theta_1$ & $\Theta_2$ &
     $f_{\sigma,I}$ & ${\cal N}_{\rm c}$ & $\chi^2_{\rm G}/\nu$ \\
\hline
 343 &    0.13~ &  437 & 33.5508 & 3572.393  & 17.35 & 0.27 & 0.17 & 0.694 & 0.0124 & 0.0087 & 1.30 & 1 & 1.03 \\ 
\hline
~4458 & $-$0.07~ &  445 & 39.5436 & 3581.850  & 16.14 & 0.11 & 0.22 & 0.501 & 0.0029 & 0.0007 & 1.14 & 2 & 1.41 \\ 
\hline
\end{tabular}
\end{center}

\end{figure*}

\renewcommand{\arraystretch}{0.93}
\begin{figure*}[t!]\footnotesize
{\small {\bf Table 1 (cont.): }}

~

\vspace*{-0.15cm}
{\small Category 8: list of 130 detached EBs with well-defined eclipse parameters.}

~
\vspace*{-0.47cm}
\begin{center}
\begin{tabular}{|r|r|c|c|l|c|c|c|c|c|c|c|c|c|}
\hline
\multicolumn{3}{|c|}{Catalog Properties} & \multicolumn{8}{c|}{Analytic Model Parameters} & \multicolumn{3}{c|}{Fit Statistics}\\
\hline
ID~  & $\langle V-I \rangle$ & ${\cal N}_I$ & $P$ & ~~~~~$t_{\rm o}$ &
     $\langle I \rangle$ & $\Delta I_1$ & $\Delta I_2$ & $\Phi_2$ & $\Theta_1$ & $\Theta_2$ &
     $f_{\sigma,I}$ & ${\cal N}_{\rm c}$ & $\chi^2_{\rm G}/\nu$ \\
\hline
   91 &    0.06~ &  411 & 24.8099 & 3619.170  & 16.80 & 0.45 & 0.32 & 0.459 & 0.0080 & 0.0114 & 1.28 & 0 & 1.02 \\ 
\hline
  170 & $-$0.03~ &  426 & 26.3720 & 3566.276  & 17.10 & 0.38 & 0.10 & 0.155 & 0.0036 & 0.0066 & 1.18 & 0 & 1.05 \\ 
\hline
  784 &    0.00~ &  444 & 44.1196 & 3633.801  & 16.72 & 0.50 & 0.45 & 0.115 & 0.0026 & 0.0035 & 1.18 & 2 & 1.02 \\ 
\hline
  866 &    0.06~ &  424 & 28.0132 & 3598.047* & 16.47 & 0.63 & 0.49 & 0.496 & 0.0046 & 0.0101 & 1.46 & 0 & 1.04 \\ 
\hline
 1056 &    0.01~ &  444 & 30.6713 & 3647.862* & 17.40 & 0.34 & 0.15 & 0.503 & 0.0037 & 0.0065 & 1.11 & 0 & 1.04 \\ 
\hline
 1530 &    0.06~ &  465 & 42.1989 & 3632.166  & 17.42 & 0.21 & 0.14 & 0.655 & 0.0044 & 0.0033 & 1.09 & 2 & 1.02 \\ 
\hline
 1968 &    0.13~ &  911 & 44.8940 & 3627.360  & 17.33 & 0.25 & 0.18 & 0.437 & 0.0035 & 0.0038 & 1.33 & 0 & 1.01 \\ 
\hline
 2142 &    0.01~ &  457 & 27.7692 & 3595.213  & 16.57 & 0.27 & 0.19 & 0.162 & 0.0047 & 0.0050 & 1.23 & 0 & 1.05 \\ 
\hline
 2277 &    0.01~ &  465 & 36.5936 & 3616.509  & 17.08 & 0.35 & 0.29 & 0.473 & 0.0039 & 0.0041 & 1.32 & 2 & 1.00 \\ 
\hline
 2708 &    0.04~ &  465 & 44.5975 & 3633.684  & 17.48 & 0.38 & 0.36 & 0.422 & 0.0066 & 0.0031 & 1.13 & 1 & 1.03 \\ 
\hline
 2780 & $-$0.12~ &  446 & 27.0438 & 3628.748* & 17.03 & 0.53 & 0.48 & 0.375 & 0.0024 & 0.0088 & 1.00 & 0 & 1.02 \\ 
\hline
 3082 & $-$0.06~ &  876 & 45.5327 & 3589.237  & 16.71 & 0.12 & 0.06 & 0.710 & 0.0024 & 0.0024 & 1.38 & 2 & 1.00 \\ 
\hline
 3177 & $-$0.12~ &  434 & 20.2216 & 3585.258  & 16.70 & 0.19 & 0.06 & 0.455 & 0.0045 & 0.0106 & 1.13 & 1 & 1.04 \\ 
\hline
 3388 & $-$0.08~ &  448 & 44.9697 & 3588.586* & 16.53 & 0.56 & 0.39 & 0.168 & 0.0019 & 0.0049 & 1.29 & 1 & 1.00 \\ 
\hline
 3557 &    0.00~ &  412 & 23.8734 & 3592.035  & 17.03 & 0.27 & 0.12 & 0.478 & 0.0048 & 0.0062 & 1.46 & 1 & 1.01 \\ 
\hline
 4031 & $-$0.10~ &  448 & 32.5105 & 3577.358  & 16.46 & 0.41 & 0.22 & 0.821 & 0.0045 & 0.0035 & 1.31 & 0 & 1.04 \\ 
\hline
 4399 &    0.03~ &  447 & 22.9067 & 3624.060* & 16.76 & 0.14 & 0.12 & 0.644 & 0.0068 & 0.0074 & 1.03 & 2 & 1.03 \\ 
\hline
 4419 &    0.07~ &  434 & 42.0950 & 3575.980  & 16.09 & 0.49 & 0.44 & 0.415 & 0.0032 & 0.0046 & 1.33 & 1 & 1.02 \\ 
\hline
 4721 & $-$0.05~ &  456 & 21.8793 & 3592.692  & 17.05 & 0.21 & 0.16 & 0.511 & 0.0059 & 0.0073 & 1.03 & 2 & 1.06 \\ 
\hline
 4737 &    0.07~ &  440 & 31.5717 & 3612.900* & 17.51 & 0.35 & 0.29 & 0.644 & 0.0036 & 0.0080 & 1.09 & 1 & 1.00 \\ 
\hline
 4804 & $-$0.02~ &  429 & 23.0375 & 3601.271  & 16.23 & 0.14 & 0.08 & 0.463 & 0.0046 & 0.0048 & 1.34 & 1 & 1.00 \\ 
\hline
 4837 & $-$0.02~ &  445 & 26.8055 & 3608.319* & 17.56 & 0.35 & 0.33 & 0.576 & 0.0050 & 0.0043 & 1.11 & 2 & 1.01 \\ 
\hline
 5145 &    0.00~ &  445 & 47.6341 & 3684.851  & 17.55 & 0.15 & 0.12 & 0.173 & 0.0036 & 0.0033 & 1.05 & 0 & 1.00 \\ 
\hline
 5153 & $-$0.04~ &  456 & 24.4530 & 3582.856* & 16.28 & 0.23 & 0.21 & 0.829 & 0.0039 & 0.0031 & 1.09 & 1 & 1.02 \\ 
\hline
 5195 & $-$0.06~ &  445 & 39.4245 & 3564.482  & 17.19 & 0.20 & 0.13 & 0.439 & 0.0064 & 0.0047 & 1.12 & 2 & 1.06 \\ 
\hline
 5492 & $-$0.04~ &  429 & 32.9664 & 3602.498  & 16.27 & 0.36 & 0.15 & 0.280 & 0.0042 & 0.0085 & 1.31 & 1 & 1.03 \\ 
\hline
 5965 & $-$0.05~ &  439 & 20.3374 & 3616.353  & 17.54 & 0.39 & 0.22 & 0.684 & 0.0041 & 0.0059 & 1.29 & 2 & 1.02 \\ 
\hline
 6187 &    0.00~ &  477 & 29.8706 & 3612.492* & 16.91 & 0.11 & 0.10 & 0.537 & 0.0042 & 0.0085 & 1.11 & 1 & 1.08 \\ 
\hline
 6555 & $-$0.02~ &  477 & 29.0548 & 3591.351  & 17.19 & 0.57 & 0.54 & 0.657 & 0.0071 & 0.0038 & 1.16 & 1 & 1.00 \\ 
\hline
 6996 & $-$0.04~ &  476 & 29.9285 & 3596.250* & 16.54 & 0.50 & 0.49 & 0.445 & 0.0041 & 0.0096 & 1.05 & 2 & 1.05 \\ 
\hline
 7380 &    0.02~ &  468 & 31.4810 & 3597.756* & 17.23 & 0.25 & 0.17 & 0.564 & 0.0036 & 0.0045 & 1.13 & 2 & 1.00 \\ 
\hline
 7560 & $-$0.01~ &  418 & 26.7461 & 3611.788* & 16.49 & 0.28 & 0.24 & 0.575 & 0.0036 & 0.0032 & 1.21 & 2 & 1.09 \\ 
\hline
 7565 &    0.13~ &  477 & 29.0955 & 3601.070  & 17.45 & 0.19 & 0.17 & 0.667 & 0.0103 & 0.0051 & 1.08 & 0 & 1.01 \\ 
\hline
 7935 & $-$0.01~ &  404 & 24.0615 & 3609.910  & 17.39 & 0.32 & 0.22 & 0.449 & 0.0045 & 0.0051 & 1.13 & 1 & 1.05 \\ 
\hline
 7975 & $-$0.05~ &  477 & 20.1226 & 3615.517* & 16.86 & 0.11 & 0.06 & 0.607 & 0.0031 & 0.0072 & 1.11 & 2 & 1.03 \\ 
\hline
 8543 &    0.00~ &  451 & 25.3380 & 3621.313  & 17.07 & 0.41 & 0.35 & 0.414 & 0.0067 & 0.0057 & 1.68 & 1 & 1.05 \\ 
\hline
 8559 & $-$0.06~ &  468 & 42.2643 & 3631.168* & 17.54 & 0.63 & 0.48 & 0.239 & 0.0019 & 0.0045 & 1.13 & 2 & 1.04 \\ 
\hline
 8783 &    0.08~ &  458 & 48.7243 & 3622.458  & 17.44 & 0.21 & 0.12 & 0.170 & 0.0041 & 0.0041 & 1.02 & 1 & 1.06 \\ 
\hline
 8903 & $-$0.05~ &  463 & 25.6605 & 3637.923* & 16.49 & 0.22 & 0.22 & 0.593 & 0.0049 & 0.0050 & 1.89 & 1 & 0.99 \\ 
\hline
 8993 & $-$0.09~ &  891 & 29.4599 & 3599.445* & 17.11 & 0.56 & 0.41 & 0.617 & 0.0024 & 0.0073 & 1.03 & 2 & 0.99 \\ 
\hline
 9159 &    0.05~ &  466 & 48.1575 & 3695.306  & 17.34 & 0.39 & 0.36 & 0.622 & 0.0038 & 0.0054 & 1.16 & 0 & 1.00 \\ 
\hline
 9386 & $-$0.07~ &  559 & 29.8063 & 3619.572* & 16.61 & 0.40 & 0.19 & 0.682 & 0.0030 & 0.0062 & 1.21 & 1 & 1.00 \\ 
\hline
 9429 &    0.11~ &  476 & 20.6681 & 3605.574* & 17.15 & 0.11 & 0.08 & 0.532 & 0.0075 & 0.0126 & 1.02 & 1 & 1.08 \\ 
\hline
 9441 &    0.05~ &  476 & 20.8997 & 3596.643  & 17.45 & 0.26 & 0.11 & 0.432 & 0.0058 & 0.0075 & 1.09 & 0 & 1.00 \\ 
\hline
 9953 & $-$0.10~ &  490 & 45.5783 & 3729.995* & 17.05 & 0.25 & 0.22 & 0.140 & 0.0025 & 0.0032 & 1.12 & 2 & 1.05 \\ 
\hline
10096 &    0.05~ &  431 & 35.2251 & 3628.920  & 17.58 & 0.28 & 0.20 & 0.317 & 0.0049 & 0.0060 & 1.20 & 1 & 1.01 \\ 
\hline
10422 & $-$0.11~ &  612 & 21.1694 & 3586.720* & 16.91 & 0.29 & 0.26 & 0.841 & 0.0032 & 0.0045 & 1.26 & 1 & 1.02 \\ 
\hline
10575 & $-$0.05~ &  559 & 29.1436 & 3592.083  & 17.14 & 0.24 & 0.12 & 0.546 & 0.0057 & 0.0049 & 1.04 & 2 & 1.04 \\ 
\hline
10953 & $-$0.12~ &  616 & 42.2762 & 3610.382  & 16.31 & 0.14 & 0.08 & 0.265 & 0.0059 & 0.0044 & 1.34 & 1 & 1.08 \\ 
\hline
11252 & $-$0.04~ &  490 & 22.2402 & 3592.677  & 17.19 & 0.59 & 0.56 & 0.496 & 0.0067 & 0.0052 & 1.15 & 1 & 1.02 \\ 
\hline
11299 &    0.07~ & 1049 & 39.5635 & 3584.911* & 17.22 & 0.49 & 0.45 & 0.214 & 0.0055 & 0.0063 & 1.29 & 1 & 1.01 \\ 
\hline
11526 & $-$0.01~ &  477 & 49.2837 & 3595.345  & 16.67 & 0.20 & 0.19 & 0.418 & 0.0035 & 0.0041 & 1.13 & 0 & 1.09 \\ 
\hline
11538 & $-$0.17~ &  477 & 33.7813 & 3600.922* & 16.28 & 0.56 & 0.17 & 0.504 & 0.0017 & 0.0092 & 1.18 & 2 & 1.02 \\ 
\hline
11636 &    0.08~ &  474 & 31.1510 & 3579.485  & 17.59 & 0.26 & 0.24 & 0.612 & 0.0054 & 0.0069 & 1.13 & 1 & 1.09 \\ 
\hline
11907 & $-$0.04~ &  440 & 45.4195 & 3629.747  & 17.20 & 0.28 & 0.22 & 0.589 & 0.0031 & 0.0047 & 1.17 & 0 & 1.00 \\ 
\hline
12170 & $-$0.04~ &  452 & 29.1744 & 3586.509  & 17.60 & 0.43 & 0.33 & 0.439 & 0.0060 & 0.0045 & 1.09 & 0 & 1.01 \\ 
\hline
12179 & $-$0.01~ &  477 & 25.1020 & 3555.863* & 16.84 & 0.31 & 0.30 & 0.485 & 0.0032 & 0.0098 & 1.14 & 0 & 0.99 \\ 
\hline
12384 &    0.12~ &  476 & 25.1786 & 3591.906  & 17.12 & 0.10 & 0.09 & 0.251 & 0.0116 & 0.0096 & 1.14 & 2 & 1.08 \\ 
\hline
12454 & $-$0.14~ &  476 & 30.1523 & 3591.689  & 17.01 & 0.20 & 0.16 & 0.649 & 0.0043 & 0.0029 & 1.11 & 2 & 1.04 \\ 
\hline
12832 & $-$0.08~ &  493 & 20.1668 & 3551.653* & 16.37 & 0.10 & 0.09 & 0.425 & 0.0047 & 0.0049 & 1.30 & 0 & 1.04 \\ 
\hline
13177 & $-$0.01~ &  477 & 40.5158 & 3542.700  & 17.11 & 0.32 & 0.17 & 0.813 & 0.0034 & 0.0058 & 1.17 & 2 & 1.00 \\ 
\hline
13260 &    0.00~ &  450 & 24.2957 & 3593.162  & 17.21 & 0.29 & 0.27 & 0.447 & 0.0095 & 0.0066 & 1.07 & 0 & 1.16 \\ 
\hline
13390 &    0.10~ &  477 & 35.7858 & 3571.607  & 17.59 & 0.35 & 0.24 & 0.630 & 0.0026 & 0.0057 & 1.15 & 1 & 1.01 \\ 
\hline
13418 & $-$0.11~ &  465 & 26.9111 & 3587.289* & 16.94 & 0.57 & 0.46 & 0.907 & 0.0025 & 0.0040 & 1.10 & 0 & 1.03 \\ 
\hline
13441 & $-$0.04~ &  493 & 30.6531 & 3593.354  & 17.26 & 0.22 & 0.17 & 0.618 & 0.0042 & 0.0044 & 1.24 & 1 & 1.03 \\ 
\hline

\end{tabular}
\end{center}
\vspace*{-0.25cm}
~~~~~~~~~~~~~~~*Epoch of primary eclipse minimum $t_{\rm o}$ appropriately adjusted to ensure $\Delta I_1$ $>$ $\Delta I_2$.  

\end{figure*}

\begin{figure*}[p]\footnotesize
{\small {\bf Table 1 (cont.): }}

~

\vspace*{-0.15cm}
{\small Category 8 (cont.): list of 130 detached EBs with well-defined eclipse parameters.}

~
\vspace*{-0.47cm}
\begin{center}
\begin{tabular}{|r|r|c|c|l|c|c|c|c|c|c|c|c|c|}
\hline
\multicolumn{3}{|c|}{Catalog Properties} & \multicolumn{8}{c|}{Analytic Model Parameters} & \multicolumn{3}{c|}{Fit Statistics}\\
\hline
ID~  & $\langle V-I \rangle$ & ${\cal N}_I$ & $P$ & ~~~~~$t_{\rm o}$ &
     $\langle I \rangle$ & $\Delta I_1$ & $\Delta I_2$ & $\Phi_2$ & $\Theta_1$ & $\Theta_2$ &
     $f_{\sigma,I}$ & ${\cal N}_{\rm c}$ & $\chi^2_{\rm G}/\nu$ \\
\hline
13482 & $-$0.08~ &  493 & 21.5789 & 3548.902* & 17.32 & 0.46 & 0.43 & 0.659 & 0.0050 & 0.0058 & 1.10 & 2 & 0.99 \\ 
\hline
13491 &    0.10~ &  477 & 21.3014 & 3578.036* & 16.82 & 0.31 & 0.30 & 0.510 & 0.0069 & 0.0069 & 1.10 & 2 & 1.07 \\ 
\hline
13726 &    0.10~ &  493 & 20.9344 & 3554.051  & 17.37 & 0.29 & 0.07 & 0.679 & 0.0050 & 0.0069 & 1.13 & 2 & 1.03 \\ 
\hline
13867 & $-$0.02~ &  831 & 21.7457 & 3604.874  & 17.33 & 0.33 & 0.21 & 0.493 & 0.0047 & 0.0051 & 1.65 & 1 & 1.00 \\ 
\hline
14171 & $-$0.02~ &  457 & 21.7291 & 3621.792  & 16.95 & 0.21 & 0.05 & 0.528 & 0.0043 & 0.0051 & 1.05 & 2 & 1.02 \\ 
\hline
14360 &    0.00~ &  540 & 30.2612 & 3651.363* & 16.20 & 0.15 & 0.13 & 0.595 & 0.0038 & 0.0072 & 1.00 & 1 & 0.99 \\ 
\hline
14895 &    0.06~ &  506 & 34.4641 & 3574.345* & 17.33 & 0.60 & 0.57 & 0.186 & 0.0040 & 0.0034 & 1.31 & 0 & 1.01 \\ 
\hline
15235 & $-$0.10~ &  726 & 28.9500 & 3553.091* & 17.17 & 0.33 & 0.27 & 0.317 & 0.0045 & 0.0047 & 1.08 & 2 & 1.05 \\ 
\hline
15244 &    0.07~ &  446 & 21.8435 & 3622.895  & 17.41 & 0.45 & 0.27 & 0.299 & 0.0064 & 0.0071 & 1.02 & 2 & 1.02 \\ 
\hline
15380 & $-$0.09~ &  325 & 27.6171 & 3624.679* & 16.79 & 0.17 & 0.12 & 0.802 & 0.0028 & 0.0045 & 1.09 & 0 & 1.05 \\ 
\hline
15788 & $-$0.06~ &  600 & 29.0140 & 3561.929* & 17.04 & 0.18 & 0.17 & 0.246 & 0.0031 & 0.0033 & 1.27 & 2 & 1.04 \\ 
\hline
15979 & $-$0.12~ &  449 & 28.1428 & 3635.572  & 16.27 & 0.22 & 0.20 & 0.637 & 0.0077 & 0.0032 & 1.09 & 1 & 1.08 \\ 
\hline
16026 &    0.00~ &  449 & 30.8480 & 3566.099  & 16.40 & 0.56 & 0.43 & 0.661 & 0.0038 & 0.0103 & 1.00 & 0 & 0.87 \\ 
\hline
16126 &    0.02~ &  600 & 22.2111 & 3609.525  & 17.49 & 0.34 & 0.20 & 0.234 & 0.0038 & 0.0075 & 1.58 & 1 & 0.99 \\ 
\hline
16350 & $-$0.10~ &  599 & 30.9801 & 3603.346* & 17.09 & 0.47 & 0.31 & 0.746 & 0.0027 & 0.0046 & 1.16 & 0 & 1.02 \\ 
\hline
16399 &    0.01~ &  456 & 39.1458 & 3653.057* & 17.45 & 0.51 & 0.28 & 0.849 & 0.0020 & 0.0039 & 1.00 & 0 & 1.00 \\ 
\hline
16418 &    0.03~ &  482 & 20.4164 & 3586.945  & 17.08 & 0.12 & 0.06 & 0.619 & 0.0070 & 0.0079 & 1.05 & 0 & 1.03 \\ 
\hline
16711 & $-$0.13~ &  449 & 26.5576 & 3632.458* & 16.25 & 0.35 & 0.25 & 0.833 & 0.0027 & 0.0056 & 1.02 & 1 & 1.04 \\ 
\hline
16964 & $-$0.10~ &  606 & 23.5422 & 3603.149  & 16.44 & 0.22 & 0.05 & 0.846 & 0.0031 & 0.0052 & 1.08 & 0 & 1.03 \\ 
\hline
17067 & $-$0.14~ &  581 & 28.1151 & 3592.517  & 16.57 & 0.19 & 0.09 & 0.666 & 0.0033 & 0.0070 & 1.24 & 0 & 1.02 \\ 
\hline
17316 & $-$0.14~ &  986 & 26.7747 & 3611.392  & 16.68 & 0.27 & 0.27 & 0.775 & 0.0056 & 0.0027 & 1.00 & 2 & 1.00 \\ 
\hline
17361 &    0.03~ &  472 & 20.9318 & 3562.002  & 17.48 & 0.36 & 0.30 & 0.551 & 0.0111 & 0.0065 & 1.00 & 0 & 0.98 \\ 
\hline
17539 &    0.06~ &  473 & 23.9094 & 3562.789* & 17.51 & 0.36 & 0.24 & 0.471 & 0.0053 & 0.0082 & 1.04 & 2 & 0.99 \\ 
\hline
17569 & $-$0.08~ &  626 & 22.2019 & 3591.019  & 16.91 & 0.29 & 0.23 & 0.523 & 0.0049 & 0.0035 & 1.18 & 2 & 1.12 \\ 
\hline
17750 & $-$0.11~ &  473 & 20.7086 & 3586.855  & 16.98 & 0.13 & 0.06 & 0.750 & 0.0066 & 0.0072 & 1.10 & 1 & 1.01 \\ 
\hline
17784 & $-$0.09~ &  626 & 25.2613 & 3575.400  & 17.11 & 0.17 & 0.12 & 0.280 & 0.0038 & 0.0031 & 1.28 & 1 & 1.01 \\ 
\hline
17822 &    0.07~ &  437 & 44.2706 & 3641.930  & 16.75 & 0.13 & 0.12 & 0.579 & 0.0072 & 0.0048 & 1.09 & 1 & 1.01 \\ 
\hline
18237 & $-$0.10~ &  588 & 20.3285 & 3616.375  & 16.03 & 0.17 & 0.14 & 0.481 & 0.0105 & 0.0092 & 1.63 & 1 & 1.14 \\ 
\hline
18582 &    0.02~ &  441 & 41.9856 & 3608.318* & 16.60 & 0.29 & 0.25 & 0.539 & 0.0047 & 0.0065 & 1.09 & 2 & 1.06 \\ 
\hline
18659 &    0.04~ &  456 & 21.4449 & 3594.435  & 17.14 & 0.28 & 0.27 & 0.665 & 0.0079 & 0.0048 & 1.19 & 1 & 1.03 \\ 
\hline
18813 &    0.11~ &  605 & 39.6989 & 3627.894* & 16.92 & 0.53 & 0.44 & 0.498 & 0.0041 & 0.0078 & 1.22 & 0 & 1.02 \\ 
\hline
18824 & $-$0.08~ &  626 & 33.5726 & 3576.478  & 16.54 & 0.21 & 0.08 & 0.439 & 0.0044 & 0.0064 & 1.13 & 1 & 1.03 \\ 
\hline
18839 & $-$0.04~ &  473 & 48.2133 & 3619.336* & 17.28 & 0.25 & 0.24 & 0.269 & 0.0037 & 0.0036 & 1.00 & 1 & 0.91 \\ 
\hline
18859 &    0.00~ &  435 & 24.5592 & 3624.022  & 17.01 & 0.55 & 0.35 & 0.505 & 0.0038 & 0.0096 & 1.08 & 0 & 1.00 \\ 
\hline
18869 & $-$0.08~ &  572 & 42.5786 & 3628.587  & 16.90 & 0.15 & 0.08 & 0.648 & 0.0039 & 0.0023 & 1.17 & 0 & 1.00 \\ 
\hline
19083 &    0.02~ &  601 & 22.2719 & 3600.521  & 16.85 & 0.21 & 0.08 & 0.517 & 0.0062 & 0.0062 & 1.26 & 2 & 1.09 \\ 
\hline
19230 &    0.04~ &  473 & 37.4005 & 3569.116  & 16.89 & 0.10 & 0.08 & 0.280 & 0.0040 & 0.0041 & 1.03 & 0 & 0.98 \\ 
\hline
19792 & $-$0.10~ &  912 & 29.6438 & 3640.465  & 16.00 & 0.10 & 0.07 & 0.519 & 0.0047 & 0.0042 & 1.52 & 0 & 1.01 \\ 
\hline
19840 &    0.02~ &  625 & 30.9709 & 3617.985  & 16.15 & 0.12 & 0.03 & 0.449 & 0.0080 & 0.0046 & 1.26 & 1 & 1.04 \\ 
\hline
20309 & $-$0.02~ &  428 & 48.2396 & 3694.899  & 17.60 & 0.48 & 0.43 & 0.443 & 0.0034 & 0.0039 & 1.00 & 0 & 1.01 \\ 
\hline
20459 & $-$0.14~ &  436 & 30.7408 & 3642.410* & 17.22 & 0.48 & 0.24 & 0.590 & 0.0033 & 0.0039 & 1.06 & 2 & 1.02 \\ 
\hline
20522 &    0.07~ &  423 & 27.2575 & 3619.621  & 17.40 & 0.19 & 0.16 & 0.457 & 0.0058 & 0.0073 & 1.14 & 1 & 1.06 \\ 
\hline
20590 &    0.06~ &  428 & 40.0925 & 3618.364* & 17.32 & 0.17 & 0.16 & 0.563 & 0.0063 & 0.0058 & 1.07 & 0 & 1.11 \\ 
\hline
20646 &    0.04~ &  437 & 26.0462 & 3620.172  & 16.91 & 0.18 & 0.17 & 0.656 & 0.0068 & 0.0066 & 1.09 & 0 & 1.02 \\ 
\hline
20746 &    0.08~ &  436 & 28.3719 & 3614.200* & 16.69 & 0.57 & 0.50 & 0.575 & 0.0045 & 0.0090 & 1.03 & 0 & 1.02 \\ 
\hline
21059 &    0.07~ &  428 & 25.7904 & 3611.136* & 17.56 & 0.24 & 0.19 & 0.674 & 0.0043 & 0.0046 & 1.06 & 0 & 0.98 \\ 
\hline
21518 &    0.02~ &  433 & 21.4601 & 3629.580  & 17.10 & 0.41 & 0.34 & 0.443 & 0.0074 & 0.0077 & 1.05 & 2 & 1.08 \\ 
\hline
21621 &    0.04~ &  444 & 38.5314 & 3576.469* & 17.20 & 0.51 & 0.37 & 0.341 & 0.0032 & 0.0075 & 1.00 & 2 & 0.96 \\ 
\hline
21881 &    0.04~ &  428 & 21.0056 & 3588.139* & 16.80 & 0.28 & 0.18 & 0.354 & 0.0050 & 0.0073 & 1.10 & 0 & 1.05 \\ 
\hline
22082 & $-$0.06~ &  436 & 33.1149 & 3586.199  & 16.76 & 0.20 & 0.08 & 0.341 & 0.0056 & 0.0027 & 1.08 & 0 & 1.04 \\ 
\hline
22553 &    0.01~ &  419 & 22.8430 & 3598.817  & 17.31 & 0.41 & 0.29 & 0.804 & 0.0042 & 0.0061 & 1.00 & 2 & 0.96 \\ 
\hline
22691 &    0.00~ &  434 & 31.8865 & 3579.266* & 16.89 & 0.53 & 0.21 & 0.235 & 0.0018 & 0.0068 & 1.26 & 0 & 0.98 \\ 
\hline
22713 & $-$0.09~ &  437 & 33.3752 & 3617.549  & 17.00 & 0.12 & 0.09 & 0.679 & 0.0037 & 0.0035 & 1.11 & 1 & 1.02 \\ 
\hline
22764 &    0.00~ &  428 & 29.7044 & 3572.606* & 16.63 & 0.60 & 0.53 & 0.155 & 0.0026 & 0.0050 & 1.13 & 0 & 1.03 \\ 
\hline
23088 &    0.09~ &  424 & 22.4312 & 3598.919  & 17.43 & 0.20 & 0.11 & 0.495 & 0.0065 & 0.0070 & 1.09 & 0 & 1.07 \\ 
\hline
23101 &    0.08~ &  427 & 46.3819 & 3616.728* & 16.79 & 0.59 & 0.22 & 0.196 & 0.0014 & 0.0094 & 1.02 & 1 & 1.06 \\ 
\hline
23368 &    0.03~ &  426 & 22.9011 & 3629.347* & 17.29 & 0.25 & 0.20 & 0.646 & 0.0037 & 0.0056 & 1.00 & 1 & 0.99 \\ 
\hline
23773 &    0.06~ &  428 & 35.2308 & 3570.982* & 16.75 & 0.45 & 0.35 & 0.372 & 0.0040 & 0.0055 & 1.17 & 0 & 1.08 \\ 
\hline
24195 &    0.13~ &  423 & 36.6221 & 3639.504  & 17.33 & 0.12 & 0.08 & 0.235 & 0.0041 & 0.0047 & 1.05 & 0 & 1.02 \\ 
\hline
24580 & $-$0.06~ &  844 & 31.1688 & 3608.707  & 16.50 & 0.18 & 0.17 & 0.456 & 0.0043 & 0.0027 & 1.19 & 1 & 1.01 \\ 
\hline
24818 & $-$0.04~ &  711 & 21.5495 & 3586.096  & 17.54 & 0.43 & 0.43 & 0.778 & 0.0050 & 0.0027 & 1.18 & 0 & 1.02 \\ 
\hline
24858 &    0.13~ &  397 & 45.0560 & 3623.442  & 17.14 & 0.18 & 0.14 & 0.937 & 0.0029 & 0.0033 & 1.03 & 1 & 1.03 \\ 
\hline
25297 &    0.04~ &  451 & 34.3872 & 3647.584  & 17.44 & 0.29 & 0.28 & 0.225 & 0.0053 & 0.0055 & 1.05 & 0 & 1.03 \\ 
\hline
25578 & $-$0.11~ &  423 & 21.7054 & 3582.818  & 16.08 & 0.54 & 0.44 & 0.298 & 0.0060 & 0.0097 & 1.34 & 2 & 1.01 \\ 
\hline
26109 & $-$0.04~ &  422 & 34.5497 & 3622.492  & 16.55 & 0.57 & 0.54 & 0.864 & 0.0035 & 0.0041 & 1.17 & 1 & 0.99 \\ 
\hline
\end{tabular}
\end{center}

\vspace*{-0.25cm}
~~~~~~~~~~~~~~~*Epoch of primary eclipse minimum $t_{\rm o}$ appropriately adjusted to ensure $\Delta I_1$ $>$ $\Delta I_2$.  

\end{figure*}

}

\afterpage{

\renewcommand{\tabcolsep}{2.4pt}
\renewcommand{\arraystretch}{0.86}
\begin{figure*}[t!]\footnotesize
\vspace*{-0.23cm}
{\small {\bf Table 2:} Physical model properties and statistics for the 130 detached EBs in the well-defined sample. We list the OGLE-III~LMC~EB identification number and the nine physical model properties: orbital period $P$ (days), epoch of primary eclipse $t_{\rm o}$ (JD\,$-$\,2450000), primary and secondary component masses $M_{\rm p}$ and $M_{\rm s}$ (\Msun), age $\tau$ (Myr), inclination $i$ ($^{\rm o}$), eccentricity~$e$, argument of periastron $\omega$ ($^{\rm o}$), and dust extinction $A_I$ (mag).  We then list other physical properties including the mass ratio $q$ = $M_2$/$M_1$ = min\{$M_{\rm p}$,\,$M_{\rm s}$\}/max\{$M_{\rm p}$,$M_{\rm s}$\}, orbital separation $a$ (\Rsun), stellar radii $R_{\rm p}$ and $R_{\rm s}$ (\Rsun), and effective temperatures $T_{\rm p}$ and $T_{\rm s}$ (K).  Finally, we list the fit statistics including the photometric correction factors $f_{\sigma, I}$ and $f_{\sigma, V}$, number of data points  ${\cal N}_I$ and  ${\cal N}_V$, and number of data points we clipped ${\cal N}_{{\rm c},I}$ and ${\cal N}_{{\rm c},V}$ in the I-band and V-band, respectively.}

\vspace*{0.05cm}
\begin{tabular}{|r|c|c|r|r|r|c|c|r|c|c|r|c|c|r|r|c|c|r|r|c|c|c|}
\hline
\multicolumn{1}{|c|}{~} & \multicolumn{9}{c|}{Independent Physical Model Properties}
          & \multicolumn{6}{c|}{Dependent Physical Properties}
          & \multicolumn{7}{c|}{Fit Statistics}                 \\
\hline
     ID~\,  & $P$ & $t_{\rm o}$ & $M_{\rm p}$\, & $M_{\rm s}$\, & 
     $\tau$~~\, & $i$ & $e$ & $\omega$~ & $A_I$ & $q$ & $a$~\, & 
     $R_{\rm p}$ &  $R_{\rm s}$ & $T_{\rm p}$~~ &  $T_{\rm s}$~~ &
     $f_{\sigma, I}$ & $f_{\sigma, V}$ & ${\cal N}_I$\, &  ${\cal N}_V$ & 
     ${\cal N}_{{\rm c},I}$ & ${\cal N}_{{\rm c},V}$ & $\chi^2/\nu$ \\
\hline
   91 & 24.8098 & 3619.171 &  3.7 &  3.7 & 190~\,\,\, & 87.3 & 0.17 & 112 & 0.23 & 
        1.00 &  70 & 4.7 & 5.3 & 12,500 & 11,900 & 
        1.28 & 1.32 &  411 &  34 & 0 & 0 & 1.11 \\
\hline
  170 & 26.3719 & 3566.280 &  9.5 &  3.0 &   2.0  & 87.1 & 0.59 & 165 & 0.37 & 
        0.31 &  86 & 3.6 & 2.2 & 25,700 & 13,100 & 
        1.18 & 1.76 &  426 &  35 & 0 & 1 & 1.02 \\
\hline
  784 & 44.1198 & 3633.792 &  8.4 &  7.5 &   6.5  & 88.3 & 0.66 & 175 & 0.36 & 
        0.89 & 132 & 3.5 & 3.2 & 23,900 & 22,600 & 
        1.18 & 2.12 &  444 &  47 & 2 & 0 & 1.04 \\
\hline
  866 & 28.0132 & 3598.058 &  5.7 &  5.5 &  65~\,\,\, & 89.1 & 0.37 &  91 & 0.35 & 
        0.97 &  87 & 5.3 & 4.6 & 16,500 & 16,900 & 
        1.46 & 1.68 &  424 &  46 & 0 & 0 & 1.07 \\
\hline
 1056 & 30.6714 & 3647.865 &  6.1 &  2.8 &  39~\,\,\, & 89.9 & 0.21 &  89 & 0.32 & 
        0.46 &  86 & 3.8 & 1.8 & 19,300 & 13,000 & 
        1.11 & 1.44 &  444 &  47 & 0 & 0 & 1.05 \\
\hline
 1530 & 42.1990 & 3632.165 &  5.4 &  2.8 &  62~\,\,\, & 88.4 & 0.27 & 333 & 0.34 & 
        0.52 & 103 & 4.2 & 1.8 & 17,200 & 12,700 & 
        1.09 & 1.00 &  465 &  67 & 2 & 0 & 1.02 \\
\hline
 1968 & 44.8939 & 3627.357 &  5.9 &  3.7 &  49~\,\,\, & 88.5 & 0.11 & 158 & 0.45 & 
        0.63 & 113 & 4.2 & 2.3 & 18,300 & 15,100 & 
        1.33 & 1.16 &  911 &  41 & 0 & 0 & 0.99 \\
\hline
 2142 & 27.7691 & 3595.207 & 10.4 &  6.8 &   7.2  & 86.2 & 0.56 & 178 & 0.44 & 
        0.66 & 100 & 4.2 & 3.0 & 26,400 & 21,700 & 
        1.23 & 1.27 &  457 &  45 & 0 & 0 & 1.03 \\
\hline
 2277 & 36.5937 & 3616.510 &  6.3 &  4.8 &  37~\,\,\, & 88.6 & 0.05 & 151 & 0.31 & 
        0.76 & 103 & 3.9 & 2.8 & 19,500 & 17,400 & 
        1.32 & 1.03 &  465 &  67 & 2 & 0 & 1.02 \\
\hline
 2708 & 44.5972 & 3633.673 &  4.0 &  3.3 & 140~\,\,\, & 89.8 & 0.36 & 251 & 0.22 & 
        0.82 & 102 & 4.1 & 2.5 & 13,400 & 13,300 & 
        1.13 & 1.20 &  465 &  67 & 1 & 0 & 1.01 \\
\hline
 2780 & 27.0440 & 3628.753 &  6.3 &  5.0 &  30~\,\,\, & 89.7 & 0.56 & 107 & 0.14 & 
        0.80 &  85 & 3.5 & 2.7 & 19,900 & 17,900 & 
        1.00 & 1.00 &  446 &  45 & 0 & 0 & 1.03 \\
\hline
 3082 & 45.5329 & 3589.236 & 11.0 &  4.8 &   2.8  & 87.7 & 0.34 & 358 & 0.37 & 
        0.44 & 135 & 4.0 & 2.3 & 27,500 & 18,400 & 
        1.38 & 1.70 &  876 &  45 & 2 & 0 & 1.00 \\
\hline
 3177 & 20.2215 & 3585.256 & 10.2 &  2.4 &   8.7  & 89.2 & 0.40 &  99 & 0.27 & 
        0.24 &  73 & 4.3 & 1.6 & 26,000 & 12,200 & 
        1.13 & 1.33 &  434 &  42 & 1 & 1 & 1.02 \\
\hline
 3388 & 44.9699 & 3588.605 & 10.7 &  7.8 &   0.6  & 88.8 & 0.60 & 152 & 0.32 & 
        0.73 & 141 & 3.7 & 3.0 & 27,600 & 23,700 & 
        1.29 & 1.11 &  448 &  48 & 1 & 0 & 1.06 \\
\hline
 3557 & 23.8734 & 3592.036 &  7.8 &  3.3 &  20~\,\,\, & 88.6 & 0.08 & 116 & 0.34 & 
        0.42 &  78 & 4.0 & 1.9 & 22,400 & 14,300 & 
        1.46 & 1.49 &  412 &  32 & 1 & 0 & 1.01 \\
\hline
 4031 & 32.5105 & 3577.365 & 12.0 &  5.6 &   2.0  & 88.2 & 0.54 & 348 & 0.29 & 
        0.47 & 111 & 4.2 & 2.5 & 28,600 & 20,100 & 
        1.31 & 1.33 &  448 &  48 & 0 & 0 & 1.04 \\
\hline
 4399 & 22.9066 & 3624.073 &  5.2 &  3.4 &  83~\,\,\, & 85.9 & 0.23 &  13 & 0.24 & 
        0.64 &  70 & 5.9 & 2.2 & 14,900 & 14,000 & 
        1.03 & 1.00 &  447 &  43 & 2 & 0 & 1.03 \\
\hline
 4419 & 42.0952 & 3575.980 & 11.5 & 11.6 &   6.0  & 89.1 & 0.22 & 127 & 0.58 & 
        0.99 & 145 & 4.5 & 4.5 & 27,600 & 27,800 & 
        1.33 & 1.11 &  434 &  41 & 1 & 0 & 1.02 \\
\hline
 4721 & 21.8791 & 3592.698 &  4.8 &  3.3 &  96~\,\,\, & 87.2 & 0.07 &  77 & 0.13 & 
        0.69 &  66 & 4.8 & 2.3 & 14,800 & 13,800 & 
        1.03 & 1.04 &  456 &  44 & 2 & 0 & 1.03 \\
\hline
 4737 & 31.5717 & 3612.891 &  4.7 &  3.3 &  89~\,\,\, & 89.2 & 0.42 &  60 & 0.32 & 
        0.70 &  84 & 4.0 & 2.2 & 15,600 & 13,800 & 
        1.09 & 1.20 &  440 &  43 & 1 & 0 & 0.99 \\
\hline
 4804 & 23.0376 & 3601.269 & 13.0 &  6.5 &   4.9  & 86.8 & 0.06 & 161 & 0.43 & 
        0.50 &  92 & 4.8 & 2.9 & 29,300 & 21,400 & 
        1.34 & 1.16 &  429 &  41 & 1 & 0 & 1.01 \\
\hline
 4837 & 26.8054 & 3608.320 &  4.7 &  3.7 &  72~\,\,\, & 88.5 & 0.14 & 330 & 0.20 & 
        0.80 &  77 & 3.3 & 2.4 & 16,400 & 14,900 & 
        1.11 & 1.21 &  445 &  41 & 2 & 0 & 1.02 \\
\hline
 5145 & 47.6339 & 3684.852 &  4.6 &  2.8 &  94~\,\,\, & 86.7 & 0.54 & 183 & 0.21 & 
        0.62 & 108 & 3.9 & 1.9 & 15,400 & 12,700 & 
        1.05 & 1.15 &  445 &  41 & 0 & 0 & 1.01 \\
\hline
 5153 & 24.4530 & 3582.859 & 10.6 & 11.2 &   0.8  & 85.5 & 0.55 &   8 & 0.39 & 
        0.95 &  99 & 3.7 & 3.9 & 27,200 & 27,900 & 
        1.09 & 1.05 &  456 &  44 & 2 & 1 & 1.12 \\
\hline
 5195 & 39.4239 & 3564.495 &  4.8 &  2.5 &  92~\,\,\, & 89.6 & 0.20 & 241 & 0.13 & 
        0.52 &  95 & 4.6 & 1.7 & 15,100 & 11,800 & 
        1.12 & 1.16 &  445 &  41 & 2 & 0 & 0.99 \\
\hline
 5492 & 32.9666 & 3602.509 & 12.6 &  5.6 &   6.4  & 89.8 & 0.42 & 144 & 0.41 & 
        0.44 & 114 & 4.9 & 2.6 & 28,700 & 19,600 & 
        1.31 & 1.25 &  429 &  41 & 2 & 0 & 1.09 \\
\hline
 5965 & 20.3374 & 3616.352 &  6.0 &  3.0 &  28~\,\,\, & 88.7 & 0.35 &  36 & 0.21 & 
        0.50 &  65 & 3.3 & 1.9 & 19,600 & 13,600 & 
        1.29 & 1.74 &  439 &  44 & 3 & 0 & 1.00 \\
\hline
 6187 & 29.8703 & 3612.495 &  2.7 &  5.5 &  69~\,\,\, & 88.0 & 0.45 &  83 & 0.23 & 
        0.49 &  82 & 1.8 & 5.3 & 12,600 & 16,200 & 
        1.11 & 1.03 &  477 &  72 & 1 & 0 & 1.03 \\
\hline
 6555 & 29.0549 & 3591.353 &  4.8 &  4.4 &  77~\,\,\, & 89.5 & 0.39 & 307 & 0.21 & 
        0.91 &  83 & 3.7 & 3.1 & 16,200 & 15,900 & 
        1.16 & 1.10 &  477 &  72 & 0 & 0 & 1.02 \\
\hline
 6996 & 29.9282 & 3596.246 &  5.8 &  5.2 &  62~\,\,\, & 89.6 & 0.41 & 101 & 0.21 & 
        0.90 &  90 & 5.1 & 3.7 & 16,900 & 17,200 & 
        1.05 & 2.09 &  476 &  72 & 2 & 0 & 1.04 \\
\hline
 7380 & 31.4810 & 3597.752 &  5.8 &  5.2 &  38~\,\,\, & 87.8 & 0.16 &  51 & 0.33 & 
        0.89 &  93 & 3.4 & 3.0 & 18,900 & 18,000 & 
        1.13 & 1.43 &  468 &  72 & 2 & 0 & 1.02 \\
\hline
 7560 & 26.7462 & 3611.776 & 10.5 &  9.8 &   0.8  & 88.0 & 0.16 &  42 & 0.45 & 
        0.93 & 102 & 3.7 & 3.5 & 27,400 & 26,400 & 
        1.21 & 1.69 &  418 &  42 & 2 & 0 & 1.14 \\
\hline
 7565 & 29.0954 & 3601.066 &  4.2 &  2.8 & 130~\,\,\, & 88.5 & 0.42 & 307 & 0.36 & 
        0.67 &  76 & 4.9 & 2.0 & 13,200 & 12,500 & 
        1.08 & 1.30 &  477 &  72 & 0 & 1 & 0.98 \\
\hline
 7935 & 24.0616 & 3609.906 &  6.0 &  3.8 &  36~\,\,\, & 88.3 & 0.10 & 141 & 0.28 & 
        0.64 &  75 & 3.5 & 2.3 & 19,300 & 15,400 & 
        1.13 & 1.75 &  404 &  38 & 1 & 0 & 1.06 \\
\hline
 7975 & 20.1227 & 3615.519 &  2.8 &  8.0 &  21~\,\,\, & 87.1 & 0.46 &  71 & 0.28 & 
        0.35 &  69 & 1.8 & 4.3 & 13,200 & 22,500 & 
        1.11 & 1.29 &  477 &  72 & 2 & 0 & 1.05 \\
\hline
 8543 & 25.3381 & 3621.308 &  5.3 &  4.1 &  68~\,\,\, & 89.6 & 0.16 & 210 & 0.25 & 
        0.78 &  77 & 4.4 & 2.7 & 16,800 & 15,700 & 
        1.68 & 1.41 &  451 &  41 & 2 & 0 & 1.02 \\
\hline
 8559 & 42.2644 & 3631.167 &  5.8 &  4.4 &  20~\,\,\, & 89.6 & 0.52 & 140 & 0.23 & 
        0.77 & 111 & 2.9 & 2.4 & 19,400 & 16,900 & 
        1.13 & 1.28 &  468 &  72 & 2 & 0 & 1.06 \\
\hline
 8783 & 48.7240 & 3622.451 &  5.5 &  2.6 &  62~\,\,\, & 88.3 & 0.55 & 177 & 0.37 & 
        0.48 & 113 & 4.2 & 1.8 & 17,200 & 12,300 & 
        1.02 & 1.34 &  458 &  41 & 1 & 0 & 1.05 \\
\hline
 8903 & 25.6604 & 3637.910 &  7.5 &  7.6 &  19~\,\,\, & 87.3 & 0.15 & 359 & 0.28 & 
        1.00 &  90 & 3.8 & 3.8 & 22,200 & 22,200 & 
        1.89 & 2.05 &  463 &  66 & 1 & 0 & 1.03 \\
\hline
 8993 & 29.4599 & 3599.449 &  6.5 &  4.7 &  25~\,\,\, & 89.6 & 0.48 &  70 & 0.18 & 
        0.73 &  90 & 3.5 & 2.6 & 20,400 & 17,400 & 
        1.03 & 1.36 &  891 &  41 & 2 & 0 & 1.00 \\
\hline
 9159 & 48.1568 & 3695.292 &  4.3 &  3.5 & 120~\,\,\, & 89.4 & 0.24 &  37 & 0.25 & 
        0.81 & 110 & 4.4 & 2.6 & 14,100 & 13,900 & 
        1.16 & 1.26 &  466 &  40 & 0 & 1 & 1.03 \\
\hline
 9386 & 29.8063 & 3619.568 & 10.7 &  5.1 &   5.3  & 88.9 & 0.42 &  49 & 0.32 & 
        0.48 & 101 & 4.2 & 2.4 & 26,800 & 18,800 & 
        1.21 & 1.22 &  559 & 115 & 1 & 2 & 1.01 \\
\hline
 9429 & 20.6681 & 3605.551 &  4.6 &  2.3 & 120~\,\,\, & 88.1 & 0.31 &  81 & 0.33 & 
        0.50 &  60 & 5.7 & 1.6 & 13,600 & 11,100 & 
        1.02 & 1.52 &  476 &  68 & 1 & 0 & 1.03 \\
\hline
 9441 & 20.8997 & 3596.647 &  6.4 &  2.5 &  34~\,\,\, & 89.9 & 0.19 & 125 & 0.39 & 
        0.39 &  66 & 3.8 & 1.7 & 19,900 & 12,300 & 
        1.09 & 1.59 &  476 &  68 & 0 & 1 & 1.01 \\
\hline
 9953 & 45.5780 & 3730.003 &  6.9 &  6.6 &   3.0  & 87.1 & 0.61 & 168 & 0.21 & 
        0.95 & 128 & 2.9 & 2.8 & 22,200 & 21,500 & 
        1.12 & 1.54 &  490 &  63 & 2 & 1 & 1.04 \\
\hline
10096 & 35.2251 & 3628.935 &  4.5 &  2.7 & 100~\,\,\, & 89.9 & 0.31 & 158 & 0.27 & 
        0.62 &  87 & 4.1 & 1.9 & 14,900 & 12,500 & 
        1.20 & 1.48 &  431 &  46 & 1 & 0 & 1.00 \\
\hline
10422 & 21.1695 & 3586.712 &  6.5 &  8.1 &   0.6  & 85.7 & 0.57 &   9 & 0.22 & 
        0.81 &  79 & 2.7 & 3.1 & 22,000 & 24,100 & 
        1.26 & 1.00 &  612 &  38 & 2 & 0 & 1.13 \\
\hline
10575 & 29.1431 & 3592.096 &  6.4 &  2.9 &  38~\,\,\, & 89.1 & 0.12 & 307 & 0.22 & 
        0.45 &  84 & 4.1 & 1.8 & 19,600 & 13,100 & 
        1.04 & 1.15 &  559 & 115 & 3 & 0 & 1.02 \\
\hline
10953 & 42.2762 & 3610.406 &  7.9 &  3.1 &  31~\,\,\, & 88.3 & 0.41 & 203 & 0.15 & 
        0.39 & 113 & 5.8 & 1.9 & 20,500 & 13,700 & 
        1.34 & 1.00 &  616 &  38 & 1 & 0 & 0.99 \\
\hline
11252 & 22.2402 & 3592.679 &  4.6 &  4.2 &  88~\,\,\, & 89.6 & 0.13 & 268 & 0.17 & 
        0.92 &  69 & 3.7 & 3.1 & 15,600 & 15,500 & 
        1.15 & 1.53 &  490 &  63 & 1 & 1 & 0.97 \\
\hline
11299 & 39.5635 & 3584.917 &  4.0 &  3.6 & 150~\,\,\, & 89.9 & 0.47 & 177 & 0.30 & 
        0.91 &  96 & 4.8 & 3.2 & 12,800 & 13,600 & 
        1.29 & 1.18 & 1049 &  41 & 1 & 0 & 1.03 \\
\hline
11526 & 49.2822 & 3595.336 &  4.5 &  6.5 &  47~\,\,\, & 88.3 & 0.15 & 150 & 0.28 & 
        0.69 & 125 & 2.7 & 5.2 & 16,600 & 18,300 & 
        1.13 & 1.12 &  477 &  61 & 1 & 0 & 1.06 \\
\hline
11538 & 33.7813 & 3600.922 & 10.4 &  9.3 &   1.3  & 88.8 & 0.67 &  90 & 0.20 & 
        0.89 & 118 & 3.7 & 3.5 & 26,900 & 25,500 & 
        1.18 & 1.38 &  477 &  61 & 2 & 0 & 1.05 \\
\hline
11636 & 31.1511 & 3579.505 &  4.0 &  2.8 & 140~\,\,\, & 90.0 & 0.23 &  40 & 0.29 & 
        0.69 &  79 & 4.4 & 2.0 & 13,300 & 12,400 & 
        1.13 & 1.34 &  474 &  60 & 1 & 0 & 1.04 \\
\hline
11907 & 45.4194 & 3629.746 &  3.6 &  4.8 &  86~\,\,\, & 88.5 & 0.25 &  56 & 0.16 & 
        0.75 & 109 & 2.4 & 4.2 & 14,500 & 15,700 & 
        1.17 & 1.27 &  440 &  41 & 0 & 0 & 1.01 \\
\hline
12170 & 29.1745 & 3586.504 &  4.0 &  3.1 & 120~\,\,\, & 89.6 & 0.17 & 237 & 0.13 & 
        0.78 &  77 & 3.6 & 2.3 & 14,300 & 13,200 & 
        1.09 & 1.21 &  452 &  88 & 0 & 0 & 1.01 \\
\hline
12179 & 25.1022 & 3555.832 &  5.5 &  4.3 &  67~\,\,\, & 88.7 & 0.46 &  92 & 0.23 & 
        0.77 &  78 & 4.9 & 2.8 & 16,600 & 16,000 & 
        1.14 & 1.77 &  477 &  61 & 0 & 1 & 1.04 \\
\hline
12384 & 25.1787 & 3591.907 &  3.9 &  2.3 & 170~\,\,\, & 87.0 & 0.41 & 193 & 0.29 & 
        0.60 &  67 & 6.4 & 1.8 & 11,600 & 11,200 & 
        1.14 & 1.18 &  476 &  61 & 2 & 0 & 1.03 \\
\hline
12454 & 30.1522 & 3591.701 &  7.1 &  3.3 &  24~\,\,\, & 88.1 & 0.29 & 323 & 0.12 & 
        0.47 &  89 & 3.8 & 2.0 & 21,300 & 14,300 & 
        1.11 & 1.37 &  476 &  61 & 3 & 0 & 1.03 \\
\hline
12832 & 20.1669 & 3551.657 &  7.7 &  8.2 &  16~\,\,\, & 85.8 & 0.12 & 170 & 0.25 & 
        0.94 &  78 & 3.7 & 4.0 & 22,600 & 23,100 & 
        1.30 & 1.21 &  493 &  90 & 0 & 0 & 1.03 \\
\hline
13177 & 40.5163 & 3542.688 &  7.3 &  3.7 &  21~\,\,\, & 88.4 & 0.55 &  24 & 0.29 & 
        0.50 & 111 & 3.8 & 2.1 & 21,700 & 15,300 & 
        1.17 & 1.32 &  477 &  61 & 2 & 0 & 1.02 \\
\hline
13260 & 24.2959 & 3593.154 &  4.1 &  3.2 & 140~\,\,\, & 89.9 & 0.20 & 246 & 0.17 & 
        0.78 &  69 & 4.9 & 2.4 & 13,000 & 13,200 & 
        1.07 & 1.51 &  450 &  56 & 0 & 0 & 1.00 \\
\hline
\end{tabular}
\end{figure*}

\begin{figure*}[p]\footnotesize
\vspace*{-0.23cm}
{\small {\bf Table 2 (cont.):}} \\
\vspace*{-0.1cm}
\begin{tabular}{|r|c|c|r|r|r|c|c|r|c|c|r|c|c|r|r|c|c|r|r|c|c|c|}
\hline
\multicolumn{1}{|c|}{~} & \multicolumn{9}{c|}{Independent Physical Model Properties}
          & \multicolumn{6}{c|}{Dependent Physical Properties}
          & \multicolumn{7}{c|}{Fit Statistics}                 \\
\hline
     ID~\,  & $P$ & $t_{\rm o}$ & $M_{\rm p}$\, & $M_{\rm s}$\, & 
     $\tau$~~\, & $i$ & $e$ & $\omega$~ & $A_I$ & $q$ & $a$~\, & 
     $R_{\rm p}$ &  $R_{\rm s}$ & $T_{\rm p}$~~ &  $T_{\rm s}$~~ &
     $f_{\sigma, I}$ & $f_{\sigma, V}$ & ${\cal N}_I$\, &  ${\cal N}_V$ & 
     ${\cal N}_{{\rm c},I}$ & ${\cal N}_{{\rm c},V}$ & $\chi^2/\nu$ \\
\hline
13390 & 35.7853 & 3571.619 &  4.0 &  5.2 &  53~\,\,\, & 88.5 & 0.41 &  62 & 0.41 & 
        0.77 &  96 & 2.5 & 3.5 & 15,700 & 17,700 & 
        1.15 & 1.26 &  477 &  61 & 1 & 0 & 1.01 \\
\hline
13418 & 26.9107 & 3587.265 &  7.9 &  6.5 &   0.5  & 87.6 & 0.71 &   7 & 0.21 & 
        0.83 &  92 & 3.1 & 2.7 & 23,900 & 22,000 & 
        1.10 & 1.14 &  465 &  61 & 1 & 0 & 1.13 \\
\hline
13441 & 30.6531 & 3593.346 &  5.7 &  3.7 &  51~\,\,\, & 87.8 & 0.19 &   6 & 0.22 & 
        0.65 &  87 & 3.9 & 2.3 & 18,100 & 15,000 & 
        1.24 & 1.48 &  493 &  90 & 1 & 0 & 1.03 \\
\hline
13482 & 21.5789 & 3548.903 &  4.9 &  4.7 &  55~\,\,\, & 88.5 & 0.26 &  17 & 0.16 & 
        0.97 &  69 & 3.1 & 3.0 & 17,100 & 17,000 & 
        1.10 & 1.21 &  493 &  90 & 2 & 0 & 1.01 \\
\hline
13491 & 21.3014 & 3578.037 &  6.2 &  5.1 &  51~\,\,\, & 87.6 & 0.02 &   0 & 0.43 & 
        0.83 &  72 & 5.0 & 3.2 & 18,000 & 17,600 & 
        1.10 & 1.09 &  477 &  61 & 2 & 1 & 1.02 \\
\hline
13726 & 20.9342 & 3554.047 &  8.7 &  2.0 &   5.3  & 89.5 & 0.32 &  30 & 0.51 & 
        0.23 &  70 & 3.6 & 1.7 & 24,400 &  9,900 & 
        1.13 & 1.29 &  493 &  90 & 2 & 0 & 1.03 \\
\hline
13867 & 21.7458 & 3604.872 &  6.6 &  4.3 &  21~\,\,\, & 88.1 & 0.06 & 101 & 0.30 & 
        0.65 &  73 & 3.4 & 2.3 & 20,700 & 16,600 & 
        1.65 & 1.89 &  831 &  40 & 1 & 0 & 1.00 \\
\hline
14171 & 21.7291 & 3621.788 & 10.6 &  2.4 &   3.6  & 88.0 & 0.10 &  66 & 0.41 & 
        0.23 &  77 & 4.0 & 1.6 & 26,900 & 12,500 & 
        1.05 & 1.21 &  457 &  39 & 2 & 0 & 0.99 \\
\hline
14360 & 30.2614 & 3651.358 &  4.1 &  8.4 &  29~\,\,\, & 87.6 & 0.35 &  66 & 0.32 & 
        0.48 &  95 & 2.3 & 6.4 & 16,000 & 20,800 & 
        1.00 & 1.32 &  540 & 136 & 1 & 0 & 1.02 \\
\hline
14895 & 34.4644 & 3574.342 &  6.2 &  6.4 &   7.5  & 89.5 & 0.52 & 186 & 0.39 & 
        0.97 & 104 & 2.9 & 2.9 & 20,600 & 21,000 & 
        1.31 & 1.41 &  506 & 131 & 0 & 0 & 1.01 \\
\hline
15235 & 28.9499 & 3553.094 &  5.9 &  4.2 &  40~\,\,\, & 88.3 & 0.29 & 174 & 0.19 & 
        0.70 &  86 & 3.7 & 2.4 & 19,000 & 16,100 & 
        1.08 & 1.52 &  726 &  99 & 2 & 0 & 1.04 \\
\hline
15244 & 21.8434 & 3622.893 &  5.1 &  3.6 &  70~\,\,\, & 89.9 & 0.32 & 175 & 0.33 & 
        0.70 &  68 & 4.0 & 2.3 & 16,700 & 14,600 & 
        1.02 & 1.19 &  446 &  38 & 2 & 1 & 1.05 \\
\hline
15380 & 27.6171 & 3624.681 &  6.3 &  8.5 &   4.5  & 85.9 & 0.53 &  25 & 0.24 & 
        0.74 &  94 & 2.8 & 3.5 & 20,900 & 24,300 & 
        1.09 & 2.08 &  325 &  82 & 0 & 0 & 1.08 \\
\hline
15788 & 29.0140 & 3561.932 &  7.2 &  6.6 &   2.4  & 86.9 & 0.41 & 181 & 0.26 & 
        0.91 &  95 & 3.0 & 2.8 & 22,700 & 21,700 & 
        1.27 & 1.18 &  600 &  40 & 3 & 0 & 1.01 \\
\hline
15979 & 28.1428 & 3635.584 & 10.0 &  5.3 &  13~\,\,\, & 88.1 & 0.45 & 296 & 0.23 & 
        0.53 &  97 & 4.8 & 2.6 & 25,300 & 18,800 & 
        1.09 & 1.30 &  449 &  38 & 1 & 0 & 1.03 \\
\hline
16026 & 30.8481 & 3566.097 &  6.2 &  5.9 &  49~\,\,\, & 88.7 & 0.48 &  61 & 0.20 & 
        0.94 &  95 & 4.8 & 4.1 & 18,300 & 18,300 & 
        1.00 & 1.24 &  449 &  38 & 0 & 0 & 1.02 \\
\hline
16126 & 22.2111 & 3609.527 &  5.0 &  4.9 &  45~\,\,\, & 86.5 & 0.50 & 146 & 0.33 & 
        0.98 &  71 & 3.0 & 2.9 & 17,600 & 17,400 & 
        1.58 & 1.00 &  600 &  40 & 1 & 1 & 0.98 \\
\hline
16350 & 30.9797 & 3603.341 &  7.1 &  5.5 &  11~\,\,\, & 88.5 & 0.45 &  30 & 0.23 & 
        0.77 &  97 & 3.3 & 2.7 & 21,900 & 19,200 & 
        1.16 & 1.30 &  599 &  40 & 0 & 0 & 1.05 \\
\hline
16399 & 39.1462 & 3653.058 &  7.7 &  3.9 &   1.4  & 88.2 & 0.63 &  29 & 0.37 & 
        0.51 & 110 & 3.1 & 2.0 & 23,400 & 16,400 & 
        1.00 & 1.08 &  456 &  39 & 0 & 0 & 1.07 \\
\hline
16418 & 20.4167 & 3586.960 &  5.6 &  2.2 &  64~\,\,\, & 87.0 & 0.19 &  17 & 0.29 & 
        0.39 &  63 & 5.0 & 1.6 & 16,800 & 11,200 & 
        1.05 & 1.23 &  482 & 128 & 0 & 3 & 0.99 \\
\hline
16711 & 26.5578 & 3632.442 & 10.5 & 10.7 &   0.5  & 86.4 & 0.57 &  18 & 0.28 & 
        0.98 & 104 & 3.7 & 3.7 & 27,400 & 27,600 & 
        1.02 & 1.67 &  449 &  38 & 3 & 0 & 1.10 \\
\hline
16964 & 23.5422 & 3603.148 & 12.7 &  3.7 &   2.9  & 85.1 & 0.63 &  29 & 0.33 & 
        0.29 &  88 & 4.5 & 2.0 & 29,200 & 15,700 & 
        1.08 & 1.28 &  606 &  47 & 0 & 0 & 1.07 \\
\hline
17067 & 28.1151 & 3592.517 &  4.8 &  8.4 &  17~\,\,\, & 87.2 & 0.43 &  55 & 0.16 & 
        0.56 &  92 & 2.5 & 4.3 & 17,700 & 23,300 & 
        1.24 & 1.25 &  581 &  36 & 0 & 0 & 1.03 \\
\hline
17316 & 26.7748 & 3611.388 & 10.4 &  6.1 &   0.6  & 87.5 & 0.51 & 328 & 0.25 & 
        0.59 &  96 & 3.6 & 2.6 & 27,200 & 21,200 & 
        1.00 & 1.10 &  986 &  39 & 2 & 0 & 1.06 \\
\hline
17361 & 20.9317 & 3562.007 &  3.6 &  3.0 & 190~\,\,\, & 89.8 & 0.31 & 284 & 0.19 & 
        0.83 &  60 & 4.5 & 2.5 & 12,100 & 12,500 & 
        1.00 & 1.04 &  472 &  54 & 0 & 0 & 0.98 \\
\hline
17539 & 23.9095 & 3562.793 &  5.0 &  3.5 &  71~\,\,\, & 88.9 & 0.13 & 110 & 0.34 & 
        0.70 &  71 & 3.8 & 2.3 & 16,600 & 14,400 & 
        1.04 & 1.24 &  473 &  54 & 2 & 0 & 1.05 \\
\hline
17569 & 22.2016 & 3590.999 &  9.0 &  4.4 &   4.8  & 88.5 & 0.19 & 281 & 0.27 & 
        0.48 &  79 & 3.6 & 2.2 & 24,900 & 17,300 & 
        1.18 & 1.06 &  626 &  66 & 2 & 1 & 1.11 \\
\hline
17750 & 20.7084 & 3586.844 &  5.9 &  2.0 &  55~\,\,\, & 86.6 & 0.41 &   9 & 0.11 & 
        0.34 &  63 & 4.7 & 1.5 & 17,700 & 10,500 & 
        1.10 & 1.20 &  473 &  54 & 1 & 0 & 0.99 \\
\hline
17784 & 25.2611 & 3575.386 &  8.0 &  3.6 &   8.8  & 87.2 & 0.36 & 189 & 0.24 & 
        0.44 &  82 & 3.5 & 1.9 & 23,300 & 15,100 & 
        1.28 & 1.50 &  626 &  66 & 1 & 0 & 1.02 \\
\hline
17822 & 44.2711 & 3641.934 &  5.3 &  3.1 &  84~\,\,\, & 88.3 & 0.25 & 299 & 0.30 & 
        0.59 & 107 & 6.2 & 2.1 & 14,800 & 13,500 & 
        1.09 & 1.29 &  437 &  45 & 1 & 0 & 1.02 \\
\hline
18237 & 20.3284 & 3616.374 &  7.4 &  4.3 &  40~\,\,\, & 88.3 & 0.10 & 254 & 0.15 & 
        0.58 &  71 & 6.9 & 2.5 & 18,400 & 16,500 & 
        1.63 & 1.18 &  588 &  70 & 1 & 0 & 0.99 \\
\hline
18582 & 41.9853 & 3608.319 &  6.5 &  4.6 &  48~\,\,\, & 90.0 & 0.15 &  66 & 0.32 & 
        0.71 & 113 & 5.6 & 2.8 & 18,000 & 16,900 & 
        1.09 & 1.16 &  441 &  54 & 2 & 0 & 1.07 \\
\hline
18659 & 21.4450 & 3594.449 &  5.9 &  4.0 &  51~\,\,\, & 87.8 & 0.35 & 317 & 0.31 & 
        0.67 &  70 & 4.2 & 2.4 & 18,200 & 15,600 & 
        1.19 & 1.61 &  456 &  50 & 1 & 0 & 1.06 \\
\hline
18813 & 39.6988 & 3627.901 &  4.4 &  4.3 & 120~\,\,\, & 88.9 & 0.32 &  91 & 0.34 & 
        0.98 & 101 & 4.8 & 4.2 & 13,900 & 14,200 & 
        1.22 & 1.24 &  605 &  47 & 0 & 0 & 1.05 \\
\hline
18824 & 33.5722 & 3576.497 &  9.2 &  3.0 &  17~\,\,\, & 89.8 & 0.23 & 115 & 0.27 & 
        0.32 & 101 & 4.7 & 1.8 & 24,000 & 13,600 & 
        1.13 & 1.23 &  626 &  66 & 1 & 0 & 1.03 \\
\hline
18839 & 48.2133 & 3619.339 &  5.0 &  3.7 &  72~\,\,\, & 88.2 & 0.37 & 182 & 0.18 & 
        0.74 & 115 & 3.9 & 2.4 & 16,500 & 14,800 & 
        1.00 & 1.02 &  473 &  54 & 1 & 0 & 0.93 \\
\hline
18859 & 24.5592 & 3624.012 &  4.8 &  4.7 &  81~\,\,\, & 88.5 & 0.36 &  89 & 0.24 & 
        0.97 &  75 & 3.9 & 3.6 & 16,000 & 16,000 & 
        1.08 & 1.36 &  435 &  45 & 0 & 0 & 1.06 \\
\hline
18869 & 42.5786 & 3628.596 &  9.0 &  2.5 &   9.9  & 88.6 & 0.35 & 310 & 0.27 & 
        0.28 & 116 & 3.9 & 1.6 & 24,500 & 12,500 & 
        1.17 & 1.50 &  572 &  43 & 0 & 0 & 0.99 \\
\hline
19083 & 22.2720 & 3600.517 &  9.6 &  2.6 &  11~\,\,\, & 89.8 & 0.03 & 345 & 0.43 & 
        0.27 &  77 & 4.3 & 1.6 & 25,200 & 12,700 & 
        1.26 & 1.39 &  601 &  70 & 3 & 0 & 1.06 \\
\hline
19230 & 37.4005 & 3569.103 &  6.5 &  4.3 &  44~\,\,\, & 86.5 & 0.35 & 178 & 0.36 & 
        0.66 & 104 & 4.8 & 2.5 & 18,800 & 16,300 & 
        1.03 & 1.12 &  473 &  54 & 0 & 0 & 0.98 \\
\hline
19792 & 29.6439 & 3640.471 & 13.9 &  5.6 &   5.5  & 87.4 & 0.07 & 294 & 0.34 & 
        0.40 & 109 & 5.2 & 2.6 & 30,000 & 19,600 & 
        1.52 & 1.43 &  912 &  55 & 0 & 0 & 1.01 \\
\hline
19840 & 30.9701 & 3617.984 & 13.3 &  2.7 &   8.6  & 89.6 & 0.27 & 253 & 0.50 & 
        0.20 & 105 & 5.7 & 1.7 & 28,700 & 13,100 & 
        1.26 & 1.16 &  625 &  66 & 1 & 0 & 1.03 \\
\hline
20309 & 48.2395 & 3694.901 &  3.9 &  3.2 & 140~\,\,\, & 89.9 & 0.11 & 143 & 0.16 & 
        0.81 & 107 & 3.7 & 2.4 & 13,800 & 13,300 & 
        1.00 & 1.00 &  428 &  40 & 0 & 0 & 1.01 \\
\hline
20459 & 30.7410 & 3642.410 &  6.7 &  3.6 &  19~\,\,\, & 89.3 & 0.18 &  37 & 0.10 & 
        0.53 &  90 & 3.4 & 2.0 & 20,900 & 15,000 & 
        1.06 & 1.09 &  436 &  45 & 2 & 0 & 1.01 \\
\hline
20522 & 27.2573 & 3619.602 &  4.2 &  2.6 & 130~\,\,\, & 88.6 & 0.14 & 117 & 0.28 & 
        0.62 &  72 & 4.9 & 1.9 & 13,400 & 12,000 & 
        1.14 & 1.52 &  423 &  41 & 1 & 0 & 1.02 \\
\hline
20590 & 40.0940 & 3618.390 &  3.9 &  2.6 & 170~\,\,\, & 88.7 & 0.11 & 336 & 0.23 & 
        0.68 &  92 & 5.3 & 2.0 & 12,100 & 12,000 & 
        1.07 & 1.07 &  428 &  41 & 0 & 0 & 1.05 \\
\hline
20646 & 26.0463 & 3620.168 &  5.0 &  3.5 &  89~\,\,\, & 86.9 & 0.25 & 358 & 0.25 & 
        0.69 &  75 & 5.5 & 2.4 & 14,800 & 14,200 & 
        1.09 & 1.53 &  437 &  42 & 0 & 0 & 1.00 \\
\hline
20746 & 28.3718 & 3614.202 &  5.8 &  5.3 &  60~\,\,\, & 89.7 & 0.30 &  68 & 0.39 & 
        0.91 &  87 & 5.1 & 3.8 & 17,000 & 17,300 & 
        1.03 & 1.00 &  436 &  45 & 0 & 0 & 1.10 \\
\hline
21059 & 25.7903 & 3611.139 &  5.5 &  4.1 &  44~\,\,\, & 87.4 & 0.28 &   9 & 0.38 & 
        0.75 &  78 & 3.4 & 2.4 & 18,300 & 16,000 & 
        1.06 & 1.60 &  428 &  41 & 0 & 0 & 0.98 \\
\hline
21518 & 21.4601 & 3629.581 &  4.7 &  3.9 &  95~\,\,\, & 89.7 & 0.09 & 174 & 0.24 & 
        0.81 &  67 & 4.6 & 2.7 & 15,000 & 14,900 & 
        1.05 & 1.00 &  433 &  42 & 3 & 0 & 1.10 \\
\hline
21621 & 38.5318 & 3576.485 &  4.8 &  4.2 &  86~\,\,\, & 88.9 & 0.44 & 122 & 0.28 & 
        0.88 & 100 & 4.1 & 3.0 & 15,700 & 15,500 & 
        1.00 & 1.39 &  444 &  42 & 2 & 0 & 0.98 \\
\hline
21881 & 21.0055 & 3588.141 &  5.6 &  6.8 &  36~\,\,\, & 86.6 & 0.30 & 140 & 0.38 & 
        0.83 &  74 & 3.3 & 4.4 & 18,800 & 20,000 & 
        1.10 & 1.29 &  428 &  41 & 0 & 0 & 1.04 \\
\hline
22082 & 33.1143 & 3586.180 & 11.2 &  2.7 &   3.7  & 89.0 & 0.41 & 234 & 0.37 & 
        0.24 & 104 & 4.2 & 1.6 & 27,600 & 13,100 & 
        1.08 & 1.50 &  436 &  45 & 0 & 0 & 1.02 \\
\hline
22553 & 22.8430 & 3598.817 &  6.8 &  5.2 &  12~\,\,\, & 87.5 & 0.52 &  17 & 0.34 & 
        0.76 &  78 & 3.2 & 2.6 & 21,400 & 18,600 & 
        1.00 & 1.62 &  419 &  36 & 2 & 0 & 0.97 \\
\hline
22691 & 31.8866 & 3579.269 & 10.6 &  4.7 &   0.7  & 88.3 & 0.63 & 126 & 0.41 & 
        0.44 & 105 & 3.7 & 2.5 & 27,400 & 18,400 & 
        1.26 & 2.26 &  434 &  37 & 0 & 0 & 1.11 \\
\hline
22713 & 33.3752 & 3617.560 &  7.3 &  3.6 &  24~\,\,\, & 87.3 & 0.29 & 352 & 0.20 & 
        0.49 &  97 & 3.9 & 2.1 & 21,500 & 15,000 & 
        1.11 & 1.27 &  437 &  42 & 1 & 0 & 1.03 \\
\hline
22764 & 29.7048 & 3572.591 &  9.8 &  9.0 &   0.6  & 88.8 & 0.59 & 162 & 0.44 & 
        0.91 & 107 & 3.5 & 3.3 & 26,500 & 25,400 & 
        1.13 & 1.26 &  428 &  37 & 1 & 0 & 1.12 \\
\hline
23088 & 22.4310 & 3598.932 &  5.6 &  2.5 &  59~\,\,\, & 89.6 & 0.02 & 110 & 0.39 & 
        0.45 &  67 & 4.3 & 1.7 & 17,400 & 11,900 & 
        1.09 & 1.15 &  424 &  40 & 0 & 0 & 1.04 \\
\hline
23101 & 46.3827 & 3616.722 & 12.0 &  3.9 &   1.0  & 89.7 & 0.68 & 128 & 0.55 & 
        0.32 & 137 & 4.1 & 2.8 & 28,900 & 15,000 & 
        1.02 & 1.39 &  427 &  37 & 2 & 0 & 1.04 \\
\hline
23368 & 22.9011 & 3629.355 &  5.1 &  6.8 &  16~\,\,\, & 87.6 & 0.29 &  39 & 0.37 & 
        0.74 &  77 & 2.6 & 3.3 & 18,200 & 21,200 & 
        1.00 & 1.00 &  426 &  37 & 1 & 0 & 1.00 \\
\hline
23773 & 35.2309 & 3570.983 &  7.3 &  5.5 &  31~\,\,\, & 89.1 & 0.25 & 144 & 0.40 & 
        0.75 & 106 & 4.6 & 3.0 & 20,800 & 18,700 & 
        1.17 & 1.91 &  428 &  37 & 0 & 0 & 1.05 \\
\hline
24195 & 36.6224 & 3639.498 &  5.0 &  2.8 &  87~\,\,\, & 86.2 & 0.43 & 173 & 0.41 & 
        0.57 &  92 & 4.9 & 1.9 & 15,200 & 12,800 & 
        1.05 & 1.41 &  423 &  37 & 0 & 0 & 1.02 \\
\hline
24580 & 31.1688 & 3608.710 & 12.5 &  5.4 &   1.6  & 88.3 & 0.22 & 252 & 0.39 & 
        0.43 & 109 & 4.3 & 2.5 & 29,300 & 19,700 & 
        1.19 & 1.29 &  844 &  31 & 1 & 0 & 1.02 \\
\hline
24818 & 21.5492 & 3586.112 &  6.7 &  4.7 &   0.9  & 88.3 & 0.48 & 337 & 0.31 & 
        0.71 &  73 & 2.8 & 2.2 & 22,200 & 18,400 & 
        1.18 & 2.07 &  711 &  41 & 0 & 0 & 1.08 \\
\hline
24858 & 45.0564 & 3623.441 &  8.2 &  6.5 &   3.2  & 83.6 & 0.77 &   4 & 0.55 & 
        0.79 & 131 & 3.3 & 2.8 & 23,900 & 21,500 & 
        1.03 & 1.19 &  397 &  48 & 1 & 0 & 1.03 \\
\hline
25297 & 34.3873 & 3647.587 &  4.1 &  3.3 & 130~\,\,\, & 87.7 & 0.45 & 182 & 0.26 & 
        0.79 &  87 & 4.3 & 2.5 & 13,600 & 13,400 & 
        1.05 & 1.19 &  451 &  53 & 0 & 0 & 1.02 \\
\hline
25578 & 21.7055 & 3582.825 &  7.5 &  7.0 &  32~\,\,\, & 88.3 & 0.37 & 150 & 0.18 & 
        0.93 &  80 & 5.1 & 4.3 & 20,600 & 20,500 & 
        1.34 & 1.78 &  423 &  33 & 2 & 0 & 1.09 \\
\hline
26109 & 34.5499 & 3622.487 &  8.5 &  7.5 &  10~\,\,\, & 88.6 & 0.62 &   9 & 0.32 & 
        0.88 & 113 & 3.8 & 3.4 & 23,900 & 22,500 & 
        1.17 & 1.27 &  422 &  35 & 1 & 0 & 1.04 \\
\hline
\end{tabular}
\end{figure*}

}

\end{document}